\pdfoutput=1
\documentclass[plaindraft]{jpp-AAS}
\usepackage{graphicx}

\usepackage[utf8]{inputenc}
\usepackage[T1]{fontenc}
\usepackage{amsmath}
\usepackage{tabularx}
\usepackage{siunitx}
\usepackage{mathtools}
\usepackage{tikz}
\usetikzlibrary{patterns,arrows,decorations.pathreplacing}
\usetikzlibrary{avoidpath}
\usetikzlibrary{decorations.markings}
\usepackage{hyperref}
\hypersetup{
    colorlinks=true,
    linkcolor=blue,
    urlcolor=blue,
    citecolor=blue
    }
\usepackage[parfill]{parskip}
\usepackage{caption}
\usepackage{subcaption}
\usepackage{upmath}
\usepackage{orcidlink}
\usepackage[version=4]{mhchem}
\usepackage{comment}

\def\Xint#1{\mathchoice
   {\XXint\displaystyle\textstyle{#1}}%
   {\XXint\textstyle\scriptstyle{#1}}%
   {\XXint\scriptstyle\scriptscriptstyle{#1}}%
   {\XXint\scriptscriptstyle\scriptscriptstyle{#1}}%
   \!\int}
\def\XXint#1#2#3{{\setbox0=\hbox{$#1{#2#3}{\int}$}
     \vcenter{\hbox{$#2#3$}}\kern-.5\wd0}}

\def\dashint{\Xint-}

\ifCUPmtlplainloaded \else
  \checkfont{eurm10}
  \iffontfound
    \IfFileExists{upmath.sty}
      {\typeout{^^JFound AMS Euler Roman fonts on the system,
                   using the 'upmath' package.^^J}%
       \usepackage{upmath}}
      {\typeout{^^JFound AMS Euler Roman fonts on the system, but you
                   dont seem to have the}%
       \typeout{'upmath' package installed. JPP.cls can take advantage
                 of these fonts, if you use 'upmath' package.^^J}%
       \providecommand\upi{\pi}%
      }
  \else
    \providecommand\upi{\pi}%
  \fi
\fi

\ifCUPmtlplainloaded \else
  \checkfont{msam10}
  \iffontfound
    \IfFileExists{amssymb.sty}
      {\typeout{^^JFound AMS Symbol fonts on the system, using the
                'amssymb' package.^^J}%
       \usepackage{amssymb}%

      }{}
  \fi
\fi

\ifCUPmtlplainloaded \else
  \IfFileExists{amsbsy.sty}
    {\typeout{^^JFound the 'amsbsy' package on the system, using it.^^J}%
     \usepackage{amsbsy}}
    {\providecommand\boldsymbol[1]{\mbox{\boldmath $##1$}}}
\fi

\newcommand{\imag}{{\rm i}}
\newcommand{\rmd}{{\rm d}}
\newcommand{\rme}{{\rm e}}

\shorttitle{Correlation Heating in Moderately Coupled Plasmas}
\shortauthor{T. E. Foster, H. Fetsch and N. J. Fisch}

\title{Fast Correlation Heating in Moderately Coupled Electron-Ion Plasmas}

\author{Thomas E. Foster\aff{1}\orcidlink{0000-0001-6620-1217}
  \corresp{\email{thomas.foster@princeton.edu}},
  Henry Fetsch\aff{1}
 \and Nathaniel J. Fisch\aff{1}}

\affiliation{\aff{1}Department of Astrophysical Sciences, Princeton University, Princeton, NJ
08544, USA}

\begin{document}

\maketitle

\begin{abstract}
If the electrons in a plasma are suddenly heated, the resulting change in Debye shielding causes the ion kinetic energy to quickly increase. For the first time, this \textit{correlation heating}, which is much faster than collisional energy exchange, is rigorously derived for a moderately coupled, electron-ion plasma. The electron-ion mass ratio is taken to be the smallest parameter in the BBGKY hierarchy, smaller even than the reciprocal of the plasma parameter. This ordering differs from conventional kinetic theory by making the electron collision rates faster than the ion plasma frequency, which allows stronger coupling and makes the ion heating a function only of the total energy supplied to the electrons. The calculation uses known formulas for correlations in a two-temperature plasma, for which a new, elementary derivation is presented. Suprathermal ions can be created more rapidly by this mechanism than by ion-electron Coulomb collisions. This means that the use of a femtosecond laser pulse could potentially help to achieve ignition in certain fast ignition approaches to ICF.%
\end{abstract}

\section{Introduction}\label{sec:intro}

Plasmas in various experimental contexts are heated by a rapid input of energy primarily to the electrons. Examples include inertial confinement fusion (ICF) plasmas \cite[]{Lindl1995}, ultracold neutral plasmas \cite[]{Killian1999}, or photoionised plasmas created by x-rays from a Z-pinch \cite[]{Rochau2014}. The large mass difference between ions and electrons means that Maxwellianisation of each species occurs much faster than collisional energy exchange between species \cite[]{Spitzer,MongomeryTidman}, so the sudden electron heating results in a transitory two-temperature state that persists until collisions eventually restore thermal equilibrium. Since this equilibration is slow, one might ask the following question: in a two-temperature plasma with massless electrons, meaning collisional energy exchange is switched off, does suddenly heating the electrons have any effect on the ion temperature?

\begin{figure}
    \centering
    \begin{subfigure}[t]{.3\textwidth}
    \centering
    \begin{tikzpicture}
    \fill[white] (0,0) rectangle (4,3);

    \fill [fill=blue, opacity=0.2] (1.5,1.25) circle (5mm);
    \fill [fill=blue, opacity=0.5] (1.5,1.25) circle (3mm);
    \fill[black] (1.5,1.25) circle (1.25mm);
    
    \fill [fill=blue, opacity=0.2] (2.5,1.75) circle (5mm);
    \fill [fill=blue, opacity=0.5] (2.5,1.75) circle (3mm);
    \fill[black] (2.5,1.75) circle (1.25mm);
    \end{tikzpicture}
    \caption{Two nearby ions are initially well shielded by small Debye clouds of cold electrons. The screened ions interact weakly.}
    \end{subfigure}\hfill
    \begin{subfigure}[t]{.3\textwidth}
    \centering
    \begin{tikzpicture}
    \fill[white] (0,0) rectangle (4,3);

    \fill [fill=purple, opacity=0.2] (1.5,1.25) circle (10mm);
    \fill [fill=purple, opacity=0.5] (1.5,1.25) circle (6mm);
    \fill[black] (1.5,1.25) circle (1.25mm);
    
    \fill [fill=purple, opacity=0.2] (2.5,1.75) circle (10mm);
    \fill [fill=purple, opacity=0.5] (2.5,1.75) circle (6mm);
    \fill[black] (2.5,1.75) circle (1.25mm);
    \end{tikzpicture}
    \caption{If the electrons are suddenly heated, the Debye spheres get larger and the ions are screened less effectively. They suddenly repel more strongly.}
    \end{subfigure}\hfill
    \begin{subfigure}[t]{.3\textwidth}
    \centering
    \begin{tikzpicture}
    \fill[white] (0,0) rectangle (4,3);

    \fill [fill=purple, opacity=0.2] (1,1) circle (10mm);
    \fill [fill=purple, opacity=0.5] (1,1) circle (6mm);
    \fill[black] (1,1) circle (1.25mm);
    
    \fill [fill=purple, opacity=0.2] (3,2) circle (10mm);
    \fill [fill=purple, opacity=0.5] (3,2) circle (6mm);
    \fill[black] (3,2) circle (1.25mm);

    \draw [-stealth, very thick](3,2) -- (3.75,2.375);
    \draw [-stealth, very thick](1,1) -- (0.25,0.625);
    
    \end{tikzpicture}
    \caption{The ions repel each other and fly apart. Their potential energy is converted into kinetic energy.}
    \end{subfigure}
    \caption{Physical explanation of correlation heating. The black circles represent ions, and the coloured circles represent the electron density around each ion due to Debye screening.}
    \label{fig:cartoon}
\end{figure}
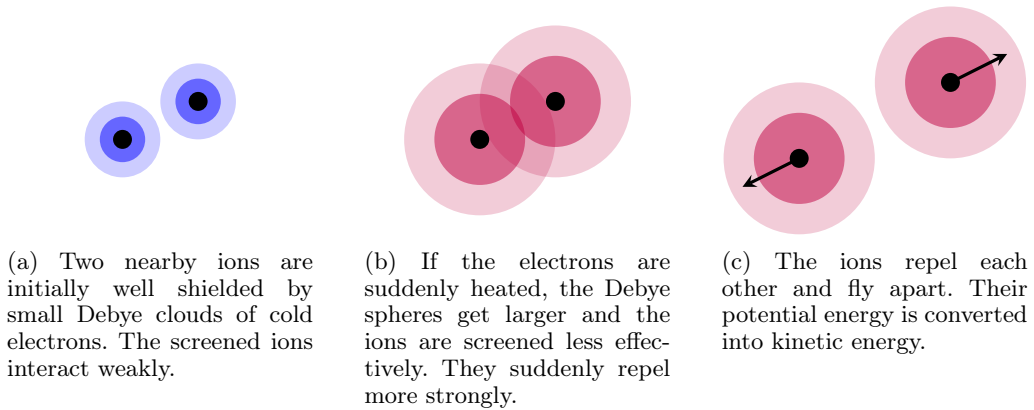

Intriguingly, the answer to this question is yes: the ion temperature rapidly jumps upwards. After the electrons are heated, the ions are shielded less effectively and nearby ions repel each other more strongly. Therefore, the ions tend to push away from each other, which converts potential energy into ion kinetic energy (see figure \ref{fig:cartoon}). This mechanism is an example of \textit{correlation heating}, in which changing the spatial correlations between particles leads to a rapid conversion of potential energy into kinetic energy. 

This correlation heating mechanism operates on ion-plasma-frequency timescales ${t\sim 1/\omega_{pi}}$, which is much faster than energy exchange due to interspecies Coulomb collisions \cite[]{More1980,More1981,More1983}.

This curious effect is similar to ionisation or `proximity' heating, in which suddenly changing the charge state of the ions causes them to push away from each other and heat up \cite[]{Hagelstein1981,More1986,BoerckerMore1986,Evans2008,Baggott2021}. Monte Carlo calculations in \cite{Hagelstein1981} show that, when lasers are used to ionise neon gas, the ion temperature could increase by more than an order of magnitude due to ionisation heating. This in turn is reminiscent of the rapid conversion of potential energy into ion kinetic energy by Coulomb explosions \cite[]{Fleischer1965}, which has been proposed as a mechanism for creating fast deuterons for nuclear fusion \cite[]{Last2001, Ron2012}.

Another example of correlation heating is disorder-induced heating of ultracold neutral plasmas \cite[e.g.][]{Murillo2001,Kuzmin2002,Gericke2003,Simien2004,Cummings2005,Killian2007,Lyon2016}. These plasmas are generated when neutral atoms in a magneto-optical trap are cooled to low temperatures, typically $\SI{1}{\milli\kelvin}$ or even lower, and suddenly ionised by lasers. The newly-formed ions quickly transition from an uncorrelated, disordered initial state to a state with correlations imposed by their strong Coulomb repulsion. This change in the particle correlations causes a tremendous increase in the ion temperature, which can climb to $\SI{1}{\kelvin}$ or more.

However, ultracold neutral plasmas are low-temperature systems in which quantum effects are important. The correlation heating discussed in this paper, caused by changing Debye shielding, is a universal phenomenon, which could occur in any hot, classical plasma in which electrons shield ions. The enormous ion temperature increase that can be achieved using ionisation heating or disorder-induced heating is a tantalising prospect for fusion. This is particularly true for fusion approaches that utilise more strongly coupled plasmas, in which there is a large reservoir of potential energy.

In this paper, we calculate the ion temperature change due to correlation heating after a sudden input of energy to the electrons in a \textit{moderately coupled} plasma, by which we mean a plasma satisfying the ordering $1\ll\Lambda/\lambda\ll(m_i/m_e)^{1/2}$. Here, $\Lambda$ is the plasma parameter (the number of particles in a Debye sphere) and $\lambda$ is the Coulomb logarithm. The assumption of moderate coupling allows a novel, rigorous expansion of the Bogoliubov-Born-Green-Kirkwood-Yvon (BBGKY) hierarchy, giving a first-principles derivation of correlation heating in a two-component plasma with electrons and ions. Previous work has investigated the effect of changing shielding in one-component plasmas, both theoretically \cite[]{Morawetz2001}, and numerically \cite[]{Gericke2003ii,Chen2004}.

Our calculations require formulas for two-particle correlations in a two-temperature plasma, first presented by \cite{Salpeter1963}. In \S\ref{sec:salpeter}, we present a streamlined derivation of these correlations that employs only basic plasma physics concepts. 

We compute the time-dependence of the ion kinetic energy following electron heating and find that it does not overshoot its final value. We also calculate the modification to the ion velocity distribution and find that correlation heating in a moderately coupled plasma is relatively inefficient at heating suprathermal ions, which are crucial for fusion. However, we argue that correlation heating is more effective at creating fast ions if the plasma is more strongly coupled or if there is a light minority ion species present. This may be particularly relevant for fast ignition approaches to ICF. Even if few fast ions are created directly, the combination of correlation heating and subsequent ion-ion collisions could allow the ions in a fusion device, including suprathermal ions, to be energised more rapidly than would be possible using ion-electron collisions alone. 

In complementary work \cite[]{Fetsch2023} that considers timescales longer that the $1/\omega_{pi}$ timescales resolved here, the same result for the total change in ion temperature when the electrons are suddenly heated is obtained. \citet[]{Fetsch2023} generalises thermodynamic relations to two-temperature plasma and obtains a \textit{quasi-Equation of State}. This is used to compute temperature changes under compression and heating of either species; for example, initially equal electron and ion temperatures are found to separate under compression. The thermodynamic method also allows calculation of the entropy of each species in a two-temperature plasma, thus providing a particularly effective way to analyse processes that are adiabatic for at least one species.

\section{Ordering Assumptions}\label{sec:assumptions}

We are interested in classical, homogeneous, non-magnetised plasmas consisting of electrons and a single ion species of charge $+Ze$. Since the ion charge is fixed, ionisation heating does not occur. This allows us to isolate the effect of correlation heating due to changes in the Debye shielding of ions. If the electron component and ion component have had time to Maxwellianise individually, due to intra-species collisions, then the velocity distribution of species $\alpha$ (with $\alpha=$ i for ions and $\alpha=$ e for electrons) is
\begin{equation}\label{eq:maxwellian}
    f_\alpha(\boldsymbol{v}) = n_\alpha\left( \frac{m_\alpha}{2\upi T_\alpha} \right)^{3/2}\exp\left( -\frac{m_\alpha \boldsymbol{v}^2}{2 T_\alpha} \right),
\end{equation}
where $n_\alpha$, $m_\alpha$ and $T_\alpha$ are the number density, particle mass, and temperature, respectively, of species $\alpha$. This defines what we mean by species-dependent temperature; the temperature of a species is the characteristic width of its Maxwellian velocity distribution.

In this section, we give the ordering assumptions used throughout the paper, which allow the plasma to exist in a two-temperature state. They also define a regime in which correlation heating is analytically tractable and they provide a fair model of real plasmas from experimental contexts involving sudden electron heating. The section concludes with parameters from two example plasmas.

\subsection{Masses}

Two-temperature plasma can exist when $\tau_{ie}$, the timescale of collisional energy exchange between species, is much greater than the timescales of Maxwellianisation for each species. The Maxwellianisation timescales can be estimated using the ion-ion collision time $\tau_{ii}$ and the electron-electron collision time $\tau_{ee}$. It is a standard result that the considerable mass difference between ions and electrons in a plasma leads to a large separation between these collisional timescales, since $\tau_{ee}/\tau_{ii}\sim\tau_{ii}/\tau_{ie}\sim(m_e/m_i)^{1/2}$ \cite[]{Spitzer,MongomeryTidman}. Our first assumption is therefore
\begin{equation}\label{ass:1}
\text{Assumption 1:\hspace{30pt}} \frac{m_e}{m_i}\ll 1\,.
\end{equation}
Our formalism treats this mass ratio as the smallest quantity in the problem, unlike conventional plasma kinetic theory in which the weak-coupling limit is taken before the mass ratio is sent to zero. It is worth pointing out that these limits do not commute; this fact is more than just a technical curiosity and will be crucial later for understanding how the system responds when the electrons are suddenly heated. 

\subsection{Temperatures}

A small electron-ion mass ratio is not sufficient to ensure that the species temperatures equilibrate slowly compared with all other timescales in the problem. There must also be a restriction on the ion and electron temperatures because collisional energy exchange can occur quickly if the difference between them is large enough. How far apart can the temperatures be?

In a plasma with large plasma parameter $\Lambda\gg 1$ and the mild restriction ${T_e/m_e\gg T_i/m_i}$, the collisional timescales $\tau_{ee}$, $\tau_{ii}$, $\tau_{ie}$, and $\tau_{ei}$ (the timescale of momentum transfer between electrons and ions) can be determined using the Landau-Spitzer formulas in table \ref{tb:times} \cite[]{Landau1936,Spitzer}. We constrain the temperatures by assuming that these timescales satisfy
\begin{equation}\label{eq:ineq1}
\tau_{ee}, \tau_{ei}\ll \tau_{ii}\ll \tau_{ie}\,.
\end{equation}
This ordering was chosen for two reasons. First, it ensures that the temperatures change slowly compared to the Maxwellianisation rates and compared to the inter-species momentum exchange rate.\footnote{It is possible that the species temperatures do not change at a rate $1/\tau_{ie}$, for a few reasons. First of all, the Spitzer equations $dT_e/dt = -2(T_e-T_i)/Z\tau_{ie}$ and $dT_i/dt = -2(T_i-T_e)/\tau_{ie}$ show that the temperature of the colder species can change more rapidly than this. If, for example, the ions are much colder than the electrons, then the electron temperature changes at a rate $1/\tau_{ie}$, while the ion temperature changes at a faster rate $(T_e/T_i)(1/\tau_{ie})$. However, taking this into account does not change the final temperature constraint \eqref{ass:2}. Secondly, theoretical and numerical studies \cite[]{Bobylev1997,Chankin2012,Meisl2013} show an enhanced temperature relaxation rate in a two-temperature plasma because slow electrons, which have a large Coulomb scattering cross section, strongly interact with the ions and exchange energy with them quickly even compared to the electron-electron collision timescale. This energisation of the electrons at small velocities also leads to deviations of the electron distribution from a Maxwellian. However, these effects take place in a boundary layer in velocity space with thickness vanishing as $m_e/m_i\to0$, so they are unimportant when the mass ratio is small. Finally, the temperature relaxation rate can also be modified by coupling with ion acoustic modes \cite[]{Dharma1998,Vorberger2010, Gregori2008}, but this effect does not lead to significant corrections to Landau-Spitzer theory when the plasma parameter is large \cite[]{Vorberger2009}.} This justifies the `two-temperature' model in which each species is Maxwellian, and has the same mean flow velocity, but with different temperatures. 

Secondly, we have chosen to make the electrons Maxwellianise faster than the ions, which will be justified shortly.
\begin{table}\centering
\begin{tabular}{ c c c }
\textbf{Timescale} & \textbf{Landau-Spitzer formula} & \textbf{Description}\\
 $\tau_{ee}$ & $\displaystyle \frac{3}{4\sqrt{\upi}}\frac{m_\rme^{1/2}T_\rme^{3/2}}{n_e e^4 \lambda_{ee}}$ & Electron Maxwellianisation timescale \\  
 $\tau_{ei}$ & $\displaystyle \frac{3}{4\sqrt{2\upi}}\frac{m_\rme^{1/2}T_\rme^{3/2}}{n_i Z^2e^4 \lambda_{ei}}$ & Electron-ion momentum transfer (friction) timescale \\
 $\tau_{ii}$ & $\displaystyle \frac{3}{4\sqrt{\upi}}\frac{m_i^{1/2}T_i^{3/2}}{n_i Z^4e^4 \lambda_{ii}}$ & Ion Maxwellianisation timescale\\
 $\tau_{ie}$ & $\displaystyle \frac{3}{4\sqrt{2\upi}}\frac{m_iT_\rme^{3/2}}{m_\rme^{1/2}n_e Z^2e^4 \lambda_{ie}}$ & Collisional energy transfer timescale
\end{tabular}
\caption{Landau-Spitzer formulas for collisional timescales, following \cite{Hazeltine}. The $\lambda_{\alpha\beta}$ are Coulomb logarithms.}
\label{tb:times}
\end{table}

Using the formulas in table \ref{tb:times} with $Z\sim 1$ and $\lambda_{ee}\sim\lambda_{ei}\sim\lambda_{ii}$, condition \eqref{eq:ineq1} means the temperatures must satisfy \cite[]{Bobylev1997}
\begin{equation}\label{ass:2}
    \text{Assumption 2:\hspace{30pt}} \left(\frac{m_e}{m_i}\right)^{1/3}\ll\frac{T_e}{T_i}\ll \left(\frac{m_i}{m_e}\right)^{1/3}.
\end{equation}
In a hydrogen plasma, $(m_i/m_e)^{1/3} \approx 12$, so this constraint on the temperatures is satisfied in most situations of interest. From now on, we order $T_e\sim T_i$.

\subsection{Plasma Parameter}

Our final assumption concerns the plasma parameter $\Lambda$, which is the number of particles in a Debye sphere and which measures the strength of interactions in a plasma.

We order the electron and ion plasma frequencies relative to the collisional timescales as follows:
\begin{equation}\label{eq:ordering}
1/\omega_{pe}\ll \tau_{ee}, \tau_{ei}\ll 1/\omega_{pi} \ll \tau_{ii}\,.
\end{equation}
This means $\Lambda$ must be large but also includes the unusual condition that the ion plasma frequency $\omega_{pi}$ is slower than the electron Maxwellianisation rate $1/\tau_{ee}$.  The electron timescales $\omega_{pe}$ and $\tau_{ee}$ are smaller that the corresponding ion timescales $\omega_{pi}$ and $\tau_{ii}$ by a factor of $(m_e/m_i)^{1/2}$, so they are taken to be the smallest timescales in the problem. This is valid as long as $\Lambda$ is not too large; the resulting regime, while narrow, is relevant for certain plasmas in Z-pinch and ICF experiments. In addition, our correlation heating calculations will be simplified because the electrons Maxwellianise before the ions respond to the heating, which means the ions respond independently of where in velocity space the electrons were energised.

The plasma frequency and Maxwellianisation timescale of a species are related by $1/\omega_{p\alpha}\sim\lambda\tau_{\alpha\alpha}/\Lambda$, where $\lambda$ is the Coulomb logarithm and
\begin{equation}
    \Lambda \sim \left( \frac{T}{n^{1/3}e^2} \right)^{3/2}.
\end{equation}
Species indices are not necessary here because we have taken $Z\sim 1$ and $T_e\sim T_i$. Therefore, \eqref{eq:ordering} is equivalent to
\begin{align}\label{ass:3}
    \text{Assumption 3:\hspace{30pt}} 1\ll \frac{\Lambda}{\lambda}\ll\left( \frac{m_i}{m_e} \right)^{1/2}.
\end{align}
We refer to this as the \textit{moderate coupling} regime because $\Lambda$ is much greater than one but cannot be arbitrarily large. The large plasma parameter will allow the BBGKY hierarchy to be truncated, giving a closed set of equations which we will solve to find the ion evolution during correlation heating.

\subsection{Example Parameters}

In table \ref{table:params}, we give parameters for two example plasmas.

The first set of parameters is for a hydrogen plasma \cite[]{Falcon2015} created in the Z Pulsed Power Facility at Sandia National Laboratories \cite[]{Sinars2020} as part of the Z Astrophysical Plasma Properties (ZAPP) collaboration \cite[]{Rochau2014}. X-rays from the Z-pinch are used to photoionise hydrogen in order to study spectral lines of hydrogen plasma under conditions typical of white dwarf photospheres.
\begin{table}\centering
\begin{tabular}{ c | c | c }
\textbf{Parameter} &\textbf{Photoionised hydrogen plasma} &\textbf{Fast ignition D-T hot spot plasma}\\[5pt]
$Z$ &$1$ &$1$\\
$n_e$               &$\SI{3e17}{\per\centi\meter\cubed}$  & $\SI{5e25}{\per\centi\meter\cubed}$ \\
$T_e$               &$\SI{1}{\electronvolt}$              & $\SI{8}{\kilo\electronvolt}$ \\
$T_i$               &$?$                                  & $\SI{0.3}{\kilo\electronvolt}$ \\[10pt]
$\Gamma_e$          & $0.2$                               & $0.01$ \\
$\Gamma_i$          & $0.2\,T_e/T_i$                      & $0.3$ \\[10pt]
$\tau_{ee}$         &$\SI{0.5}{\pico\second}$             & $\SI{0.9}{\femto\second}$ \\
$\tau_{ei}$         &$\SI{0.3}{\pico\second}$             & $\SI{0.7}{\femto\second}$ \\
$\tau_{ii}$         &$20(T_i\lambda_{ee}/T_e\lambda_{ii})^{3/2}\SI{}{\pico\second}$ & $\SI{1}{\femto\second}$ \\
$\tau_{ie}$         &$\SI{600}{\pico\second}$             & $\SI{3000}{\femto\second}$\\[10pt]
$1/\omega_{pe}$  &$\SI{0.03}{\pico\second}$            & $\SI{0.003}{\femto\second}$ \\
$1/\omega_{pi}$  &$\SI{1}{\pico\second}$               & $\SI{0.2}{\femto\second}$
\end{tabular}
\caption{Example parameters from two experimental contexts involving fast electron heating. The coupling strengths $\Gamma_\alpha = (4\upi n_\alpha/3)^{1/3}q_\alpha^2/T_\alpha$ are the ratio of the typical potential energy to the typical kinetic energy of a particle of species $\alpha$. The collision timescales are calculated using the formulas in table \ref{tb:times} together with the conventional definitions of the Coulomb logarithms \cite[]{Richardson2019}.}
\label{table:params}
\end{table}
The plasma exists for about $\SI{100}{\nano\second}$ and, after an initial period where the plasma is overionised, the electron temperature and density eventually stabilise around the given values. The ion temperature is not measured but should be lower than the electron temperature, meaning the ions may be weakly or strongly coupled.

The second set of parameters in table \ref{table:params} is for an ICF hot spot plasma (deuterium-tritium) in the fast ignition scenario \cite[]{Tabak2006}. Conventionally, ICF uses stagnation of converging flows to heat a central region up to ignition. In fast ignition, however, it is envisioned that the target would first be imploded with a uniform density, before ignition would be triggered in a central hot spot with a laser pulse \cite[]{Tabak1994}. This laser pulse would heat the hot spot indirectly by creating fast electrons in the surrounding plasma that rapidly propagate inwards, although other schemes, for example involving ion beams, have also been proposed \cite[]{Bychenkov2001,Logan2006}. Fast ignition offers various advantages over the conventional approach to ICF, including higher gain, lower driver energy, and relaxation of symmetry requirements \cite[]{Badziak2007}. Fast ignition requirements in two-temperature plasmas have been studied, for example in \cite{Eliezer2015}.

Both example plasmas have coupling strengths that are less than one without being extremely small. Also, both plasmas have electron-electron and electron-ion collision times of the same order as the ion plasma frequency, instead of being much longer as is usually the case in very weakly coupled systems. Since the exact parameters may be varied in the experiment, the plasma in these ICF and Z-pinch contexts may be well-described by a moderate-coupling ordering.

\section{Prerequisite: Pair Correlations in a Two-Temperature Plasma}\label{sec:correlations}

Expressions for the steady-state pair correlations in a two-temperature plasma will be an essential ingredient for the analytical work in this paper.

Pair correlations describe the relative positions of particles in a system. In a thermal equilibrium plasma, assuming large plasma parameter $\Lambda\gg 1$, the pair correlations take a well-known form \cite[]{KrallTrivelpiece},
\begin{equation}\label{eq:eqmcorr}
    G_{\alpha\beta}(\boldsymbol{r}) = -\frac{q_\alpha q_{\beta}}{T}\frac{\rme^{-k_D r}}{r}\,,
\end{equation}
where $G_{\alpha\beta}(\boldsymbol{r})$ is the spatial pair correlation between species $\alpha$ and species $\beta$ (see appendix \ref{app:defs} for the definitions and notation that we use for pair correlations and distribution functions). Here, $k_D$ is the inverse Debye length, defined by $k_D^2 = k_e^2 + k_i^2$ with
\begin{align}\label{eq:debye}
k_\alpha^2 &= \frac{4\upi n_\alpha q_\alpha^2}{T_\alpha}\,.
\end{align}
The generalisation of \eqref{eq:eqmcorr} to two-temperature plasma was first found by \cite{Salpeter1963} in order to study Thomson scattering of radio waves in the ionosphere, which has hotter electrons than ions. Salpeter's results corrected earlier work \cite[]{Kadomtsev1958,Renau1962,RenauCamnitzFlood1962,Renau1964} and were later rederived using various other methods \cite[]{EckerKroll1964,SchramKegel1965,BoerckerMore1986,Seuferling1989,Fetsch2023}. There has been recent interest in extending Salpeter's formulas to more strongly coupled plasmas \cite[]{Shaffer2017,Triola2022}.

In this section, we present two derivations of the steady-state pair correlations in a two-temperature plasma. The first derivation is simpler than other published methods and is based on Salpeter's original approach, using arguments in the style of \cite{DebyeHuckel1923}. The second derivation is more rigorous and systematic, and introduces the formalism that will be needed later to obtain analytical results for correlation heating.

\subsection{Debye-H\"uckel Method}\label{sec:salpeter}

The key idea is to exploit the fact that the mass ratio $m_e/m_i$ is a small parameter. As long as the ions are not too hot, which is ensured by assumption \eqref{ass:2}, the electron thermal speed is much greater than the ion thermal speed; compared to the electrons, the ions move extremely slowly and are essentially stationary. We therefore begin by finding the steady-state response of the electrons to a set of fixed ions. This will determine the electron-ion and ion-ion correlations.

Assume the electron distribution is a uniform Maxwellian $f_{e0}(\boldsymbol{v})$ plus a small perturbation $f_{e1}(\boldsymbol{v},\boldsymbol{r})$ with vanishing spatial average. The perturbation satisfies the linearised Vlasov equation,
\begin{equation}
    \left( \frac{\partial}{\partial t} + \boldsymbol{v\cdot\nabla} \right) f_{e1} + \frac{e}{m_e}(\boldsymbol{\nabla}\varphi)\boldsymbol{\cdot}\frac{\partial f_{e0}}{\partial \boldsymbol{v}} = 0\,.
\end{equation}
If the average potential is taken to be zero, there is a simple steady-state solution
\begin{equation}\label{eq:electronresponse}
    f_{e1} = \frac{e\varphi}{T_e}f_{e0}\,.
\end{equation}
Following \cite{Salpeter1963}, we have disregarded any terms that depend on time or that do not involve the potential $\varphi(\boldsymbol{r})$. Equation (\ref{eq:electronresponse}) is the usual linearised adiabatic response formula. We could also justify this result by arguing that the electrons reach thermal equilibrium around essentially fixed ions, which means the ion potential can be treated as an external potential acting on the system consisting of the electrons only. Then the Boltzmann distribution for the electrons $f_e(\boldsymbol{r},\boldsymbol{v}) \propto 
f_{e0}(\boldsymbol{v})\exp(e\varphi(\boldsymbol{r})/T_e)$ leads to \eqref{eq:electronresponse}, assuming weak interactions $e\varphi\ll T_e$. It will be convenient to work with Fourier transforms (our convention is given in appendix \ref{app:defs}) from now on, so we have 
\begin{equation}\label{eq:fourier}
\widetilde{f}_{e1}(\boldsymbol{k},\boldsymbol{v}) = \frac{e\widetilde{\varphi}(\boldsymbol{k})}{T_e}f_{e0}(\boldsymbol{v})\,.
\end{equation}
To completely specify the electron response to the ions, we need to calculate the potential $\widetilde{\varphi}$ for any given ion distribution. It satisfies Poisson's equation
\begin{equation}
k^2\widetilde{\varphi} = 4\upi\,(\widetilde{\rho}_e + \widetilde{\rho}_i)\,,
\end{equation}
where $\rho_e$ and $\rho_i$ are the charge densities of the electrons and ions respectively. The only part of the electron distribution with spatial variation is $f_{e1}$, so $\widetilde{\rho}_e(\boldsymbol{k}) = \int f_{e1}(\boldsymbol{k},\boldsymbol{v})\,\rmd\boldsymbol{v}$ for nonzero $\boldsymbol{k}$. Then \eqref{eq:fourier} gives
\begin{equation}\label{eq:density}
    \widetilde{\rho}_{e} = -\frac{n_ee^2\widetilde{\varphi}}{T_e}\,.
\end{equation}
Poisson's equation becomes
\begin{equation}\label{eq:poisson}
 (k^2+k_e^2)\widetilde{\varphi} = 4\upi \widetilde{\rho}_i\,,
\end{equation}
where $k_e$ is defined by equation \eqref{eq:debye}. Therefore, if the ion charge density is
\begin{equation}
\rho_i(\boldsymbol{r}) = \sum_{\mathrm{ions\, }i} Ze\,\delta(\boldsymbol{r}-\boldsymbol{r}_i)\,,
\end{equation}
corresponding to a collection of fixed ions, then the resulting potential is
\begin{equation}
    \varphi(\boldsymbol{r}) = \sum_{\mathrm{ions }\,i}\frac{Ze}{|\boldsymbol{r}-\boldsymbol{r}_i|}\,\rme^{-k_e|\boldsymbol{r}-\boldsymbol{r}_i|} + \mathrm{const.}\, ,
\end{equation}
where the constant ensures that the average potential is zero. We can think of the ions as interacting via a screened Coulomb potential (Yukawa potential)
\begin{equation}\label{eq:yukawa}
\varphi^{(s)}(\boldsymbol{r}) = \frac{Ze}{r}\,\rme^{-k_er},
\end{equation}
where $r=|\boldsymbol{r}|$, with shielding length set by the electron temperature. 

For any given ion distribution, we can use \eqref{eq:fourier} and \eqref{eq:poisson} to calculate the electron response. The resulting electron charge density is
\begin{equation}\label{eq:electrondensity}
\widetilde{\rho}_{e} = -\left(\frac{k_e^2}{k^2+k_e^2}\right)\widetilde{\rho}_i\,.
\end{equation}
It is worth noting a subtle point which will be important later. In writing \eqref{eq:electronresponse}, we assumed that the electron distribution is a steady-state solution of the linearised Vlasov equation. In reality, the electron distribution will fluctuate rapidly as the electrons move around and take part in plasma oscillations; by ignoring these fluctuations, we have essentially averaged the electron distribution over a time period much longer than the fluctuation timescale and much shorter than the timescale of ion motion (these timescales are well-separated because of the small mass ratio $m_e/m_i$). The fluctuating part of the electron distribution depends on the fluctuating part of the potential, which is independent of the slowly-varying ion distribution. Therefore, the electron fluctuations are independent of the ion distribution and they do not affect the electron-ion or ion-ion correlations. They will, however, be relevant for the electron-electron correlation.

Knowing the response of electrons to fixed ions, we can now work out the ion-ion correlation. The ions are a system of particles with temperature $T_i$ interacting via the screened potential in \eqref{eq:yukawa}, and equilibrium statistical mechanics can be used to calculate the pair correlations for such a system.\footnote{There is an assumption at work here: we're assuming that the electron temperature is the same for all ion microstates, or equivalently that the majority of the modes of the system have isothermal electrons. In reality, if a fluctuation causes the ions to move closer together or further apart on average, the associated change in field energy will modify the electron and ion temperatures. However, in a  system with large plasma parameter $\Lambda\gg 1$, the electron temperature change will be small in $1/\Lambda$. Therefore, the pair correlations will be unaffected to lowest order in $1/\Lambda$.}

Instead of using statistical mechanics and calculating a partition function, we will follow \cite{DebyeHuckel1923} and assume that the ion-ion correlation is proportional to the excess ion density around a single, fixed ion at the origin. Around the fixed ion, the Boltzmann response formula gives
\begin{equation}
\rho_i(\boldsymbol{r}) \propto\, \exp\left( -\frac{Ze\varphi(\boldsymbol{r})}{T_i} \right), 
\end{equation}
or
\begin{equation}
    \rho_i(\boldsymbol{r}) = Zen_i\left( 1 - \frac{Ze\varphi(\boldsymbol{r})}{T_i} \right),
\end{equation}
assuming $Ze\varphi/T_i\ll 1$. Now Poisson's equation \eqref{eq:poisson} becomes
\begin{equation}
    (k^2+k_e^2+k_i^2)\widetilde{\varphi} = 4\upi Ze,
\end{equation}
and the solution is 
\begin{equation}
\widetilde{\varphi}(\boldsymbol{k}) = \frac{4\upi Ze}{k^2+k_e^2+k_i^2}\,.
\end{equation}
In real space, this is
\begin{equation}
\varphi(\boldsymbol{r}) = \frac{Ze}{r}\, \rme^{-k_Dr},
\end{equation}
where $k_D$ is now the total inverse Debye length because both the ions and the electrons are taking part in the shielding of the test ion at the origin.
Then, the ion and electron densities around this fixed ion are
\begin{align}
\rho_i(\boldsymbol{r}) &= Zen_i\left( 1 - \frac{Z^2e^2}{T_i}\frac{\rme^{-k_Dr}}{r} \right),\\
\rho_e(\boldsymbol{r}) &= -en_e\left( 1 + \frac{Ze^2}{T_e}\frac{\rme^{-k_Dr}}{r} \right).
\end{align}
We can read off the ion-ion correlation, which is
\begin{align}
    G_{ii}(\boldsymbol{r}) = -\frac{Z^2e^2}{T_i}\frac{\rme^{-k_D r}}{r}\,.
\end{align}
The excess electron density around a fixed ion also tells us the electron-ion correlation because the ions move so slowly that they can all be approximated as stationary. So, we can also read off
\begin{align}\label{eq:electronion}
    G_{ei}(\boldsymbol{r}) = \frac{Ze^2}{T_e}\frac{\rme^{- k_D r}}{r}\,.
\end{align}
Now we just need the electron-electron correlation. The Debye-H\"uckel method of fixing a particle at the origin and letting the other particles reach thermal equilibrium worked well for finding the electron density around any given ion, since the ions all appear virtually stationary to the electrons. However, the ion density around an electron cannot be found this way. 

Nevertheless, the fact that the electrons form identical screening clouds around each ion suggests that we should be able to calculate the electron spatial distribution from the ion spatial distribution, which is encoded in the ion-ion correlations. The electron and ion spatial distributions are related by \eqref{eq:electrondensity}; taking the modulus squared then the thermal average gives
\begin{equation}\label{eq:chargefluc}
\langle|\widetilde{\rho}_e|^2\rangle = \left(\frac{k_e^2}{k^2+k_e^2}\right)^2\langle|\widetilde{\rho}_i|^2\rangle\,.
\end{equation}
This equation links fluctuations in the electron charge density with fluctuations in the ion charge density. It is well-known that charge density fluctuations are linked to pair correlations. Using the relation
\begin{align}\label{eq:fluc}
\langle|\widetilde{\rho}_\alpha|^2\rangle &= Vn_\alpha q_\alpha^2 \left( 1 + n_\alpha \widetilde{G}_{\alpha\alpha} \right),
\end{align}
which is proved in appendix \ref{app:fluc}, we find
\begin{equation}
\widetilde{G}_{ee}(\boldsymbol{k}) \stackrel{?}{=} \frac{4\upi Z e^2}{T_e}\frac{k_e^2}{(k^2+k_D^2)(k^2+k_e^2)}\,,
\end{equation}
or, in real space,
\begin{equation}
G_{ee}(\boldsymbol{r}) \stackrel{?}{=} -\frac{e^2}{T_e}\frac{T_e}{T_i}\left( \frac{\rme^{-k_Dr}}{r} - \frac{\rme^{-k_er}}{r} \right).\label{eq:slowgee}
\end{equation}
However, this result must be wrong! If the ion charge is taken to zero $Z\to 0$ with $n_e$ fixed, then the ions become unaffected by Coulomb forces and uniformly distributed. Equation \eqref{eq:slowgee} gives $G_{ee} = 0$ in this limit. But that isn't right! The electrons screen each other, and have nonzero charge fluctuations $\langle|\widetilde{\rho}_e|^2\rangle$, even when $Z\to 0$.

The problem is that, as discussed earlier, we ignored the rapid fluctuations in the electron distribution that correspond to the electrons moving around and taking part in plasma oscillations. It is precisely these fluctuations that are missing from the pair correlation. They are independent of the ions, so the missing term in the electron-electron correlation is the correlation that remains when $Z\to 0$. In this limit, the ions become a uniform neutralising background and we are left with a one-component plasma of electrons at temperature $T_e$. The pair correlation for such a system is, using \eqref{eq:eqmcorr},
\begin{equation}
G_{ee}^{(Z\to 0)}(\boldsymbol{r}) = -\frac{e^2}{T_e}\frac{\rme^{-k_er}}{r}\,.
\end{equation}
This should be added to the part of the electron-electron correlation arising from their interaction with the ions, which is given in \eqref{eq:slowgee}. 

In summary, the correct pair correlations in a two-temperature plasma are
\begin{align}
    G_{ii}(\boldsymbol{r}) &= -\frac{Z^2e^2}{T_i}\frac{\rme^{-k_D r}}{r}\, ,\label{eq:ioncorr}\\
    G_{ei}(\boldsymbol{r}) &= \frac{Ze^2}{T_e}\frac{\rme^{- k_D r}}{r}\, ,\\
    G_{ee}(\boldsymbol{r}) &= -\frac{e^2}{T_e}\left[ \frac{T_i}{T_e}\frac{\rme^{-k_D r}}{r} + \left( \frac{T_e - T_i}{T_e}\right)\frac{\rme^{-k_e r}}{r}  \right].\label{eq:electroncorr}
\end{align}
When the temperatures $T_e$ and $T_i$ are equal, these results reduce to the usual equilibrium formula \eqref{eq:eqmcorr}.

This derivation is simple and intuitive. However, just like the thermal equilibrium correlations \cite[]{KrallTrivelpiece}, Salpeter's two-temperature correlations can be derived using the BBGKY hierarchy. We now present this alternative, more systematic method, which introduces the formalism that we will use to analyse the response of a two-temperature plasma to sudden electron heating.

\subsection{Kinetic Method}\label{sec:bbgky}

\subsubsection{BBGKY Equations}

The first two equations in the BBGKY hierarchy for a classical, unmagnetised plasma are \cite[][p.102]{Balescu}
\begin{align}
    &\label{eq:f}\left(\frac{\partial}{\partial t} + \mathcal{L}(1)\right) f_\alpha(1) = -\sum_\beta \left( \int \rmd(2)\; \mathcal{V}_{\alpha\beta}(12)\,g_{\alpha\beta}(12) \right),\\
    &\label{eq:g}\left(\frac{\partial}{\partial t} + \mathcal{L}(12)\right)g_{\alpha\beta}(12) + \sum_\gamma \left(\int \rmd(3)\; \mathcal{V}_{\alpha\gamma}(13)\,f_\alpha(1)\,g_{\gamma\beta}(32) \right)\\
    &\nonumber\phantom{\left(\frac{\partial}{\partial t} + \mathcal{L}(12)\right)g_{\alpha\beta}(12)\,} + \sum_\gamma \left(\int \rmd(3)\; \mathcal{V}_{\gamma\beta}(32)\,f_\beta(2)\,g_{\alpha\gamma}(13) \right)\\
    &\nonumber\phantom{\left(\frac{\partial}{\partial t} + \mathcal{L}(12)\right)g_{\alpha\beta}(12)\,} = -\mathcal{V}_{\alpha\beta}(12)\, \bigl[f_\alpha(1)f_\beta(2) + g_{\alpha\beta}(12)\bigr]\\
    &\nonumber\phantom{\left(\frac{\partial}{\partial t} + \mathcal{L}(12)\right)g_{\alpha\beta}(12)\,=\,}- \sum_\gamma \left(\int \rmd(3)\, \bigl[\mathcal{V}_{\alpha\gamma}(13) + \mathcal{V}_{\beta\gamma}(23)\bigr]\,h_{\alpha\beta\gamma}(123) \right).
\end{align}
In appendix \ref{app:defs} we give the definitions of the distribution functions $f_\alpha$, $g_{\alpha\beta}$ and $h_{\alpha\beta\gamma}$, and a detailed explanation of our notation. In summary, if the position and velocity of particle $1$ are $\boldsymbol{r}_1$ and $\boldsymbol{v}_1$ respectively, we write $(1)$ for the phase-space coordinates $(\boldsymbol{r}_1, \boldsymbol{v}_1)$ and $\int \rmd(1) = \int \rmd\boldsymbol{r}_1\rmd\boldsymbol{v}_1$ for integration over these coordinates. When a function $f$ depends on the phase-space coordinates of multiple particles, we write $f(12\ldots)$. 

The operators $\mathcal{V}$ and $\mathcal{L}$ are defined as follows. The symmetric operator $\mathcal{V}_{\alpha\beta}(12)$ describes the effect of Coulomb interactions between particles $1$ and $2$:
\begin{equation}
    \mathcal{V}_{\alpha\beta}(12) = q_\alpha q_\beta
    \frac{\boldsymbol{r}_1 - \boldsymbol{r}_2}{|\boldsymbol{r}_1-\boldsymbol{r}_2|^3}\boldsymbol{\cdot}\left(\frac{1}{m_\alpha}\frac{\partial}{\partial\boldsymbol{v}_1} - \frac{1}{m_\beta}\frac{\partial}{\partial\boldsymbol{v}_2}\right).
\end{equation}
 The $\mathcal{L}(1)$ operator describes the evolution of a distribution function as particle $1$ moves in the mean electric field:
\begin{equation}
\mathcal{L}(1)\,a(1) = \biggl(\boldsymbol{v}_1\boldsymbol{\cdot}\frac{\partial}{\partial\boldsymbol{r}_1} + \sum_\beta\int \rmd(2)\;\mathcal{V}_{\alpha\beta}(12)f_\beta(2)\biggr)a(1)\,,
\end{equation}
for any function $a(1)$. Finally, $\mathcal{L}(12) = \mathcal{L}(1)+\mathcal{L}(2)$.

When a two-temperature plasma has a large plasma parameter $\Lambda\gg1$, two simplifications to the BBGKY equations \eqref{eq:f}--\eqref{eq:g} are possible. First, we can neglect $g_{\alpha\beta}(12)$ compared with $f_\alpha(1)f_\beta(2)$ on the right-hand side of \eqref{eq:g}. This means we ignore initial correlations between particles as they interact with each other and produce new correlations, which is valid as long as the particles are not too close together.\footnote{The correlations at small distances $r\ll k_D^{-1}$ give contributions to the plasma energy that are higher-order in $1/\Lambda$ than the correlations at distances $r\sim k_D$ where $g_{\alpha\beta}(12)\ll f_\alpha(1)f_\beta(2)$. Similarly, for correlation heating at leading order, the $r\ll k_D^{-1}$ regions are not important so we assume $g_{\alpha\beta}(12)\ll f_\alpha(1)f_\beta(2)$ throughout.} Secondly, terms involving the triple correlation $h_{\alpha\beta\gamma}(123)$ in \eqref{eq:g} may be dropped, which means we neglect the effect of three-particle collisions.\footnote{In a two-temperature plasma, it is not necessarily obvious that collisions do not affect the pair correlations, as certain collision times may be shorter than the ion plasma frequency timescale on which the ion-ion correlation evolves. We would not be able to neglect collisional terms in the $g_{ii}$ equation of size $\sim g_{ii}/\tau_{ee}$ or $ g_{ii}/\tau_{ei}$. However, there are no such terms because these fast timescales depend on the electron mass, which does not enter the $g_{ii}$ equation. The only collision terms are $\int \rmd(3)\; \mathcal{V}_{i\gamma}(13)\,h_{ii\gamma}(123) \sim g_{ii}(12)/\tau_{i\gamma}$, so the ion-ion correlation evolves due to ion-ion and ion-electron collisions, which occur slowly compared to the ion plasma frequency.}

With these two simplifications, we have a closed pair of equations for $f_{\alpha}(1)$ and $g_{\alpha\beta}(12)$:
\begin{align}
    &\label{eq:bbgky1}\left(\frac{\partial}{\partial t} + \mathcal{L}(1)\right) f_\alpha(1) = -\sum_\beta \left( \int \rmd(2)\; \mathcal{V}_{\alpha\beta}(12)\,g_{\alpha\beta}(12) \right),\\
    &\label{eq:bbgky2}\left(\frac{\partial}{\partial t} + \mathcal{L}(12)\right)g_{\alpha\beta}(12)\begin{aligned}[t]
    {}&+\sum_\gamma \left(\int \rmd(3)\; \mathcal{V}_{\alpha\gamma}(13)\,f_\alpha(1)\,g_{\gamma\beta}(32) \right)\\
    &+ \sum_\gamma \left(\int \rmd(3)\; \mathcal{V}_{\gamma\beta}(32)\,f_\beta(2)\,g_{\alpha\gamma}(13) \right)\\
    &= -\mathcal{V}_{\alpha\beta}(12)\,f_\alpha(1)f_\beta(2)\,.
     \end{aligned}
\end{align}
We now assume the plasma is homogeneous, which means $f_\alpha(1)$ is independent of position and $g_{\alpha\beta}(12)$ depends on $\boldsymbol{r}_1$ and $\boldsymbol{r}_2$ only through the relative separation $\boldsymbol{r} = \boldsymbol{r}_1 - \boldsymbol{r}_2$. Then, taking the Fourier transform of \eqref{eq:bbgky2} with respect to the relative separation $\boldsymbol{r}$ gives an equation for $\widetilde{g}_{\alpha\beta}(\boldsymbol{k},\boldsymbol{v}_1, \boldsymbol{v}_2) = \widetilde{g}_{\alpha\beta}(12)$,
\begin{align}\label{eq:bbgky3}
    &\left(\frac{\partial}{\partial t} + \widetilde{\mathcal{L}}(12)\right)\widetilde{g}_{\alpha\beta}(12) \begin{aligned}[t]
    {}&+ \sum_\gamma \left(\int \rmd[3]\; \widetilde{\mathcal{V}}_{\alpha\gamma}(13)\,f_\alpha(1)\,\widetilde{g}_{\gamma\beta}(32) \right)\\
    &+ \sum_\gamma \left(\int \rmd[3]\; \widetilde{\mathcal{V}}_{\gamma\beta}(32)\,f_\beta(2)\,\widetilde{g}_{\alpha\gamma}(13) \right)\\
    &= -\widetilde{\mathcal{V}}_{\alpha\beta}(12)\,f_\alpha(1)f_\beta(2)\,.
    \end{aligned}
\end{align}
Here,
\begin{align}
    \widetilde{\mathcal{V}}_{\alpha\beta}(12) &= -
    \frac{4\upi q_\alpha q_\beta }{k^2}\,\imag\boldsymbol{k}\boldsymbol{\cdot}\left(\frac{1}{m_\alpha}\frac{\partial}{\partial\boldsymbol{v}_1} - \frac{1}{m_\beta}\frac{\partial}{\partial\boldsymbol{v}_2} \right),\\
        \widetilde{\mathcal{L}}(12) &= \imag\boldsymbol{k}\boldsymbol{\cdot}(\boldsymbol{v}_1 - \boldsymbol{v}_2)
\end{align}
are the Fourier-space versions of the operators introduced earlier and $\int \rmd[3] = \int \rmd\boldsymbol{v}_3$ (see appendix \ref{app:defs}).

\subsubsection{Electron-Ion Correlation}

As before, we need to find a way to exploit the small mass ratio. The only equation that involves both $m_e$ and $m_i$ is the equation for the mixed pair correlation $\widetilde{g}_{ei}$. We therefore begin with this equation by expanding the operators in $m_e/m_i\ll 1$. Estimating the sizes of the various terms in \eqref{eq:bbgky3}, we have
\begin{equation}
    \widetilde{\mathcal{L}}(12)\,\widetilde{g}_{ei}(12) = \imag\boldsymbol{k}\boldsymbol{\cdot}(\boldsymbol{v}_1 - \boldsymbol{v}_2)\,\widetilde{g}_{ei}(12) \approx \imag\boldsymbol{k\cdot v}_1\, \widetilde{g}_{ei}(12)
\end{equation}
and
\begin{equation}
    \int \rmd[2]\; \widetilde{\mathcal{V}}_{\alpha\beta}(12)\,f_\alpha(1) \sim \frac{q_\alpha q_\beta}{ k_D\sqrt{m_\alpha T_\alpha}}f_\alpha(1)\,.
\end{equation}
The second estimate means
\begin{align}
    \int \rmd[3]\;\widetilde{\mathcal{V}}_{e\gamma}(13)\,f_e(1)\,\widetilde{g}_{i\gamma}(32)&\gg \int \rmd[3]\;\widetilde{\mathcal{V}}_{\gamma i}(32)\,f_i(2)\,\widetilde{g}_{e\gamma}(13)\,,
\end{align}
because $m_eT_e \ll m_iT_i$ by assumption \eqref{ass:2}. Writing out the operators explicitly and dividing by $\imag\boldsymbol{k\cdot v}_1$, we find
\begin{align}\label{eq:geireplace}
    \widetilde{g}_{ei}(12) + \frac{4\upi e^2}{T_e}\frac{1}{k^2}f_e(1)\int \rmd[3]\;[\widetilde{g}_{ei}(32) - Z \widetilde{g}_{ii}(32)]     = \frac{4 \upi Ze^2}{T_e}\frac{1}{k^2}f_e(1)f_i(2)\,.
\end{align}
Integrating $\rmd[1]$ then gives an expression for $\int \rmd[3]\, \widetilde{g}_{ei}(32)$, which allows us to solve \eqref{eq:geireplace} for $\widetilde{g}_{ei}(12)$ at steady state. We obtain
\begin{equation}\label{eq:geisol}
    \widetilde{g}_{ei}(12) = \frac{1}{n_e} \frac{Zk_e^2}{k^2+k_e^2}f_e(1)\left(f_i(2) + \int \rmd[3]\, \widetilde{g}_{ii}(32) \right),
\end{equation}
which relates the electron-ion correlation to the ion-ion correlation. We will now use this relation to eliminate $\widetilde{g}_{ei}$ from the equations for $\widetilde{g}_{ee}$ and $\widetilde{g}_{ii}$.

\subsubsection{Electron-Electron and Ion-Ion Correlations}

The electron-electron and ion-ion correlations both satisfy an equation of the form
\begin{align}
    &\left(\frac{\partial}{\partial t} + \widetilde{\mathcal{L}}(12)\right)\widetilde{g}_{\alpha\alpha}(12) \begin{aligned}[t]
    {}&+ \sum_\gamma \left(\int \rmd[3]\; \widetilde{\mathcal{V}}_{\alpha\gamma}(13)\,f_\alpha(1)\,\widetilde{g}_{\gamma \alpha}(32) \right)\\
    &+ \sum_\gamma \left(\int \rmd[3]\; \widetilde{\mathcal{V}}_{\gamma \alpha}(32)\,f_\alpha(2)\,\widetilde{g}_{\alpha\gamma}(13) \right)\\
    &= -\widetilde{\mathcal{V}}_{\alpha\alpha}(12)\,f_\alpha(1)f_\alpha(2)\,.
    \end{aligned}
\end{align}
Writing the operators out explicitly, at steady state we have
\begin{align}\label{eq:gaaequation}
    \imag \boldsymbol{k}\boldsymbol{\cdot}(\boldsymbol{v}_1-\boldsymbol{v}_2)\,\widetilde{g}_{\alpha\alpha}(12) &+ \sum_\gamma\frac{4\upi q_\alpha q_\gamma}{T_\alpha}\frac{\imag\boldsymbol{k\cdot v}_1}{k^2}f_\alpha(1)\int \rmd[3]\; \widetilde{g}_{\gamma\alpha}(32)\\
    &\nonumber- \sum_\gamma\frac{4\upi q_\alpha q_\gamma}{T_\alpha}\frac{\imag\boldsymbol{k\cdot v}_2}{k^2}f_\alpha(2)\int \rmd[3]\; \widetilde{g}_{\alpha\gamma}(13)\\    
    \nonumber\phantom{\left(\frac{\partial}{\partial t} + \mathcal{L}(1) + \mathcal{L}(2)\right)g_{\alpha\alpha}(12)\, } &= -\frac{4 \upi q_\alpha^2}{T_\alpha}\frac{1}{k^2}\,\imag\boldsymbol{k}\boldsymbol{\cdot}(\boldsymbol{v}_1 - \boldsymbol{v}_2)\,f_\alpha(1)f_\alpha(2)\,.
\end{align}
This equation can easily be solved if we assume that $\widetilde{g}_{\alpha\beta}(\boldsymbol{k},\boldsymbol{v}_1,\boldsymbol{v}_2)$ depends on the wavevector $\boldsymbol{k}$ only through its magnitude $k$. Then, 
for \eqref{eq:gaaequation} to hold for all $\boldsymbol{k}$ and for all $\boldsymbol{v}_1$ and $\boldsymbol{v}_2$, we must have 
\begin{align}\label{eq:gaasolved}
    \widetilde{g}_{\alpha\alpha}(12) + \sum_\gamma\frac{4\upi q_\alpha q_\gamma}{T_\alpha}\frac{1}{k^2}f_\alpha(1)\int \rmd[3]\; \widetilde{g}_{\gamma\alpha}(32) = -\frac{4 \upi q_\alpha^2}{T_\alpha}\frac{1}{k^2}f_\alpha(1)f_\alpha(2)\,.
\end{align}%
Equation \eqref{eq:gaasolved} is almost identical to \eqref{eq:geireplace}, which determined $\widetilde{g}_{ei}$. The method used to solve that equation can be applied here, leading to
\begin{equation}
    \widetilde{g}_{ee}(12) = -\frac{1}{n_e} \frac{k_e^2}{k^2+k_e^2}f_e(1)\left(f_e(2) - Z\int \rmd[3]\, \widetilde{g}_{ie}(32) \right)
\end{equation}
for the electrons and
\begin{equation}\label{eq:ionsol}
    \widetilde{g}_{ii}(12) = -\frac{1}{n_i} \frac{k_i^2}{k^2+k_i^2}f_i(1)\left(f_i(2) - \frac{1}{Z}\int \rmd[3]\, \widetilde{g}_{ei}(32) \right)
\end{equation}
for the ions. Using our solution \eqref{eq:geisol} for $\widetilde{g}_{ei}$ in terms of $\widetilde{g}_{ii}$, together with ${\widetilde{g}_{ie}(\boldsymbol{k},\boldsymbol{v}_1,\boldsymbol{v}_2) = \widetilde{g}_{ei}(-\boldsymbol{k},\boldsymbol{v}_2,\boldsymbol{v}_1}$, gives
\begin{align}\label{eq:geesol}
    \widetilde{g}_{ee}(12) = \left[\frac{1}{n_e^2}\left( \frac{Zk_e^2}{k^2+k_e^2} \right)^2 \left( n_i + \int \rmd[34]\, \widetilde{g}_{ii}(34) \right) - \frac{1}{n_e}\frac{k_e^2}{k^2+k_e^2}\right]f_e(1)f_e(2)\,.
\end{align}
Here we can clearly see that the electron-electron correlation consists of two parts, one due to their interactions with the ions and one that is independent of the ions, as explained in \S\ref{sec:salpeter}.

\label{strongcoupling}Next we will calculate $\widetilde{g}_{ii}$, but it is first worth highlighting what has been achieved so far. Given any model for the ion-ion correlation, equations \eqref{eq:geisol} and \eqref{eq:geesol} determine the electron-ion and electron-electron correlations. These relations hold as long as the electrons interact weakly, so they constrain the electron correlations in a plasma with weakly or strongly coupled ions. This is possible because the electrons form completely symmetric, identical screening clouds around the ions, so we expect that knowing how the ions are distributed should tell us how the electrons are distributed. \cite{BoerckerMore1986} obtained the same relations by using the the electron Helmholtz free energy to construct an effective ion-ion interaction; this procedure is rigorously justified in \cite{Rosenfeld1994}.

We now solve \eqref{eq:ionsol}, which assumes the ions interact weakly, for $\widetilde{g}_{ii}$. Eliminating $\widetilde{g}_{ei}$ using \eqref{eq:geisol} gives
\begin{equation}\label{eq:ionclosed}
    \widetilde{g}_{ii}(12) = -\frac{1}{n_i} \frac{k_i^2}{k^2+k_i^2}f_i(1)\left(\frac{k^2}{k^2+k_e^2}f_i(2) - \frac{k_e^2}{k^2+k_e^2}\int \rmd[3]\, \widetilde{g}_{ii}(32) \right).
\end{equation}
Integrating $\rmd[1]$ and substituting into \eqref{eq:ionclosed} leads to
\begin{equation}
\widetilde{g}_{ii}(12) = -\frac{1}{n_i}\frac{k_i^2}{k^2+k_D^2}f_i(1)f_i(2)\,.
\end{equation}
We have calculated the ion-ion correlation, which now determines both the other correlations via \eqref{eq:geisol} and \eqref{eq:geesol}. The full set of pair correlations is
\begin{align}
&\label{eq:giifourier}\widetilde{g}_{ii}(12) = -\frac{4\upi Z^2 e^2}{T_i}\frac{1}{k^2+k_D^2}f_i(1)f_i(2)\,,\\
&\label{eq:geifourier}\widetilde{g}_{ei}(12) = \frac{4\upi Z e^2}{T_e}\frac{1}{k^2+k_D^2}f_e(1)f_i(2)\,,\\
&\label{eq:geefourier}\widetilde{g}_{ee}(12) = -\frac{4\upi e^2}{T_e}\left[\frac{T_i}{T_e}\frac{1}{k^2+k_D^2} + \left( \frac{T_e-T_i}{T_e} \right)\frac{1}{k^2+k_e^2}\right]f_e(1)f_e(2)\,.
\end{align}
Of course, these are consistent with \eqref{eq:ioncorr}--\eqref{eq:electroncorr} upon integrating over velocities and inverse-Fourier transforming. We now have the spatial and velocity parts of the pair correlations; the velocity dependence enters through a product of Maxwellians, as for systems at thermal equilibrium. 

\section{Response of Plasma to Sudden Electron Heating}\label{sec:response}

\subsection{Correlation Heating Mechanism}

The main objective of this paper is to answer the following question. Suppose some energy $Q$ per particle is instantaneously supplied to the electrons in a moderately coupled two-temperature plasma. The hotter electrons will now shield the ions less; what effect does this have on the ion temperature?

The answer is there is a small but rapid increase in the ion temperature, on the ion-plasma-frequency timescale $t\sim 1/\omega_{pi}$. To understand why, consider the extreme case of very cold electrons which screen the ions perfectly. The ions will be distributed randomly, as they do not interact. Suppose the ions are also very cold, so they are essentially stationary.

Now imagine the electrons are suddenly heated, which means their Debye length increases. As a result, the ions are not so well screened and will repel each other. This acceleration increases their average kinetic energy. The effect is particularly noticeable for pairs, or larger clusters, of ions that were initially quite close together: the sudden removal of their shielding clouds means they repel each other and fly apart. The ions were initially stationary but have now been given some kinetic energy, so their temperature must have increased. This mechanism is most effective at producing fast ions if the electrons shielding is stripped away rapidly.

\begin{figure}
\vspace{4mm}
    \centering
    \mbox{}\hfill
    \begin{subfigure}[t]{.45\textwidth}
    \centering
    \begin{tikzpicture}
    \draw[thick] (0,0) rectangle (4,4);
    \draw[clip] (0,0) rectangle (4,4);
    \pgfmathsetseed{24122015}
    \foreach \p in {1,...,700}
    { \fill[black]  (2.5*rand+2.5,2.5*rand+2.5) circle (0.05);
    }
    \end{tikzpicture}
    \caption{Before heating: random, uncorrelated positions}
    \end{subfigure}\hfill
    \begin{subfigure}[t]{.45\textwidth}
    \centering
    \includegraphics{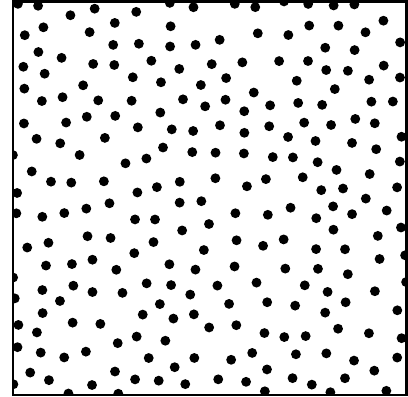}
    \caption{After heating: repulsion leads to more ordered, correlated positions}
\end{subfigure}
\hfill\mbox{}
\caption{Cartoon illustration of ion positions before and after ionisation and disorder-induced heating}
\label{fig:correlations}
\end{figure}
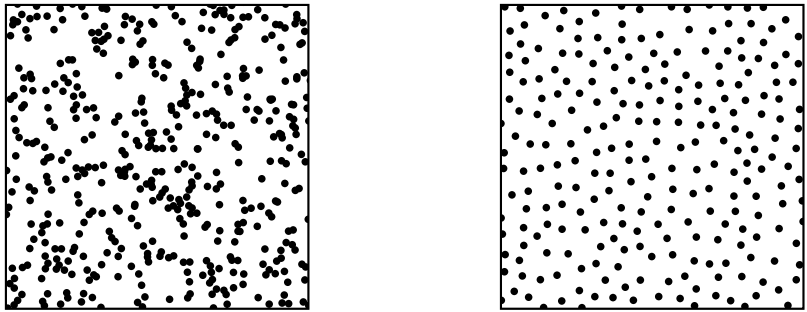

Disorder-induced heating of ultracold neutral plasmas has exactly the same underlying mechanism, except the starting point is a neutral gas rather than well-shielded ions \cite[see e.g.][]{Acciarri2022}. In this context, a common explanation is that the ions transition from an initial state of random, uncorrelated positions to a final state with correlations imposed by their Coulomb repulsion. This is depicted in figure \ref{fig:correlations}; their repulsion after the electrons are heated means the ions push away from each other and are generally not found close together any more. This means their spatial correlation has changed and become more ordered, which leads to heating; the timescale for the ion-ion correlation to evolve is $t\sim 1/\omega_{pi}$, which is the time it takes a thermal ion to traverse a Debye length.

This temperature increase can also be understood as a consequence of the Second Law of Thermodynamics; it is required to compensate for the decrease in ion entropy when they become more spatially ordered. For an entropic perspective on correlation heating, see \cite{Fetsch2023}.

\subsection{Multiple Timescales}\label{sec:multipletimescales}

The response of the plasma occurs in stages over multiple timescales. 

Before the heating at $t=0$, the pair correlations in Fourier space are given by \eqref{eq:giifourier}\nobreakdash--\eqref{eq:geefourier}:
\begin{align}
&\label{eq:giifourier2}\widetilde{g}_{ii}(12) = -\frac{4\upi Z^2 e^2}{T_{i0}}\frac{1}{k^2+k_{D0}^2}f_{i0}(1)f_{i0}(2)\,,\\
&\label{eq:geifourier2}\widetilde{g}_{ei}(12) = \frac{4\upi Z e^2}{T_{e0}}\frac{1}{k^2+k_{D0}^2}f_{e0}(1)f_{i0}(2)\,,\\
&\label{eq:geefourier2}\widetilde{g}_{ee}(12) = -\frac{4\upi e^2}{T_{e0}}\left[\frac{T_{i0}}{T_{e0}}\frac{1}{k^2+k_{D0}^2} + \left( \frac{T_{e0}-T_{i0}}{T_{e0}} \right)\frac{1}{k^2+k_{e0}^2}\right]f_{e0}(1)f_{e0}(2)\,,
\end{align}
where the `$0$' subscript denotes the initial value of a quantity. This includes the initial electron and ion velocity distributions, which are $f_{e0}$ and $f_{i0}$ respectively. 

First, the electrons are instantaneously given a larger speed. Their velocity distribution will change in a way that depends on the exact mechanism of the heating, but the spatial pair correlations cannot change because this instantaneous heating takes place with fixed particle positions.

\begin{figure}
\vspace{4mm}
    \centering
    \begin{tikzpicture}[scale=0.75]

    \draw[line width=2mm, -stealth, lightgray] (9, 9) -- (9, 0.5);

    \draw(9.02,8.8)  node[scale=3,label={[align=left]$\boldsymbol{t=0}$}](){};
     
    \draw(2.7,8.2)  node[scale=3,label={[align=center]Electrons instantaneously heated}](){};
    \draw[thick] (6,8.97) -- (9,8.97);
    \draw  (8.9,8.97)  node[scale=5](O){.};
      
    \draw(2.7,6.55)  node[scale=3,label={[align=center]Electron shielding clouds expand\\ and  electron correlations adjust}](){};
    \draw[thick] (6,7.5) -- (9,7.5);
    \draw  (8.9,7.5)  node[scale=5](O){.};
    \draw(8,7.15)  node[scale=3,label={[align=center]$\sim 1/\omega_{pe}$}](){};
    
    \draw(2.8,5.35)  node[scale=3,label={[align=center]Electrons re-Maxwellianise}](){};
    \draw[thick] (5.6,6) -- (9,6);
    \draw  (8.9,6)  node[scale=5](O){.};
    \draw(8,5.65)  node[scale=3,label={[align=center]$\sim \tau_{ee}$}](){};
    
    \draw(2.8,3.55)  node[scale=3,label={[align=center]Ion correlations adjust and\\ ions heat up}](){};
    \draw[thick] (5.6,4.5) -- (9,4.5);
    \draw  (8.9,4.5)  node[scale=5](O){.};
    \draw(8,4.15)  node[scale=3,label={[align=center]$\sim 1/\omega_{pi}$}](){};
    
     \draw(2.8,2.35)  node[scale=3,label={[align=center]Ions re-Maxwellianise}](){};
     \draw[thick] (5.3,3) -- (9,3);
    \draw  (8.9,3)  node[scale=5](O){.};
    \draw(8,2.65)  node[scale=3,label={[align=center]$\sim \tau_{ii}$}](){};
     
    \draw(2.8,0.65)  node[scale=3,label={[align=center]Thermal equilibration,\\ ion-electron collisions}](){};
     \draw[thick] (5.3,1.5) -- (9,1.5);
    \draw  (8.9,1.5)  node[scale=5](O){.};
    \draw(8,1.15)  node[scale=3,label={[align=center]$\sim \tau_{ie}$}](){};
    
    \end{tikzpicture}
    \caption{Timeline of the system's response to sudden electron heating.}
    \label{fig:timeline}
\end{figure}
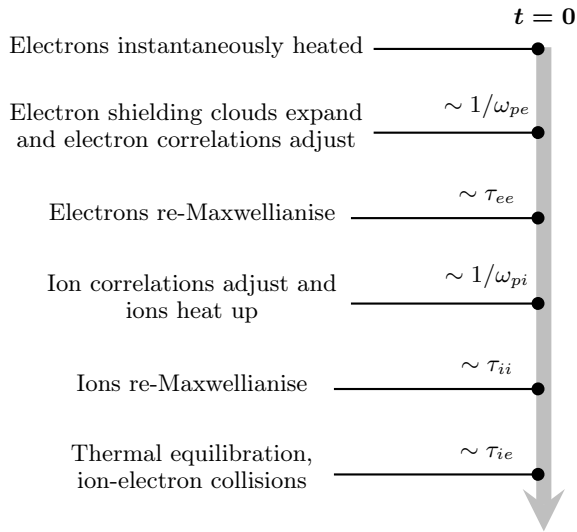

Next, the electron-electron and electron-ion correlations will adjust on an electron-plasma-frequency timescale $t\sim 1/\omega_{pe}$. The increase in temperature makes the electron Debye length larger, so the electron shielding clouds must rapidly expand.

Then, over the electron-electron collision timescale $t\sim \tau_{ee}$, the electron velocity distribution will become Maxwellian again, with new temperature $T_{e1}$. The electron correlations $g_{ee}$ and $g_{ei}$ will continually adjust as this occurs and will approach their steady-state solutions given constant $f_i$ and $g_{ii}$. At the end of this stage of the process, the electrons are Maxwellian, and the steady-state solutions for the correlations are given in \eqref{eq:geisol} and \eqref{eq:geesol}\footnote{In fact, for arbitrary initial conditions it is not quite true that the correlations approach these steady-state forms. There will always be ballistic terms that do not decay and that depend on the initial state. However, at long times these ballistic terms do not contribute to any velocity integrals; they phase-mix away. Since we are only ever interested in integrals of the correlations over at least one of the velocities, we ignore this subtlety.}: we must have
\begin{align}
\label{eq:geisol2}&\widetilde{g}_{ei}(12) = \frac{1}{n_e} \frac{Zk_{e1}^2}{k^2+k_{e1}^2}f_{e1}(1)\left(f_{i0}(2) + \int \rmd[3]\, \widetilde{g}_{ii}(32) \right)\\
    &\widetilde{g}_{ee}(12) = 
    \left[\frac{1}{n_e^2}\left( \frac{Zk_{e1}^2}{k^2+k_{e1}^2} \right)^2 \left( n_i + \int \rmd[34]\, \widetilde{g}_{ii}(34) \right) - \frac{1}{n_e}\frac{k_e^2}{k^2+k_{e1}^2}\right]f_{e1}(1)f_{e1}(2)\,,\label{eq:geesol2}
\end{align}
where $k_{e1}$ is the new inverse-Debye length of the electrons at temperature $T_{e1}$ and $f_{e1}$ is their new velocity distribution. Note that these two formulas assumed Maxwellian electrons but did not require Maxwellian ions, so they do not involve an ion temperature.

The ion-ion correlation changes on the ion-plasma-frequency timescale $t\sim 1/\omega_{pi}$, so the ions have not yet had time to adjust. We still have
\begin{align}\label{eq:gii0}
\widetilde{g}_{ii}(12) = -\frac{4\upi Z^2 e^2}{T_{i0}}\frac{1}{k^2+k_{D0}^2}f_{i0}(1)f_{i0}(2)\,.
\end{align}
The assumption of instantaneous heating is an idealisation that is not actually necessary. Our calculations are valid as long as the electrons are heated fast enough that they can Maxwellianise before the ions respond.

Next, over the ion-plasma-frequency timescale, the ion-ion correlation will change. The ions are screened less by the hotter electrons, so they push each other further apart.

This is the critical timescale on which collisionless heating occurs. At any moment, just as in \eqref{eq:geisol2} and \eqref{eq:geesol2}, the electron distribution $f_e$ and correlations $g_{ei}$ and $g_{ee}$ may be assumed to have relaxed to their steady-state solutions given constant $f_i$ and $g_{ii}$, because the electron evolution timescales ($1/\omega_{pe}$ and $\tau_{ee}$) are much shorter than the timescale of interest ($1/\omega_{pi}$).

Finally, on the ion-ion collisional timescale $t\sim \tau_{ii}$ the ion distribution will become Maxwellian again. This step is important because correlation heating causes the ion distribution to deviate from a Maxwellian. The correlations all adjust continually as this occurs, but we will show that there is no further change in the ion kinetic energy during this last stage. If the final electron and ion temperatures are $T_{e2}$ and $T_{i2}$ respectively, then at the end we must have correlations in their steady-state form again:
\begin{align}
&\label{eq:giifourier3}\widetilde{g}_{ii}(12) = -\frac{4\upi Z^2 e^2}{T_{i2}}\frac{1}{k^2+k_{D2}^2}f_{i2}(1)f_{i2}(2)\\
&\label{eq:geifourier3}\widetilde{g}_{ei}(12) = \frac{4\upi Z e^2}{T_{e2}}\frac{1}{k^2+k_{D2}^2}f_{e2}(1)f_{i2}(2)\\
&\label{eq:geefourier3}\widetilde{g}_{ee}(12) = -\frac{4\upi e^2}{T_{e2}}\left[\frac{T_{i2}}{T_{e2}}\frac{1}{k^2+k_{D2}^2} + \left( \frac{T_{e2}-T_{i2}}{T_{e2}} \right)\frac{1}{k^2+k_{e2}^2}\right]f_{e2}(1)f_{e2}(2)\,.
\end{align}
There is also a much longer timescale of temperature equilibration. However, we are not interested in processes that occur over such long times. Figure \ref{fig:timeline} summarises the entire correlation heating process.

\subsection{Ion Evolution}

To calculate how much the ion temperature increases, we start with equation \eqref{eq:bbgky1} for the ion distribution, repeated here:
\begin{align}
\left(\frac{\partial}{\partial t} + \mathcal{L}(1)\right) f_i(1) = -\sum_\beta \left( \int \rmd(2)\; \mathcal{V}_{\beta i}(21)\,g_{\beta i}(21) \right)\,.
\end{align}
On the right-hand side, particles $1$ and $2$ and their corresponding species labels were swapped for convenience. In a homogeneous plasma this is equivalent to
\begin{equation}
\frac{\partial f_i(1)}{\partial t} = -\int \frac{\rmd\boldsymbol{k}}{(2\upi)^3}\int \rmd[2] \left(\widetilde{\mathcal{V}}_{ei}(21)^*\,\widetilde{g}_{ei}(21) + \widetilde{\mathcal{V}}_{ii}(21)^*\,\widetilde{g}_{ii}(21) \right).
\end{equation}
As explained, we use the steady-state solution for $g_{ei}$ given Maxwellian electrons and fixed $f_i$ and $g_{ii}$, equation \eqref{eq:geisol}, to eliminate the electrons from this equation. We obtain
\begin{align}\label{eq:fiequation}
\frac{\partial f_i(1)}{\partial t} &= -\int \frac{\rmd\boldsymbol{k}}{(2\upi)^3}\int \rmd[2] \left( \frac{4\upi Z^2e^2}{m_i}\frac{\imag\boldsymbol{k}}{k^2+k_e^2}\boldsymbol{\cdot}\frac{\partial}{\partial\boldsymbol{v}_1}\widetilde{g}_{ii}(21) \right).
\end{align}
In real space, this reads
\begin{equation}
    \frac{\partial f_i(1)}{\partial t} = - \int \rmd(2)\, \mathcal{V}^s(12)\,g_{ii}(12),
\end{equation}
where we have defined an effective interaction operator
\begin{align}
    \mathcal{V}^s(12) &= -\frac{1}{m_i}\frac{\partial}{\partial(\boldsymbol{r}_1 - \boldsymbol{r}_2)}\left( \frac{Z^2e^2}{|\boldsymbol{r}_1 - \boldsymbol{r}_2|}\,\rme^{-k_e|\boldsymbol{r}_1 - \boldsymbol{r}_2}| \right)
    \boldsymbol{\cdot}\left(\frac{\partial}{\partial\boldsymbol{v}_1} - \frac{\partial}{\partial\boldsymbol{v}_2}\right).
\end{align}
Ion-ion correlations drive changes in the ion distribution function as if the ions were a gas of particles interacting via the screened Coulomb potential ${\varphi^{(s)}(\boldsymbol{r}) = Z^2e^2 \exp(-k_er)/r}$.

The ion distribution function changes when the ion correlation has not yet reached its steady-state form. To calculate how much the ion kinetic energy changes, we will need to know how $g_{ii}$ evolves with time.

When the ions are weakly coupled, the BBGKY equation \eqref{eq:bbgky3} for $\widetilde{g}_{ii}$ is
\begin{align}
    &\left(\frac{\partial}{\partial t} + \widetilde{\mathcal{L}}(12)\right)\widetilde{g}_{ii}(12) \begin{aligned}[t]
    {}&+ \sum_\gamma \left(\int \rmd[3]\; \widetilde{\mathcal{V}}_{i\gamma}(13)\,f_i(1)\,\widetilde{g}_{\gamma i}(32) \right)\\
    &+ \sum_\gamma \left(\int \rmd[3]\; \widetilde{\mathcal{V}}_{\gamma i}(32)\,f_i(2)\,\widetilde{g}_{i\gamma}(13) \right)\\
    &= -\widetilde{\mathcal{V}}_{ii}(12)\,f_i(1)f_i(2)\,.
    \end{aligned}
\end{align}
The electron-ion correlation appears in the three-particle interaction terms, for example
\begin{align}
    \sum_\gamma \left(\int \rmd[3]\; \widetilde{\mathcal{V}}_{i\gamma}(13)\,f_i(1)\,\widetilde{g}_{\gamma i}(32) \right) = &- \frac{4\upi Z^2e^2}{m_i}\frac{\imag\boldsymbol{k}}{k^2}\boldsymbol{\cdot}\frac{\partial f_i(1)}{\partial\boldsymbol{v}_1} \int \rmd[3]\,\widetilde{g}_{ii}(32)\\
    &\nonumber+ \frac{4\upi Ze^2}{m_i}\frac{\imag\boldsymbol{k}}{k^2}\boldsymbol{\cdot}\frac{\partial f_i(1)}{\partial\boldsymbol{v}_1} \int \rmd[3]\,\widetilde{g}_{ei}(32)\,.
\end{align}
Substituting \eqref{eq:geisol2} in for $\widetilde{g}_{ei}$ as before gives
\begin{align}
    \sum_\gamma \left(\int \rmd[3]\; \widetilde{\mathcal{V}}_{i\gamma}(13)\,f_i(1)\,\widetilde{g}_{\gamma i}(32) \right) = &- \frac{4\upi Z^2e^2}{m_i}\frac{\imag\boldsymbol{k}}{k^2+k_e^2}\boldsymbol{\cdot}\frac{\partial f_i(1)}{\partial\boldsymbol{v}_1} \int \rmd[3]\,\widetilde{g}_{ii}(32)\\
    &\nonumber+ \frac{4\upi Z^2e^2}{m_i}\frac{k_e^2}{k^2+k_e^2}\frac{\imag\boldsymbol{k}}{k^2}\boldsymbol{\cdot}\frac{\partial f_i(1)}{\partial\boldsymbol{v}_1} f_i(2)\,.
\end{align}
Using the same substitution in the other terms, we therefore find
\begin{align}
    \left(\frac{\partial}{\partial t} + \widetilde{\mathcal{L}}(12)\right)\widetilde{g}_{ii}(12) &- \frac{4\upi Z^2e^2}{m_i}\frac{\imag\boldsymbol{k}}{k^2+k_e^2}\boldsymbol{\cdot}\frac{\partial f_i(1)}{\partial\boldsymbol{v}_1} \int \rmd[3]\;\widetilde{g}_{ii}(32)\\
    &\nonumber- \frac{4\upi Z^2e^2}{m_i}\frac{\imag\boldsymbol{k}}{k^2+k_e^2}\boldsymbol{\cdot}\frac{\partial f_i(2)}{\partial\boldsymbol{v}_2} \int \rmd[3]\;\widetilde{g}_{ii}(13)\\
    &\nonumber= + \frac{4\upi Z^2e^2}{m_i}\frac{\imag\boldsymbol{k}}{k^2+k_e^2}\boldsymbol{\cdot}\left(\frac{\partial }{\partial\boldsymbol{v}_1} - \frac{\partial }{\partial\boldsymbol{v}_2}\right) f_i(1)f_i(2)\,.
\end{align}
Returning to real space, we now have a system of equations determining the evolution of $f_i$ and $g_{ii}$:
\begin{align}
&\label{eq:ionevol}\frac{\partial f_i(1)}{\partial t} = - \int \rmd(2)\, \mathcal{V}^s(12)\,g_{ii}(12)\,,\\
\label{eq:giieff}&\left(\frac{\partial}{\partial t} + \mathcal{L}(12)\right)g_{ii}(12) \begin{aligned}[t]
{}&+ \int \rmd(3)\; \mathcal{V}^s(13)\,f_i(1)\,g_{ii}(32)\\
&+ \int \rmd(3)\; \mathcal{V}^s(32)\,f_i(2)\,g_{ii}(13) = -\mathcal{V}^s(12)\,f_i(1)f_i(2)\,.
\end{aligned}
\end{align}
The ion distribution function and ion-ion pair correlation evolve according to \eqref{eq:ionevol} and \eqref{eq:giieff}. The electrons do not enter these equations anywhere, except through the temperature $T_e$ which controls the screening length $k_e$. These equations are precisely the equations that would describe a gas of particles interacting via screened Yukawa potentials, although here the screening length $k_e$ is also a dynamical variable that changes with time. The ions therefore behave like a weakly coupled Yukawa One-Component Plasma (YOCP). This result, which is consistent with the physical picture developed in \S\ref{sec:salpeter}, will allow us to calculate the temperature changes during correlation heating in a very simple way.

\section{Temperature Changes Due to Correlation Heating}\label{sec:temps}

In this section we calculate how the temperatures of each species change in response to a sudden input of energy to the electrons. 

First, we review the sequence of temperatures defined in \S\ref{sec:multipletimescales}. The electron and ion temperatures start out at $T_{e0}$ and $T_{i0}$ respectively, before each electron's kinetic energy is suddenly increased by $Q$. In general, the electron distribution may not be Maxwellian immediately after this heating; however, their total kinetic energy must change from $3N_eT_{e0}/2$ to $3N_eT_{e0}/2 + N_e Q$. Next, the electron shielding clouds expand and they  reach a new Maxwellian distribution at temperature $T_{e1}$, all while the ion temperature remains $T_{i0}$. Then, the ions move further apart (the ion-ion correlation changes) and their kinetic energy increases due to correlation heating. Eventually, the ions re-Maxwellianise and the final temperatures are $T_{e2}$ and $T_{i2}$.

We expect the changes in particle correlations to cause temperature changes that are small in $1/\Lambda$ because this heating relies on conversion of potential energy into kinetic energy, and the potential energy in a moderately coupled plasma is a factor of $1/\Lambda$ smaller than the kinetic energies. This is consistent with the fact that, for an ideal plasma with $\Lambda\to\infty$, the species temperatures are completely decoupled and there should be no correlation heating. So, $T_{e1}$ and $T_{e2}$ are close to $T_{e0}+2Q/3$, while $T_{i2}$ is close to $T_{i0}$.

\subsection{Conservation of Energy}

The BBGKY equations for a system of particles with interaction potential $\varphi_{\alpha\beta}$ between species $\alpha$ and species $\beta$ conserve energy. The total plasma energy $U$ is the sum of the kinetic energy of each species and the potential energy, which depends on the relative positions of the particles and must therefore be calculated from the pair correlations. It is given by \cite[]{LandauLifshitz}
\begin{equation}
\label{eq:energy2}
U = \sum_\alpha V\int \rmd\boldsymbol{v}\,\tfrac{1}{2}m_\alpha\boldsymbol{v}^2 f_\alpha(\boldsymbol{v}) + \frac{1}{2}\sum_{\alpha}\sum_{\beta} Vn_\alpha n_\beta\int \rmd\boldsymbol{r}\, \varphi_{\alpha\beta}(\boldsymbol{r})\, G_{\alpha\beta}(\boldsymbol{r})\,,
\end{equation}
where $V$ is the system volume. For example, the energy of a steady-state two-temperature plasma, which is found using the pair correlations \eqref{eq:ioncorr}--\eqref{eq:electroncorr} and Coulomb potential $\varphi_{\alpha\beta}(\boldsymbol{r}) = q_\alpha q_\beta/r$, is:
\begin{equation}\label{eq:internalenergy}
U = \frac{3}{2}N_eT_e + \frac{3}{2}N_iT_i -\frac{V}{8\upi}\left[T_i(k_D^3 - k_e^3) + T_ek_e^3\right].
\end{equation}
This calculation was carried out in \cite{More1980}, although it seems the correct formula \eqref{eq:internalenergy} was first published in \cite{Triola2022}. An alternative derivation appears in \cite{Fetsch2023}. Note that, when $Z\sim 1$ and $T_e\sim T_i$, the potential energy correction is $\sim NT/\Lambda$ as expected.

\subsection{Ion Temperature $T_{i2}$}\label{sec:iontemp}

A central aim of this paper is to calculate the ion temperature after correlation heating. It is not possible to find $T_{i2}$ just using conservation of the total plasma energy, because the final energy depends on two unknowns, $T_{e2}$ and $T_{i2}$; one conservation equation is not enough to determine both temperatures.

However, at the end of \S\ref{sec:response} we found that the ion distribution function and ion-ion pair correlation satisfy the same equations as a weakly coupled YOCP with screening length $k_e$. Over the course of the ion correlation adjustment and subsequent re-Maxwellianisation, the electron temperature changes from $T_{e1}$ to $T_{e2}$, so this screening length is not constant. However, the variation in $k_e$ is small in $1/\Lambda$ and is therefore a higher-order effect which can be neglected for the purposes of obtaining a leading-order expression for the ion heating.

\begin{figure}
\centering
\includegraphics[scale=0.6]{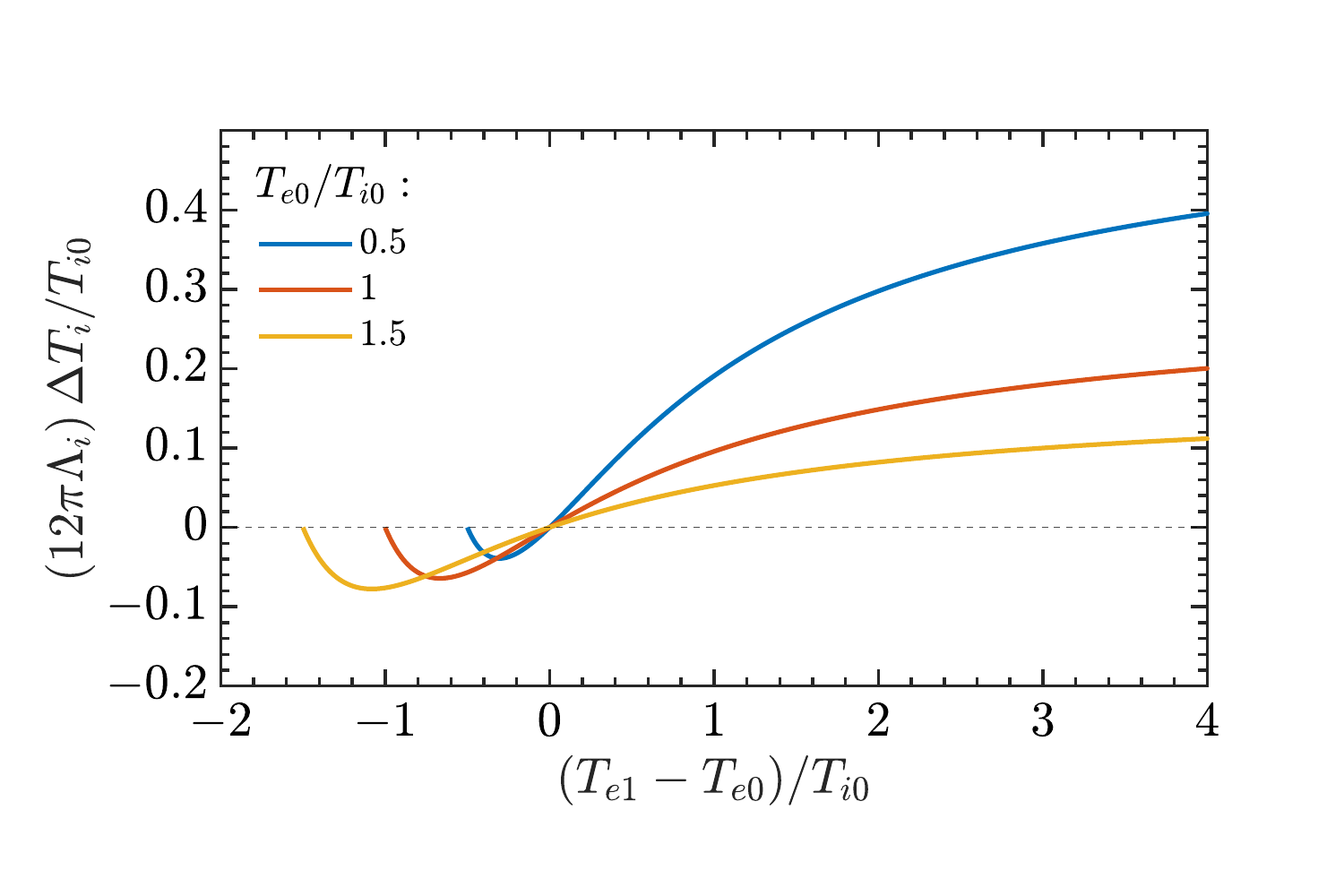}
\vspace*{-5mm}
\caption{Plot of fractional change in ion temperature $\Delta T_i/T_{i0}$ against the change in electron temperature, for three different initial electron temperatures. Here, $Z=1$ and we normalise using ${\Lambda_i = n_i / k_{i0}^3}$, which means the actual temperature change is roughly a factor of the plasma parameter smaller than the values on these curves.  Note that the curves do not continue to arbitrarily negative abscissae because $T_{e1}$ cannot be reduced below zero.} 
\label{DeltaTi}
\end{figure}

Thus, the screening length can be treated as fixed, $k_e=k_{e1}$. The ion equations \eqref{eq:ionevol} and \eqref{eq:giieff} then have a conserved energy,
\begin{equation}\label{eq:ionenergy}
E = \int \rmd\boldsymbol{v}\, \tfrac{1}{2}\,m_i\boldsymbol{v}^2 f_i(\boldsymbol{v}) + \frac{1}{2}Vn_i^2\int \rmd\boldsymbol{r} \left(\frac{Z^2e^2}{r}\,\rme^{-k_{e1}r}\right) G_{ii}(\boldsymbol{r})\,,
\end{equation}
which is the energy the ions would have if they were a YOCP. Initially, before the ion correlation adjusts, we have $G_{ii}(\boldsymbol{r},t=0)=-(Z^2e^2/T_{i0})(\rme^{-k_{D0}r}/r)$ and the energy is
\begin{equation}
    E_1 = \frac{3}{2}N_iT_{i0} - \frac{1}{8\upi}VT_{i0}\frac{k_{i0}^4}{k_{D0}+k_{e1}}\,.
\end{equation}
After correlation heating, the ion correlation is $G_{ii}(\boldsymbol{r},t=\infty) = -(Z^2e^2/T_{i2})(\rme^{-k_{D2}r}/r)$, which is equivalent to $G_{ii}(\boldsymbol{r},t=\infty) = -(Z^2e^2/T_{i0})(\rme^{-k_{D1}r}/r)$ up to corrections small in $1/\Lambda$. The final energy is therefore
\begin{equation}
    E_2 = \frac{3}{2}N_iT_{i2} - \frac{1}{8\upi}VT_{i0}\frac{k_{i0}^4}{k_{D1}+k_{e1}}\,.
\end{equation}
Equating $E_1$ and $E_2$ allows us to determine the change in ion temperature, ${T_{i2} - T_{i1} = \Delta T_i}$:
\begin{equation}\label{eq:ionke}
\Delta T_i = \frac{Z^2e^2}{3}k_{i0}^2\left(\frac{1}{k_{D1}+k_{e1}} - \frac{1}{k_{D0}+k_{e1}} \right).
\end{equation}
This is one of our main results.
\begin{itemize}
    \item When the electrons are not heated at all, $k_{D1}=k_{D0}$ and so $\Delta T_i = 0$. Nothing happens.
    \item If the electrons are heated then $k_{e1}<k_{e0}$. In this case $\Delta T_i > 0$, and the ions also heat up. On the other hand, if the electrons are suddenly cooled, then the ions also cool.
    \item As expected, the change in the ion temperature is small in a moderately coupled plasma because $\Delta T_i \sim e^2 k_D \sim T/\Lambda$. However, the temperature change is not small in the mass ratio $m_e/m_i$; it is nonzero, even though we took $m_e/m_i \to 0$ in the derivation. 
    \item If the electrons become much hotter than the ions, meaning $T_e \gg T_i$, then the ion heating approaches a finite limit that depends on the initial conditions:
    \begin{equation}
        \Delta T_i = \frac{Z^2e^2}{3}k_{i0}^2\left( \frac{1}{k_{i0}} - \frac{1}{k_{D0}} \right).
    \end{equation}
    \item On the other hand, if the electrons are suddenly cooled so that $T_e\ll T_i$, then ${\Delta T_i \to 0}$. Physically, the reason is that cooling the electrons causes them to form small shielding clouds around each of the ions, which become perfectly screened. Then, the ions no longer interact and so must continue moving in a straight line at a constant speed; there is no heating.
    \end{itemize}
Expression \eqref{eq:ionke} is plotted in figure \ref{DeltaTi} for different ratios of initial electron temperature to initial ion temperature. The figure shows that $\Delta T_i$ approaches a constant, maximum achievable value when the electron heating is large, and that it vanishes when the electrons are cooled to zero temperature. It shows that correlation heating has the greatest effect on the ions when the electrons are started at a low temperature.

\subsection{Electron Temperature $T_{e1}$}

We now work out how the electron temperature changes. This occurs in two stages: first, they are suddenly heated and they re-Maxwellianise before the ions respond. At this point the electron temperature is $T_{e1}$. Then, once the ion correlations have adjusted, the electrons will reach a new temperature $T_{e2}$.

The potential energy is easiest to calculate using Fourier-space pair correlations. The energy formula \eqref{eq:energy2} is equivalent to
\begin{align}
U = \sum_\alpha V\int \rmd\boldsymbol{v}\,\tfrac{1}{2}m_\alpha\boldsymbol{v}^2 f_\alpha(\boldsymbol{v}) + \frac{1}{2}\sum_{\alpha}\sum_{\beta} Vn_\alpha n_\beta\int \frac{\rmd\boldsymbol{k}}{(2\upi)^3}\, \widetilde{\varphi}_{\alpha\beta}(\boldsymbol{k})\, \widetilde{G}_{\alpha\beta}(\boldsymbol{k})\,,
\end{align}
with $\widetilde{\varphi}_{\alpha\beta}(\boldsymbol{k}) = 4\upi q_\alpha q_\beta / k^2$ for the Coulomb potential. This expression involves all three correlations $G_{ee}$, $G_{ei}$ and $G_{ii}$ in the potential energy. However, we have already seen that the electron-electron and electron-ion correlations are determined by the slowly-varying ion correlation. Assuming each species is Maxwellian, we now use \eqref{eq:geisol2} and \eqref{eq:geesol2} integrated over velocities,
\begin{align}
    \widetilde{G}_{ei}(\boldsymbol{k}) &= \frac{1}{n_i}\frac{k_e^2}{k^2+k_e^2}\left( 1 + n_i\widetilde{G}_{ii}(\boldsymbol{k}) \right),\\
    \widetilde{G}_{ee}(\boldsymbol{k}) &=  \frac{1}{n_i} \left(\frac{k_e^2}{k^2+k_e^2}\right)^2 \left( 1 + n_i\widetilde{G}_{ii}(\boldsymbol{k}) \right) - \frac{1}{n_e}\frac{k_e^2}{k^2+k_e^2}\,,
\end{align}
to express the potential energy in terms of the ion correlation only. We find
\begin{align}\label{eq:energy}
    U = \frac{3}{2}N_eT_e + \frac{3}{2}N_iT_i &- \frac{1}{2}N_e e^2 \left(\frac{3Z}{2}+1\right)k_e+ 2\upi n_i^2 VZ^2e^2\int \frac{\rmd\boldsymbol{k}}{(2\upi)^3}\frac{k^2\,\widetilde{G}_{ii}(\boldsymbol{k})}{(k^2+k_e^2)^2}\,.
\end{align}
This formula for the energy of a two-temperature plasma with arbitrary ion correlation is interesting in its own right. It uses the fact that the electrons interact weakly but does not assume the same about the ions, as explained on p.\pageref{strongcoupling}.

It is common to model the ions as a YOCP in theoretical and computational work \cite[e.g. ][]{Hamaguchi1997,Murillo2001,Gericke2003ii,Murillo2009,Lyon2013,Langin2016,Sprenkle2022}; this picture was already useful for understanding the steady-state properties of a two-temperature plasma in \S\ref{sec:salpeter}. However, when such a plasma is not at steady state, the electron temperature does not necessarily remain constant, so the screening length is a dynamical variable which should be evolved along with the ion distribution function. Conservation of energy, using the potential energy in \eqref{eq:energy}, allows the electron temperature to be calculated at any given time from the ion-ion correlation, without reference to the fast electron dynamics. The fact that the effective interaction range of the ions is a dynamical variable that changes with time could be important when studying strongly coupled plasmas.

Now using the initial ion correlation $G_{ii}(\boldsymbol{r},t=0) = -(Z^2e^2/T_{i0})(\rme^{-k_{D0}r}/r)$ in \eqref{eq:energy}, we find the initial energy is
\begin{equation}
U_0 = \frac{3}{2}N_e T_{e0} + \frac{3}{2}N_i T_{i0} - \frac{1}{2}N_ee^2\left( \frac{3Z}{2} + 1 \right)k_{e0} - \frac{1}{4}N_iZ^2e^2k_{i0}^2\frac{2k_{D0}+k_{e0}}{(k_{D0}+k_{e0})^2},
\end{equation}
and the energy after electron heating, but before the ion correlations adjust, is
\begin{equation}\label{eq:u1}
    U_1 = \frac{3}{2}N_e T_{e1} + \frac{3}{2}N_i T_{i0} - \frac{1}{2}N_ee^2\left( \frac{3Z}{2} + 1 \right)k_{e1} - \frac{1}{4}N_iZ^2e^2k_{i0}^2\frac{2k_{D0}+k_{e1}}{(k_{D0}+k_{e1})^2}\,.
\end{equation}
In the potential energy term, $k_{e1}$ depends on $T_{e1}$, which is the unknown that we are trying to find. However, the potential energy term is small in $1/\Lambda$, so we can use the leading order formula
\begin{equation}
k_{e1} = \left(\frac{4\upi n_e e^2}{T_{e0}+2Q/3}\right)^{1/2},
\end{equation}
since the difference between $T_{e1}$ and $T_{e0}+2Q/3$ is higher-order in $1/\Lambda$. 

We find the unknown temperature $T_{e1}$ using conservation of energy. Equating the final energy after the ion adjustment to the energy just after heating, $U_1 = U_0 + N_e Q$, gives
\begin{align}\label{eq:t1t0}
T_{e1} - T_{e0} = \frac{2}{3} Q &- \frac{1}{2}N_ee^2\left( \frac{3Z}{2} + 1 \right)(k_{e0}-k_{e1})\\
\nonumber&+ \frac{1}{4}N_iZ^2e^2k_{i0}^2\left(\frac{2k_{D0}+k_{e1}}{(k_{D0}+k_{e1})^2} - \frac{2k_{D0}+k_{e0}}{(k_{D0}+k_{e0})^2} \right).
\end{align}
Expression \eqref{eq:t1t0} is not particularly enlightening, but it can be used to prove that ${T_{e1}-T_{e0} < 2Q/3}$, so the electrons do not heat up quite as much as they would if there were an ideal gas (the opposite is true if they are cooled).

\subsection{Electron Temperature $T_{e2}$}

The last unknown temperature to calculate is the final electron temperature $T_{e2}$. Now that $T_{i2}$ is known, conservation of the total plasma energy can be used to calculate $T_{e2}$.

Using \eqref{eq:energy} again, 
the final energy after the ions have adjusted is
\begin{equation}
    U_2 = \frac{3}{2}N_e T_{e2} + \frac{3}{2}N_i T_{i2} - \frac{1}{2}N_ee^2\left( \frac{3Z}{2} + 1 \right)k_{e2} - \frac{1}{4}N_iZ^2e^2k_{i2}^2\frac{2k_{D2}+k_{e2}}{(k_{D2}+k_{e2})^2},
\end{equation}
where $k_{D2}^2 = k_{i2}^2 + k_{e2}^2$. Approximating $k_{i2}=k_{i0}$, $k_{e2} = k_{e1}$ and $k_{D2} = k_{D1}$ to leading order in the potential energy term, we equate $U_2 = U_1$ and obtain
\begin{equation}
T_{e2}-T_{e1} = -\frac{Z^2e^2}{6}k_{i0}^2k_{e1}\left(\frac{1}{(k_{D1}+k_{e1})^2} - \frac{1}{(k_{D0}+k_{e1})^2} \right)\,.
\end{equation}
If $k_{e1}<k_{e0}$, so the electrons are heated initially, then this expression is negative and the electron temperature drops back down slightly (that is, by an amount small in $1/\Lambda$).

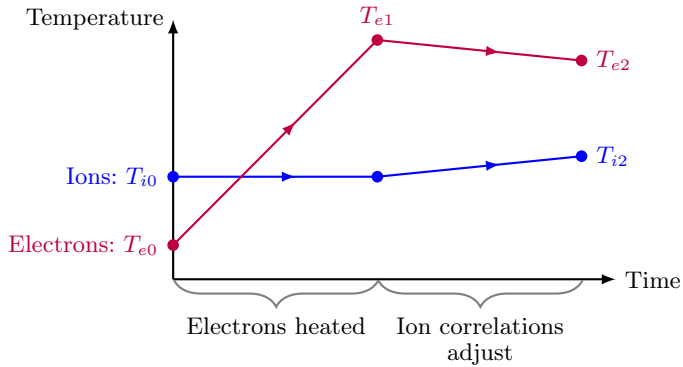
\begin{figure}
\vspace{4mm}
    \centering
\begin{tikzpicture}[scale=0.9]

    \draw(-3,-1) edge[thick,-latex] node[at end, left]{Temperature} (-3,2.8);
     \draw(-3,-1) edge[thick,-latex] node[at end, right]{Time} (3.5,-1);

\draw [thick,gray,decorate,decoration={brace,amplitude=10pt,mirror},xshift=0.4pt,yshift=-0.4pt](-3,-1) -- (0,-1) node[black,midway,yshift=-0.6cm] {\footnotesize Electrons heated};

\draw [thick,gray,decorate,decoration={brace,amplitude=10pt,mirror},xshift=0.4pt,yshift=-0.4pt](0,-1) -- (3,-1) node[black,midway,yshift=-0.8cm, text width=3cm,align=center] {\footnotesize Ion correlations adjust};

\node[circle,fill=blue,scale=0.5,label={[blue]left:Ions: $T_{i0}$}] at (-3,0.5) {};
\node[circle,fill=blue,scale=0.5] at (0,0.5) {};
\node[circle,fill=blue,scale=0.5,label={[blue]right:$T_{i2}$}] at (3,0.8) {};

    \begin{scope}[thick,blue, decoration={
    markings,
    mark=at position 0.6 with {\arrow{latex}}}
    ] 
    \draw[postaction={decorate}] (-3,0.5) -- (0,0.5);
`   \end{scope}
    \begin{scope}[thick,blue, decoration={
    markings,
    mark=at position 0.6 with {\arrow{latex}}}
    ] 
    \draw[postaction={decorate}] (0,0.5) -- (3,0.8);
`   \end{scope}

    \node[circle,fill=purple,scale=0.5,label={[purple]left:Electrons: $T_{e0}$}] at (-3,-0.5) {};
\node[circle,fill=purple,scale=0.5,label={[purple]above:$T_{e1}$}] at (0,2.5) {};
\node[circle,fill=purple,scale=0.5,label={[purple]right:$T_{e2}$}] at (3,2.2) {};
    
    \begin{scope}[thick,purple, decoration={
    markings,
    mark=at position 0.6 with {\arrow{latex}}}
    ] 
    \draw[postaction={decorate}] (-3,-0.5) -- (0,2.5);
`   \end{scope}

    \begin{scope}[thick,purple, decoration={
    markings,
    mark=at position 0.6 with {\arrow{latex}}}
    ] 
    \draw[postaction={decorate}] (0,2.5) -- (3,2.2);
`   \end{scope}
    
    \end{tikzpicture}
    \caption{Temperature changes for each species during correlation heating.}
    \label{fig:temps}
\end{figure}
In figure \ref{fig:temps} we summarise the temperature changes that occur during correlation heating. If the electrons are initially heated then the ions heat up due to correlation heating, while the electrons cool during the process. The opposite is true if the electrons are initially cooled; in either case, correlation heating causes the two temperatures to move towards each other.

\section{Time Evolution of the Ion Kinetic Energy}\label{sec:timeevol}

In the previous section, we calculated the change in ion temperature due to correlation heating, assuming the final state has pair correlations given by the Salpeter formulas for a steady-state, two-temperature plasma. In this section, we find the time-dependence of the ion kinetic energy on the ion-plasma-frequency timescale $t\sim 1/\omega_{pi}$.

This calculation is interesting for two reasons. First, it shows that the ion heating occurs entirely on this fast timescale. This means that, when the ions re-Maxwellianise on the ion-ion collision timescale $t\sim \tau_{ii}$, they do not undergo any further heating.

Secondly, it allows us to investigate whether the kinetic energy oscillates or overshoots its final value. Such an overshoot could facilitate fusion reactions within a short window of time. Indeed, in strongly coupled plasmas, it has been observed in simulations \cite[]{Morozov2003,Pohl2004,Niffenegger2011,Sprenkle2022} and experiments \cite[]{Chen2004,Laha2006,Lyon2011,Langin2016} that the kinetic energy does oscillate about its final value if the correlations are suddenly modified. However, we are able to prove that, in a moderately coupled plasma, the kinetic energy may not be a monotonic function of time but it does not overshoot. A similar problem was solved for a YOCP in \cite{Morawetz2001} using the Kadanoff-Baym equations of quantum statistical theory.

To evolve the ion-ion correlation on this timescale, we need to solve \eqref{eq:giieff}, repeated here,
\begin{align}
\label{eq:giief2}\left(\frac{\partial}{\partial t} + \mathcal{L}(12)\right)g_{ii}(12) &+ \int \rmd(3)\; \mathcal{V}^s(13)\,f_i(1)\,g_{ii}(32)\\
\nonumber&+ \int \rmd(3)\; \mathcal{V}^s(32)\,f_i(2)\,g_{ii}(13) = -\mathcal{V}^s(12)\,f_i(1)f_i(2)\,,
\end{align}
with fixed ion distribution $f_i = f_{i0}$ and screening length $k_e = k_{e1}$. 

The reason the ion distribution can be fixed is that it changes on the much slower ion-ion collision timescale; we invoke the usual Bogoliubov timescale hierarchy\footnote{The assumption that the species temperatures are not too far apart is also important here. If the electrons were significantly hotter than the ions then the system could support long-lived ion acoustic modes and wave-particle interactions would need to be included in a consistent theory.}. Of course, we are claiming that the ions heat up on the plasma-frequency timescale $t\sim 1/\omega_{pi}$, so the distribution function $f_i$ cannot actually be constant! However, we will show that this change is small in $1/\Lambda$ and is therefore a higher-order effect that can be neglected (remember that in \eqref{eq:giief2}, terms which are small in $1/\Lambda$ have already been dropped).

The reason the electron temperature, and therefore the screening length $k_e$, can be fixed is similar and has already been discussed: the variation is small in $1/\Lambda$ so can be neglected.

To proceed, we will use a Green's function approach to solve
\begin{align}\label{eq:source}
    \left(\frac{\partial}{\partial t} + \mathcal{L}(12)\right)g_{ii}(12) &+ \int \rmd(3)\; \mathcal{V}^s(13)\,f_{i0}(1)\,g_{ii}(32)\\
    \nonumber&+ \int \rmd(3)\; \mathcal{V}^s(32)\,f_{i0}(2)\,g_{ii}(13) = S(12)\,,
\end{align}
where $\mathcal{V}^s$ is assumed to have fixed screening length $k_e=k_{e1}$ from now on, and 
 the source term is $S(12) = -\mathcal{V}^s(12)f_{i0}(1)f_{i0}(2)$.

Since $\mathcal{L}(12) = \mathcal{L}(1) + \mathcal{L}(2)$, we can write \eqref{eq:source} as
\begin{align}\label{eq:operator}
    &\left(\frac{\partial}{\partial t} + \mathcal{O}(1) + \mathcal{O}(2)\right)g_{ii}(12) = S(12)\,,
\end{align}
where the operator $\mathcal{O}(1)$ is defined by
\begin{align}
\mathcal{O}(1)\,a(1) = \mathcal{L}(1)\,a(1) + \int \rmd(3)\; \mathcal{V}^s(13)\,f_{i0}(1)\,a(3)
\end{align}
for any function $a(1)$.

Let $U(12,t)$ be the solution of
\begin{align}\label{eq:linv}
\left( \frac{\partial}{\partial t} + \mathcal{O}(1) \right)U(12,t) = \delta(t)\delta(12)\,,
\end{align}
where $\delta(12) = \delta(\boldsymbol{v}_1-\boldsymbol{v}_2)\delta(\boldsymbol{r}_1-\boldsymbol{r}_2)$ and $U(12,t)=0$ for $t<0$. This means $U(12,t)$ is the Green's function for the linearised Vlasov equation with Yukawa interactions.

Equation \eqref{eq:linv} is first-order in time, so $U(12,t)$ can also be characterised as the function that satisfies
\begin{align}
\left( \frac{\partial}{\partial t} + \mathcal{O}(1) \right)U(12,t) = 0
\end{align}
for $t>0$, with initial condition $\lim_{t\to0^+}U(12,t) = \delta(12)$.

Then, the Green's function for \eqref{eq:source} or \eqref{eq:operator} is $U(13,t)U(24,t)$. This means
\begin{align}
\left( \frac{\partial}{\partial t} + \mathcal{O}(1) + \mathcal{O}(2)\right)U(13,t)U(24,t) = \delta(t)\delta(13)\delta(24)\,.
\end{align}
To prove this, we simply observe that
\begin{align}
\left( \frac{\partial}{\partial t} + \mathcal{O}(1) + \mathcal{O}(2)\right)U(13,t)U(24,t) = 0
\end{align}
for $t>0$ and $\lim_{t\to 0^+}U(13,t)U(24,t) = \delta(13)\delta(24)$.

Therefore, the general solution to \eqref{eq:source} is
\begin{align}\label{eq:giisoln}
    g_{ii}(12,t) &= \int \rmd(34)\,g_{ii}(34,t=0)U(13,t)U(24,t)\\
    \nonumber &\phantom{= }+ \int_0^t \rmd t' \int \rmd(34)\, S(34)U(13,t-t')U(24,t-t')\,.
\end{align}
The source term is time-independent, so we can also write
\begin{align}\label{eq:greensoln}
    g_{ii}(12,t) &= \int \rmd(34)\,g_{ii}(34,t=0)U(13,t)U(24,t)\\
    \nonumber &\phantom{= }+ \int_0^t \rmd t' \int \rmd(34)\, S(34)U(13,t')U(24,t')\,.
\end{align}
We still need to determine the Green's function $U(12,t)$. It can found using Fourier and Laplace transforms, following the usual solution for the Vlasov equation with one species \cite[see pp.44--46 of][]{Ichimaru} but with the Coulomb potential $\widetilde{\varphi}=4\upi Z^2e^2/k^2$ replaced with $\widetilde{\varphi}^{(s)} = 4\upi Z^2e^2/(k^2+k_{e1}^2)$. Its Fourier transform is
\begin{align}\label{eq:u}
\widetilde{U}(\boldsymbol{k}, \boldsymbol{v}_1, \boldsymbol{v}_2, t) = \int_{\mathcal{B}}\frac{\rmd\omega}{2\upi}\,&\rme^{-\imag\omega t}\biggl(\frac{\imag\,\delta(\boldsymbol{v}_1-\boldsymbol{v}_2)}{\omega - \boldsymbol{k\cdot v}_1}\\
\nonumber&+ \frac{1}{n_i}\frac{k_{i0}^2}{k^2+k_e^2}\frac{1}{\epsilon(\omega,\boldsymbol{k})}\frac{\imag\boldsymbol{k\cdot v}_1}{\omega - \boldsymbol{k\cdot v}_1}\frac{1}{\omega - \boldsymbol{k\cdot v}_2}f_{i0}(\boldsymbol{v}_1)\biggr).
\end{align}
Here, the contour $\mathcal{B}$ is the usual Bromwich contour for inverse Laplace transforms, meaning it is a straight line in the complex plane from $i\sigma-\infty$ to $i\sigma+\infty$, where $\sigma$ is is chosen so that all the poles of the integrand lie below the contour. The dielectric function $\epsilon(\omega,\boldsymbol{k})$ is given by
\begin{equation}\label{eq:dielectric}
\epsilon(\omega ,\boldsymbol{k}) = 1 - \frac{1}{n_{\imag}}\frac{k_{i0}^2}{k^2+k_{e1}^2} \int_{\mathcal{L}} \mathrm{d}\boldsymbol{v} \, \frac{\boldsymbol{k}\cdot \boldsymbol{v}}{\omega -\boldsymbol{k}\cdot\boldsymbol{v}} \, f_{i0}(\boldsymbol{v})\,,
\end{equation}
where $\mathcal{L}$ is the usual Landau contour \cite[]{Landau1965}. In Fourier space, \eqref{eq:greensoln} reads
\begin{align}\label{eq:giiksoln}
    \widetilde{g}_{ii}(\boldsymbol{k},\boldsymbol{v}_1,\boldsymbol{v}_2,t) &= \int \rmd\boldsymbol{v}_3\rmd\boldsymbol{v}_4\,\widetilde{g}_{ii}(\boldsymbol{k},\boldsymbol{v}_3,\boldsymbol{v}_4,t=0)\widetilde{U}(\boldsymbol{k},\boldsymbol{v}_1,\boldsymbol{v}_3,t)\widetilde{U}(-\boldsymbol{k},\boldsymbol{v}_2,\boldsymbol{v}_4,t)\\
    \nonumber&\phantom{= }+ \int_0^t \rmd t' \int \rmd\boldsymbol{v}_3\rmd\boldsymbol{v}_4\, \widetilde{S}(\boldsymbol{k},\boldsymbol{v}_3,\boldsymbol{v}_4)\widetilde{U}(\boldsymbol{k},\boldsymbol{v}_1,\boldsymbol{v}_3,t')\widetilde{U}(-\boldsymbol{k},\boldsymbol{v}_2,\boldsymbol{v}_4,t')\,.
\end{align}
To find the heating that results from this change in correlations, we should use this solution for $g_{ii}$ to evolve $f_i$ according to
\begin{align}\label{eq:dfidt}
\frac{\partial f_i(\boldsymbol{v}_1,t)}{\partial t} &= -\frac{T_{i0}k_{i0}^2}{n_im_i}\boldsymbol{\cdot}\int \frac{\rmd\boldsymbol{k}}{(2\upi)^3} \frac{1}{k^2+k_{e1}^2}\imag\boldsymbol{k}\boldsymbol{\cdot}\frac{\partial}{\partial\boldsymbol{v}_1}\int \rmd\boldsymbol{v}_2\,\widetilde{g}_{ii}(\boldsymbol{k},\boldsymbol{v}_1,\boldsymbol{v}_2,t)\,,
\end{align}
which is just \eqref{eq:fiequation} with fixed screening length $k_e = k_{e1}$. In fact, we are really interested in the change in the ion kinetic energy density $K_i = \int \rmd\boldsymbol{v}_1 \tfrac{1}{2}m_i\boldsymbol{v}_1^2 f_i(\boldsymbol{v}_1,t)$, given by
\begin{align}
    \frac{\rmd K_i}{\rmd t} = \frac{T_{i0}k_{i0}^2}{n_i}\int \frac{\rmd\boldsymbol{k}}{(2\upi)^3}\frac{1}{k^2+k_{e1}^2}\int \rmd\boldsymbol{v}_1 \imag \boldsymbol{k\cdot v}_1\int \rmd\boldsymbol{v}_2\, \widetilde{g}_{ii}(\boldsymbol{k},\boldsymbol{v}_1,\boldsymbol{v}_2,t)\,.
\end{align}
Using \eqref{eq:giiksoln} and \eqref{eq:u}, this becomes
\begin{equation}
\frac{\rmd K_i}{\rmd t} = \frac{\rmd K_i^{(1)}}{\rmd t} + \frac{\rmd K_i^{(2)}}{\rmd t}
\end{equation}
where
\begin{align}\label{eq:K1}
\frac{\rmd K_i^{(1)}}{\rmd t} = \,&\frac{T_{i0}k_{i0}^4}{n_i^2}\int \frac{\rmd\boldsymbol{k}}{(2\upi)^3}\frac{1}{k^2+k_{D0}^2}\frac{1}{k^2+k_{e1}^2}\int \rmd\boldsymbol{v}_3\rmd\boldsymbol{v}_4 \,f_i(\boldsymbol{v}_3)f_i(\boldsymbol{v}_4)\\
\nonumber&\times\int \rmd\boldsymbol{v}_1\rmd\boldsymbol{v}_2\,(\imag\boldsymbol{k\cdot v}_1) \int_\mathcal{B}\frac{\rmd\omega}{2\upi}\int_{\mathcal{B}'}\frac{\rmd\omega'}{2\upi}\, \rme^{-\imag(\omega+\omega')t}\\
\nonumber&\times\left( \frac{\delta(\boldsymbol{v}_1-\boldsymbol{v}_3)}{\omega - \boldsymbol{k\cdot v}_1} + \frac{1}{n_i}\frac{k_{i0}^2}{k^2+k_{e1}^2}\frac{1}{\epsilon(\omega,\boldsymbol{k})}\frac{\boldsymbol{k\cdot v}_1}{\omega - \boldsymbol{k\cdot v}_1}\frac{1}{\omega - \boldsymbol{k\cdot v}_3} f_{i0}(\boldsymbol{v}_1)\right)\\
\nonumber&\times\left( \frac{\delta(\boldsymbol{v}_2-\boldsymbol{v}_4)}{\omega' + \boldsymbol{k\cdot v}_2} - \frac{1}{n_i}\frac{k_{i0}^2}{k^2+k_{e1}^2}\frac{1}{\epsilon(\omega',\boldsymbol{k})}\frac{\boldsymbol{k\cdot v}_2}{\omega' + \boldsymbol{k\cdot v}_2}\frac{1}{\omega' + \boldsymbol{k\cdot v}_4} f_{i0}(\boldsymbol{v}_2)\right),
\end{align}
\begin{align}\label{eq:K2}
\frac{\rmd K_i^{(2)}}{\rmd t} = \,&\frac{T_{i0}k_{i0}^4}{n_i^2}\int \frac{\rmd\boldsymbol{k}}{(2\upi)^3}\frac{1}{(k^2+k_{e1}^2)^2}\int \rmd\boldsymbol{v}_3\rmd\boldsymbol{v}_4 \,f_i(\boldsymbol{v}_3)f_i(\boldsymbol{v}_4)(\imag\boldsymbol{k}\boldsymbol{\cdot}(\boldsymbol{v}_3-\boldsymbol{v}_4))\\
\nonumber&\times\int \rmd\boldsymbol{v}_1\rmd\boldsymbol{v}_2\,(\imag\boldsymbol{k\cdot v}_1) \int_0^t \rmd t'\int_\mathcal{B}\frac{\rmd\omega}{2\upi}\int_{\mathcal{B}'}\frac{\rmd\omega'}{2\upi}\, \rme^{-\imag(\omega+\omega')t'}\\
\nonumber&\times\left( \frac{\delta(\boldsymbol{v}_1-\boldsymbol{v}_3)}{\omega - \boldsymbol{k\cdot v}_1} + \frac{1}{n_i}\frac{k_{i0}^2}{k^2+k_{e1}^2}\frac{1}{\epsilon(\omega,\boldsymbol{k})}\frac{\boldsymbol{k\cdot v}_1}{\omega - \boldsymbol{k\cdot v}_1}\frac{1}{\omega - \boldsymbol{k\cdot v}_3} f_{i0}(\boldsymbol{v}_1)\right)\\
\nonumber&\times\left( \frac{\delta(\boldsymbol{v}_2-\boldsymbol{v}_4)}{\omega' + \boldsymbol{k\cdot v}_2} - \frac{1}{n_i}\frac{k_{i0}^2}{k^2+k_{e1}^2}\frac{1}{\epsilon(\omega',\boldsymbol{k})}\frac{\boldsymbol{k\cdot v}_2}{\omega' + \boldsymbol{k\cdot v}_2}\frac{1}{\omega' + \boldsymbol{k\cdot v}_4} f_{i0}(\boldsymbol{v}_2)\right).
\end{align}
    The first term $\rmd K_i^{(1)}/\rmd t$ involves the initial correlation, and therefore the parameter $k_{D0}$, whereas the second does not. While it is common in plasma kinetic theory to ignore the initial condition when evolving pair correlations, because any memory of the initial state is quickly lost, in situations involving ultrafast relaxation this Markovian approximation is not valid \cite[]{Semkat1999,Morawetz2001}. Correlation heating arises because the initial ion-ion correlation is not correct for the new electron temperature and it occurs in the brief time before memory of the initial state is lost. If there is no electron heating, so that $k_{D0} = k_{D1}$, then $\rmd K_i^{(1)}/\rmd t$ and $\rmd K_i^{(2)}/\rmd t$ must cancel. 

\begin{figure}
\centering
\includegraphics{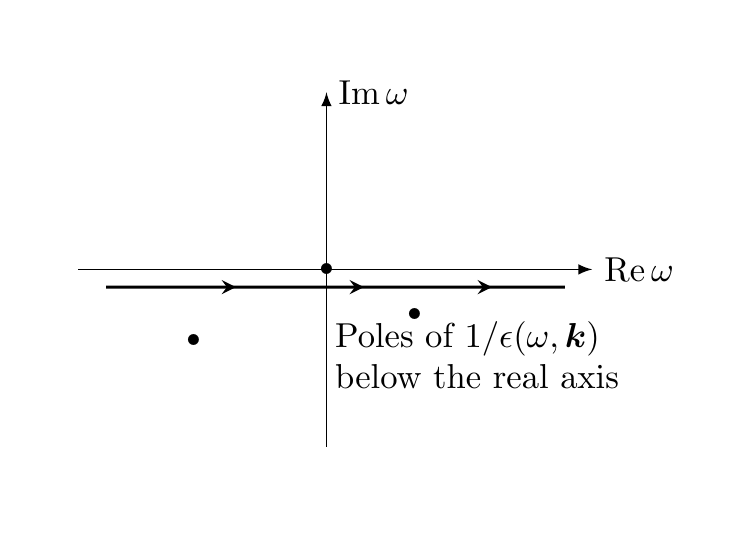}
\vspace*{-8mm}
\caption{Contour $\mathcal{C}$ used to define $\mathcal{F}(t,\boldsymbol{k})$.}
\label{contourD}
\end{figure}
\begin{figure}
\centering
\includegraphics{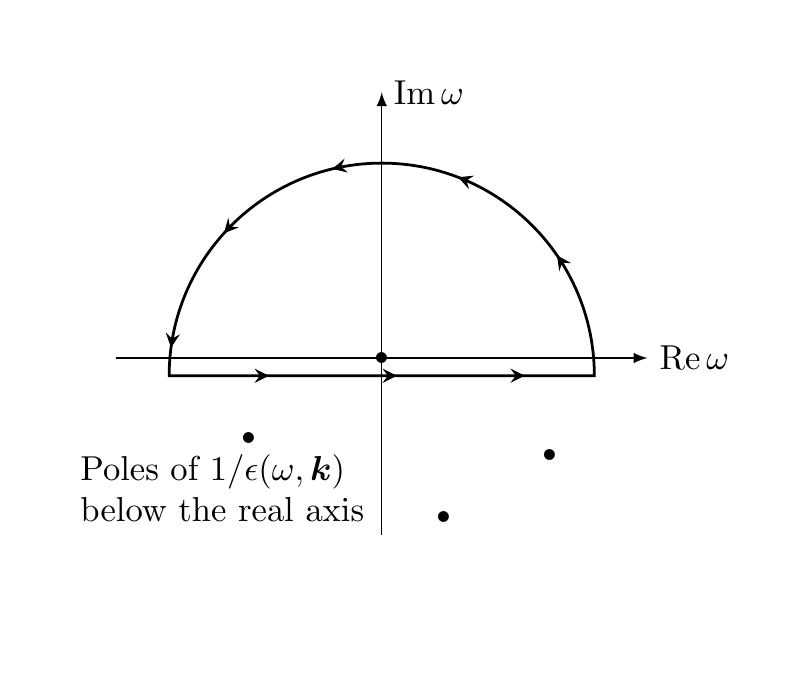}
\vspace*{-8mm}
\caption{Completing the contour in the upper half plane.}
\label{ContourF}
\end{figure}
In appendix \ref{app:integral}, the integrals in \eqref{eq:K1} and \eqref{eq:K2} are simplified separately and the result is
\begin{align}\label{eq:dKidt}
\frac{\rmd K_i}{\rmd t} = -\frac{\rmd}{\rmd t}\biggl[\frac{1}{2}T_{i0}(k_{D0}^2-k_{D1}^2)\int \frac{\rmd\boldsymbol{k}}{(2\upi)^3}\frac{(k^2 + k_{D1}^2)|\mathcal{F}(t,\boldsymbol{k})|^2}{(k^2+k_{e1}^2)(k^2+k_{D0}^2)}\biggr].
\end{align}
Indeed, when $k_{D0}=k_{D1}$ this vanishes. Here, the function $\mathcal{F}(t,\boldsymbol{k})$ is defined by
\begin{equation}\label{eq:fdef}
\mathcal{F}(t,\boldsymbol{k}) = \int_\mathcal{C}\frac{\rmd\omega}{2\upi}\, \rme^{-\imag\omega t}
\frac{1}{\omega}  \frac{\epsilon(\omega,\boldsymbol{k})-1}{\epsilon(\omega,\boldsymbol{k})}\,,
\end{equation}
where the contour $\mathcal{C}$ passes below the pole at the origin $\omega = 0$ but above all the poles of $1/\epsilon$, as shown in figure \ref{contourD}.

We can check that this is consistent with our ion temperature change in \S\ref{sec:iontemp}. Integrating \eqref{eq:dKidt} gives the change in ion kinetic energy density,
\begin{align}\label{eq:positive}
\Delta K_i(t) = -\frac{1}{2}T_{i0}(k_{D0}^2-k_{D1}^2)\int \frac{\rmd\boldsymbol{k}}{(2\upi)^3}\frac{(k^2 + k_{D1}^2)\left(|\mathcal{F}(t,\boldsymbol{k})|^2 - |\mathcal{F}(0,\boldsymbol{k})|^2\right)}{(k^2+k_{e1}^2)(k^2+k_{D0}^2)}.
\end{align}
Since the contour $\mathcal{C}$ can run entirely below the real axis, at large times the integrand in \eqref{eq:fdef} is exponentially suppressed. Therefore, $\mathcal{F}(\infty, \boldsymbol{k})=0$ and the total change over this timescale is
\begin{align}\label{eq:totalchange}
\Delta K_i(\infty) = \frac{1}{2}T_{i0}(k_{D0}^2-k_{D1}^2)\int \frac{\rmd\boldsymbol{k}}{(2\upi)^3}\frac{(k^2 + k_{D1}^2)|\mathcal{F}(0,\boldsymbol{k})|^2}{(k^2+k_{e1}^2)(k^2+k_{D0}^2)}\,.
\end{align}
To evaluate this integral we need
\begin{equation}
\mathcal{F}(0,\boldsymbol{k}) = \int_\mathcal{C}\frac{\rmd\omega}{2\upi}
\frac{1}{\omega}  \frac{\epsilon(\omega,\boldsymbol{k})-1}{\epsilon(\omega,\boldsymbol{k})}\,.
\end{equation}
Now that there is no longer an exponential factor, the integrand decays as $|\omega|\to \infty$. So, completing the contour in the upper half-plane as shown in figure \ref{ContourF} encloses a single pole at the origin $\omega = 0$:
\begin{equation}
\mathcal{F}(0,\boldsymbol{k}) = \imag \left( \frac{\epsilon(0,\boldsymbol{k})-1}{\epsilon(0,\boldsymbol{k})}\right) = \imag\left(\frac{k_{i0}^2}{k^2 + k_{D1}^2}\right).
\end{equation}
Using this in \eqref{eq:totalchange}, we find
\begin{align}
\Delta K_i(\infty) = \frac{1}{8\upi}T_{i0}k_{i0}^4\left( \frac{1}{k_{D1}+k_{e1}} - \frac{1}{k_{D0} + k_{e1}} \right).\label{eq:ionke2}
\end{align}
This change in ion kinetic energy density is exactly equivalent to the ion temperature change calculated in \eqref{eq:ionke}. Therefore, the ions only change their kinetic energy on the extremely fast ion-plasma-frequency timescale.
\begin{figure}
\centering
\includegraphics[scale=0.6]{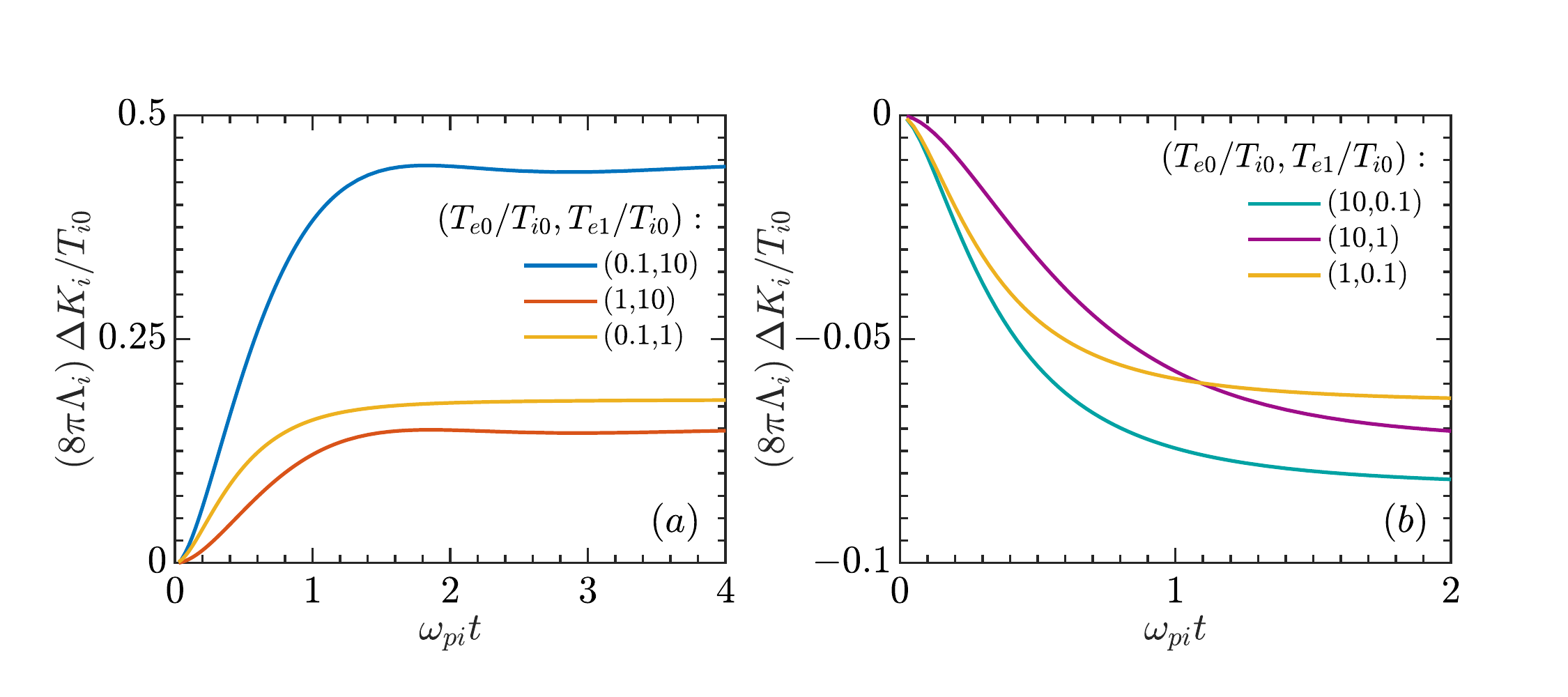}
\vspace*{-5mm}
\caption{Change in ion kinetic energy for different combinations of $T_{e0}/T_{i0}$ and $T_{e1}/T_{i0}$, normalised using $\Lambda_i = n_i/k_{i0}^3$.  In ($a$), the electrons are heated, while in ($b$) they are cooled. } 
\label{fig:tempvstime}
\end{figure}

Equation \eqref{eq:positive} is equivalent to
\begin{align}\label{eq:DeltaKiSoln}
\Delta K_i(t) = \Delta K_i(\infty) - \frac{1}{2}T_{i0}(k_{D0}^2-k_{D1}^2)\int \frac{\rmd\boldsymbol{k}}{(2\upi)^3}\frac{(k^2 + k_{D1}^2)|\mathcal{F}(t,\boldsymbol{k})|^2}{(k^2+k_{e1}^2)(k^2+k_{D0}^2)} \,.
\end{align}
Although $\mathcal{F}(t,\boldsymbol{k})$ is still an unknown function, the fact that the integrand is positive means that, when the electrons are heated, so $k_{D0}>k_{D1}$, we must have ${\Delta K_i(t)<\Delta K_i(\infty)}$. Therefore, the ion temperature cannot overshoot its final value. A similar result for the ultrafast relaxation of a weakly coupled one-component plasma with zero initial correlation has been found using molecular dynamics simulations \cite[]{Zwicknagel1999}.

Expression \eqref{eq:DeltaKiSoln} is plotted in figure \ref{fig:tempvstime} for different combinations of electron temperature before and after heating. In ($a$), the two curves with $T_{e1} = 10T_{i0}$ are not monotonically increasing; at later times, $\Delta K_i$ contains a slight wobble. 

\section{Modification to the Ion Distribution}\label{sec:distribution}

Fusion reactions involve suprathermal ions that have velocities well above the thermal speed. So, to increase the fusion rate, it is desirable to increase the number of fast ions in the tail of the distribution. In this section, we investigate how efficient correlation heating is at creating suprathermal ions, by calculating the change in the ion velocity distribution resulting from correlation heating on the ion-plasma-frequency timescale. There is no reason to expect the ions to remain Maxwellian; molelular dynamics simulations of an initially-uncorrelated YOCP in \cite{Murillo2006} showed strong deviations from a Maxwellian in the early stages of the ultrafast relaxation of the pair correlation.

\begin{figure}
\centering
\includegraphics[scale=0.6]{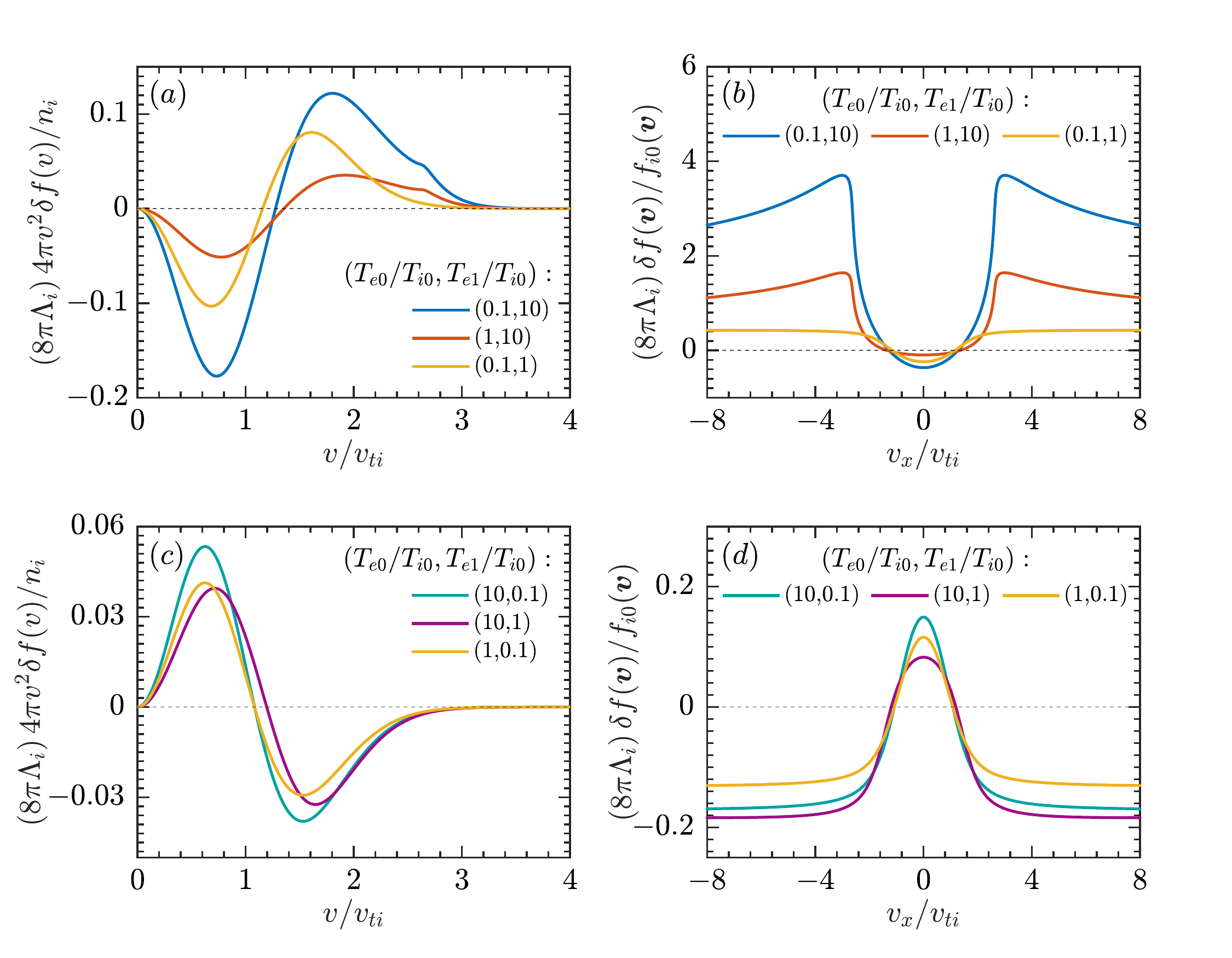}
\vspace*{-5mm}
    \caption{($a$) Change in ion speed distribution $4\upi v^2 \delta f(v)$, normalised using ${\Lambda_i = n_i/k_{i0}^3}$, for three different combinations of $T_{e0}/T_{i0}$ and $T_{e1}/T_{i0}$. In each case, the electrons are heated. ($b$) Cross section, in the plane $v_y=v_z=0$, of $\delta f(\boldsymbol{v})/f_{i0}(\boldsymbol{v})$ for the same three cases. ($c$) Similar to ($a$) except now three cases are presented in which the electrons are cooled. ($d$) Similar to ($b$) except the electrons are cooled.}
\label{fig:distributions}
\end{figure}

Using the same approach as in the previous section, by integrating \eqref{eq:dfidt} from $t=0$ to $t=\infty$ we can write the change in ion distribution ${\delta f(\boldsymbol{v}) = f_i(\boldsymbol{v},t=\infty) - f_i(\boldsymbol{v},t=0)}$ as
\begin{equation}
\delta f(\boldsymbol{v}) = \delta f^{(1)}(\boldsymbol{v}) + \delta f^{(2)}(\boldsymbol{v}),
\end{equation}
where
\begin{align}\label{eq:delta1}
\delta f^{(1)}(\boldsymbol{v}_1) = \,&-\frac{T_{i0}k_{i0}^4}{n_i^2m_i}\int \frac{\rmd\boldsymbol{k}}{(2\upi)^3}\frac{1}{k^2+k_{D0}^2}\frac{1}{k^2+k_{e1}^2}\imag\boldsymbol{k}\boldsymbol{\cdot}\frac{\partial}{\partial \boldsymbol{v}_1}\int \rmd\boldsymbol{v}_3\rmd\boldsymbol{v}_4 \,f_i(\boldsymbol{v}_3)f_i(\boldsymbol{v}_4)\\
\nonumber&\times\int \rmd\boldsymbol{v}_2\, \int_0^\infty \rmd t\int_\mathcal{B}\frac{\rmd\omega}{2\upi}\int_{\mathcal{B}'}\frac{\rmd\omega'}{2\upi}\, \rme^{-\imag(\omega+\omega')t}\\
\nonumber&\times\left( \frac{\delta(\boldsymbol{v}_1-\boldsymbol{v}_3)}{\omega - \boldsymbol{k\cdot v}_1} + \frac{1}{n_i}\frac{k_{i0}^2}{k^2+k_{e1}^2}\frac{1}{\epsilon(\omega,\boldsymbol{k})}\frac{\boldsymbol{k\cdot v}_1}{\omega - \boldsymbol{k\cdot v}_1}\frac{1}{\omega - \boldsymbol{k\cdot v}_3} f_{i0}(\boldsymbol{v}_1)\right)\\
\nonumber&\times\left( \frac{\delta(\boldsymbol{v}_2-\boldsymbol{v}_4)}{\omega' + \boldsymbol{k\cdot v}_2} - \frac{1}{n_i}\frac{k_{i0}^2}{k^2+k_{e1}^2}\frac{1}{\epsilon(\omega',\boldsymbol{k})}\frac{\boldsymbol{k\cdot v}_2}{\omega' + \boldsymbol{k\cdot v}_2}\frac{1}{\omega' + \boldsymbol{k\cdot v}_4} f_{i0}(\boldsymbol{v}_2)\right),
\end{align}
\begin{align}\label{eq:delta2}
\delta f^{(2)}(\boldsymbol{v}_1) = \,&-\frac{T_{i0}k_{i0}^4}{n_i^2m_i}\int \frac{\rmd\boldsymbol{k}}{(2\upi)^3}\frac{1}{(k^2+k_{e1}^2)^2}\imag\boldsymbol{k}\boldsymbol{\cdot}\frac{\partial}{\partial \boldsymbol{v}_1}\int \rmd\boldsymbol{v}_3\rmd\boldsymbol{v}_4 \,f_i(\boldsymbol{v}_3)f_i(\boldsymbol{v}_4)\\
\nonumber&\times(\imag\boldsymbol{k}\boldsymbol{\cdot}(\boldsymbol{v}_3-\boldsymbol{v}_4))\int \rmd\boldsymbol{v}_2\,\int_0^\infty \rmd t \int_0^t \rmd t'\int_\mathcal{B}\frac{\rmd\omega}{2\upi}\int_{\mathcal{B}'}\frac{\rmd\omega'}{2\upi}\, \rme^{-\imag(\omega+\omega')t'}\\
\nonumber&\times\left( \frac{\delta(\boldsymbol{v}_1-\boldsymbol{v}_3)}{\omega - \boldsymbol{k\cdot v}_1} + \frac{1}{n_i}\frac{k_{i0}^2}{k^2+k_{e1}^2}\frac{1}{\epsilon(\omega,\boldsymbol{k})}\frac{\boldsymbol{k\cdot v}_1}{\omega - \boldsymbol{k\cdot v}_1}\frac{1}{\omega - \boldsymbol{k\cdot v}_3} f_{i0}(\boldsymbol{v}_1)\right)\\
\nonumber&\times\left( \frac{\delta(\boldsymbol{v}_2-\boldsymbol{v}_4)}{\omega' + \boldsymbol{k\cdot v}_2} - \frac{1}{n_i}\frac{k_{i0}^2}{k^2+k_{e1}^2}\frac{1}{\epsilon(\omega',\boldsymbol{k})}\frac{\boldsymbol{k\cdot v}_2}{\omega' + \boldsymbol{k\cdot v}_2}\frac{1}{\omega' + \boldsymbol{k\cdot v}_4} f_{i0}(\boldsymbol{v}_2)\right).
\end{align}
These integrals are simplified in appendix \ref{app:integral2} and the result is
\begin{align}\label{eq:deltasolution}
\delta f(\boldsymbol{v}) = \frac{T_{i0}k_{i0}^2}{2n_im_i}(k_{D0}^2-k_{D1}^2)\,\frac{\partial}{\partial \boldsymbol{v}}\boldsymbol{\cdot}\left[\frac{\boldsymbol{v}f_{i0}(\boldsymbol{v})}{\boldsymbol{v}^2}\int \frac{\rmd\boldsymbol{k}}{(2\upi)^3}\frac{1-|\epsilon(0,\boldsymbol{k})/\epsilon(\boldsymbol{k}\cdot\boldsymbol{v}, \boldsymbol{k})|^2}{(k^2+k_{D0}^2)(k^2+k_{D1}^2)}\right].
\end{align}
At this point it is reassuring that, by multiplying by $\tfrac{1}{2} m_i\boldsymbol{v}^2$, integrating over $\boldsymbol{v}$, and making use of the integral 
\begin{equation}
\int \rmd\boldsymbol{v}\, \frac{f_{i0}(\boldsymbol{v})}{|\epsilon(\boldsymbol{k}\cdot\boldsymbol{v}, \boldsymbol{k})|^2} = n_i\left(\frac{k^2+k_{e1}^2}{k^2+k_{D1}^2}\right),
\end{equation}
the same result is obtained for the total ion kinetic energy change that we found in \eqref{eq:ionke} and \eqref{eq:ionke2}. The expression \eqref{eq:deltasolution} is plotted in figure \ref{fig:distributions} for different combinations of the initial and final electron temperatures $T_{e0}$ and $T_{e1}$. We see that heating the electrons has a larger effect on the ion distribution than cooling them.  In addition, the effect is larger when the electrons are initially colder. 

An interesting feature is that, if the electrons are made hotter than the ions, then $\delta f(\boldsymbol{v})$ develops a small bump. This bump is located around the ion sound speed, which is on the order of $v=v_{\mathrm{th}i}\,k_{i0}/k_{e1}$, where $v_{\mathrm{th}i}=(2T_{i0}/m_i)^{1/2}$. The reason is that, when the electrons are made very hot, correlation heating leads to the excitation of long-lived ion-acoustic modes, which damp partially on the ions. The ion distribution therefore flattens slightly at speeds in resonance with these modes, which appears as a bump in $\delta f(\boldsymbol{v})$.

We also see that, at large velocities, $\delta f(\boldsymbol{v})/f_{i0}(\boldsymbol{v})$ has a constant limit. It is simple to justify this behaviour analytically. If $|\boldsymbol{v}|$ is large enough, we can replace the dielectric function $\epsilon(\boldsymbol{k\cdot v},\boldsymbol{k})\to 1$.\footnote{In this context, `large enough' means $|\boldsymbol{v}|/v_{\mathrm{th}i}\gg k_{i0}/k_{e1}$. This can be shown using \eqref{eq:epsilonclosedform} for the dielectric function together with the asymptotic series \eqref{eq:Zasymptotics}.} Then
 \begin{align}
\delta f(\boldsymbol{v}) = \,&\frac{T_{i0}k_{i0}^2}{2n_im_i}(k_{D0}^2-k_{D1}^2)\,\frac{\partial}{\partial \boldsymbol{v}}\boldsymbol{\cdot}\left[\frac{\boldsymbol{v}f_{i0}(\boldsymbol{v})}{\boldsymbol{v}^2}\int \frac{\rmd\boldsymbol{k}}{(2\upi)^3}\frac{1-|\epsilon(0,\boldsymbol{k})|^2}{(k^2+k_{D0}^2)(k^2+k_{D1}^2)}\right].
\end{align}
 At large velocities, the dominant contribution comes from the derivative acting on $f_i(\boldsymbol{v}):$
 \begin{align}
\frac{\partial}{\partial \boldsymbol{v}}\boldsymbol{\cdot}\left[\frac{\boldsymbol{v}f_{i0}(\boldsymbol{v})}{\boldsymbol{v}^2}\right] \sim -\frac{m_i}{T_{i0}}f_{i0}(\boldsymbol{v})\,,
\end{align}
so
 \begin{align}\label{eq:largevchange}
\delta f(\boldsymbol{v}) = \,\frac{1}{2}\frac{k_{i0}^2}{n_i}(k_{D0}^2-k_{D1}^2)f_{i0}(\boldsymbol{v})\int \frac{\rmd\boldsymbol{k}}{(2\upi)^3}\frac{|\epsilon(0,\boldsymbol{k})|^2-1}{(k^2+k_{D0}^2)(k^2+k_{D1}^2)}\,.
\end{align}
This $\boldsymbol{k}$-integral may be evaluated using $\epsilon(0,\boldsymbol{k}) = (k^2+k_{D1}^2)/(k^2+k_{e1}^2)$ but the result is not very enlightening. The important conclusion is that $\delta f(\boldsymbol{v})\propto f_{i0}(\boldsymbol{v})$, which means the number of ions at large velocities is increased by a constant factor.

This result has the following implication: using correlation heating to quickly supply a given amount of energy to the ions in a moderately coupled plasma leads to an ion distribution with a depleted tail relative to a Maxwellian at the same energy. Over a longer timescale, the tail will be filled out by collisions.

To see this, suppose the initial ion distribution is
\begin{align}
    \nonumber f_{i0}(\boldsymbol{v}) = n_i\left( \frac{m_i}{2\upi T_{i0}} \right)^{3/2}\exp\left( -\frac{m_i\boldsymbol{v}^2}{2 T_{i0}} \right).
\end{align}
If the distribution remains Maxwellian but the temperature is increased by a small amount $\delta T_i \sim T_i/\Lambda$, then the modification to the distribution at large velocities is
\begin{align}    
    \delta f(\boldsymbol{v}) \sim \left( \frac{m_i\boldsymbol{v}^2}{2T_{i0}^2}\delta T_{\imag}\right)f_{i0}(\boldsymbol{v})
\end{align}
for $|\boldsymbol{v}|\gg (T_{i0}/m_i)^{1/2}$. For sufficiently large velocities, this modification is always greater than the change in \eqref{eq:largevchange} that can be achieved using correlation heating. This is not surprising: widening a Gaussian distribution increases the area under the tail more effectively than multiplication of the distribution by a constant factor. Therefore, correlation heating does not efficiently energise suprathermal ions on the ion-plasma-frequency timescale in a moderately-coupled plasma. In the next section, we argue that this conclusion may not hold in strongly coupled systems.

\section{Discussion}\label{sec:discussion}

\subsection{Key Formulas}

If some energy $Q$ is suddenly supplied to the electrons, how much energy and power is transferred to the ions as a result of the change in shielding? The calculations in \S\ref{sec:temps} allow us to answer this question quantitatively for moderately coupled plasmas. If the electrons become sufficiently hot, the amount of energy transferred is independent of $Q$. In this limit, the average energy supplied per ion is
\begin{equation}
\textrm{Energy in eV per ion} \approx 30\,\frac{ Z^3\,n_{26}^{3/2}}{T_{1}^{1/2}}\left( 1 - \frac{1}{\sqrt{1+(1/Z)(1/\tau)}} \right).
\end{equation}
The time taken for the ion kinetic energy to increase is roughly $t\approx1.5/\omega_{pi}$ (see figure \ref{fig:tempvstime}). So, the power delivered to the ions is
\begin{equation}
\textrm{Power in eV per fs per ion} \approx 76\,\frac{ Z^4\,n_{26}^2}{A^{1/2}\,T_{1}^{1/2}}\left( 1 - \frac{1}{\sqrt{1+(1/Z)(1/\tau)}} \right).
\end{equation}
Here, $Z$ is the ion charge state, $n_{26}$ is the ion number density normalised to $\SI{e26}{\per\centi\meter\cubed}$, $A$ is the ion mass normalised to the proton mass, $T_1$ is the initial ion temperature in $\SI{}{\kilo\electronvolt}$ and $\tau = T_{e0}/T_{i0}$ is the ratio between the initial electron and ion temperatures. These formulas give a practical way to estimate the maximum amount of correlation heating possible in any given system.

\subsection{How Large Can Correlation Heating Be?}\label{sec:howlarge}

The moderate coupling ordering $1\ll \Lambda/\lambda\ll(m_i/m_e)^{1/2}$ is required for the rigorous analytical description of correlation heating presented in this paper. However, since the ion heating is $\Delta T \sim T/\Lambda$, this ordering precludes substantial heating, which might be possible in more strongly coupled plasmas. It is therefore important to consider the maximum heating attainable when the moderate coupling restriction is removed.

The two key energy scales in a plasma are the temperature $T$ (assuming $T_e\sim T_i$) and the typical potential energy between nearby particles, $\varphi\sim e^2n^{1/3}$. The plasma parameter scales as $\Lambda \sim (T/\varphi)^{3/2}$, so in a moderately coupled plasma $\Delta T \sim (\varphi/T)^{3/2}\, T$. As the coupling strength $\varphi/T$ increases, correlation heating gets larger. A natural question to ask is whether this scaling continues to hold in strongly coupled plasmas in which $\varphi\gg T$ initially. Is it possible to get significant heating?

Consider the simplest case of electrons heated to a large, essentially infinite, temperature. They become a uniform neutralising background and have no influence over the subsequent ion relaxation. As the ion-ion correlation adjusts, the ion energy is conserved,
\begin{equation}
U = \frac{3}{2}N_iT_i + \frac{1}{2}Vn_i^2\int \rmd\boldsymbol{r} \left(\frac{Z^2e^2}{r}\right)G_{ii}(\boldsymbol{r}) = \mathrm{const.}\,,
\end{equation}
as in \eqref{eq:ionenergy}. This means \cite[]{Gericke2003}
\begin{equation}\label{eq:DeltaTi}
\Delta T_i =  - \frac{1}{3}n_i\int \rmd\boldsymbol{r} \left(\frac{Z^2e^2}{r}\right)\Delta G_{ii}(\boldsymbol{r}),
\end{equation}
where $\Delta G_{ii}(\boldsymbol{r})$ is the change in the ion-ion correlation. It is clear that, to obtain heating, we want $\Delta G_{ii}<0$ (for most values of $\boldsymbol{r}$, at least). Since the ion-ion correlation $G_{ii}(\boldsymbol{r})$ is negative for most $\boldsymbol{r}$, this means its absolute value must increase. Therefore, the ions have to become more ordered, or more correlated, to heat up.

This means it is impossible for ions that were initially strongly coupled, $\varphi\gg T$, to heat up so much that they become weakly coupled, $\Delta T \gg \varphi$. If this were the case, the ions would have a smaller $|G_{ii}|$ in their weakly coupled final state than in their initial state, so they would become less ordered and \eqref{eq:DeltaTi} implies $\Delta T_i<0$: a contradiction. Correlation heating occurs as the ions become more ordered; however, an increase in temperature encourages spatial disorder and opposes further heating. This negative feedback places an upper limit on how hot the ions can become. Therefore, the scaling $\Delta T\sim (\varphi/T)^{3/2}T$ must be modified in strongly coupled plasmas. For comparison, in disorder-induced heating the ion temperature generally rises until $\varphi/T\sim$ $1$--$4$ \cite[]{Murillo2001,Simien2004,Chen2004,Cummings2005}.

Strongly-coupled ions have an equilibrium pair correlation that approximately represents a sphere that perfectly excludes other ions, with radius on the order of the inter-particle spacing \cite[]{Ichimaru1982}. In this picture, the electron temperature does not affect the ion correlation, which implies $|\Delta G_{ii}|\ll |G_{ii}|$ so $\Delta T \ll \varphi$. However, significant heating $\Delta T \gg T$ could still be achievable.

Equilibrium correlations  of strongly coupled YOCPs and two-temperature plasmas have been found numerically. The change in ion correlation when the electron temperature is modified, $\Delta G_{ii}$, is visible in plots of the pair correlation (radial distribution function) in \cite{Daughton2000} and \cite{Shaffer2017} for different electron temperatures. These results suggest significant heating could occur. For example, if the correlation changes by roughly $1\%$ in a plasma with coupling strength $\varphi/T\sim 100$ (as in \cite{Daughton2000} for an electron temperature change by a factor of $9$) then $\Delta T \sim T$ may be possible. We conclude that correlation heating could be large if the electrons and ions are initially strongly coupled.

\subsection{Application of Correlation Heating}

In magnetic confinement fusion, it is common for plasma to exist in a two-temperature state because of heating or cooling mechanisms that target one species preferentially and drive the electron and ion temperatures apart. 

For example, devices may be operated in a \textit{hot-ion mode} in which the ion temperature is maintained significantly above the electron temperature \cite[]{Clarke1980,Kolmes2021,Jin2021} by mechanisms such as ion cyclotron heating, neutral beam heating, or alpha channelling \cite[]{Fisch1992,Fisch1994,Fisch1999}. The supershot plasma regime in the Tokamak Fusion Test Reactor (TFTR) \cite[]{Strachan1987}  and the Hot Ion H-mode regime in the Joint European Torus (JET) \cite[]{Keilhacker1999} were both characterised by high fusion rates and much hotter ions than electrons. As another example, electrons are cooled by Bremsstrahlung, which is particularly significant in reactors operating at the high temperatures required for advanced fuels such as proton-\ce{^{11}}Boron \cite[]{Rider1995,Rider1997}. In such reactors, Bremsstrahlung makes hotter ions than electrons a requirement \cite[]{Cai2022}.

In inertial confinement fusion, a two-temperature state can arise when the electron temperature increases so rapidly that collisional energy exchange is too slow to enforce equal species temperatures. ICF plasmas receive a sudden input of energy from powerful lasers and may have high coupling strengths, so correlation heating is most relevant in this context.

The fact that increasing the electron temperature causes the ion temperature to also rise can be captured by equation of state models that allow different electron and ion temperatures while also incorporating strong-coupling effects \cite[e.g.][]{Fetsch2023}. The novel possibility unlocked by correlation heating is fast, collisionless ion heating, which can be used (once, unless the electrons cool back down) to step the ion temperature upwards faster than would be possible just using ion-electron Coulomb collisions.

This could be important in fast ignition approaches to ICF, in which the hotspot must be heated to ignition in about $\SI{20}{\pico\second}$ \cite[]{Tabak1994,Atzeni1999}. The heating needs to be rapid so that there is no time for pressure equilibration within the fuel, meaning it needs to be faster than the hotspot confinement time \cite[]{Tabak2006}. It is usually assumed that the ions heat up on the ion-electron collision timescale.

Correlation heating could be leveraged to allow faster heating of the ions, which would be most effective if laser heating of the electrons occurred on an ion-plasma-frequency timescale. In order for this to be a more efficient way to achieve ignition, it is important that energy is supplied to the suprathermal tail of the ion distribution. The calculations in \S\ref{sec:distribution} for a moderately coupled plasma suggest that this does not occur. However, this result may not hold in all cases.

The simple picture given in \S\ref{sec:intro} of ions flying apart after they are suddenly stripped of their shielding clouds suggests that suprathermal ions could be created if a group of ions all work together to push another ion in one direction. This would occur when there are significant fluctuations in the number of ions in any given region, before the heating. Large fluctuations, corresponding to a large clumps of well-shielded ions in close proximity, would have large potential energy immediately after the electron screening gets removed. This potential energy would get evenly shared between the ions when they fly apart. In this picture, correlation heating is similar to localised Coulomb explosions wherever there are small-scale density fluctuations in the plasma. 

This picture suggests that fast ion creation could be significantly enhanced if the plasma is initially strongly coupled, so that it is dense and so that the electrons are cold enough to provide efficacious shielding before they are heated. On the other hand, if the plasma is too strongly coupled then the ions can start to form a regular spatial structure, which makes spatial inhomogeneities that create fast ions less likely.

Furthermore, correlation heating could be very effective at accelerating light minority ions to suprathermal speeds. If the repulsion between an initially stationary pair of ions is suddenly switched on, which is a simple model for the removal of strong shielding, then the particles fly apart but the majority of the interaction energy goes to the light ion. Note that the arguments in \S\ref{sec:howlarge}, leading to the conclusion that correlation heating cannot heat strongly coupled ions enough that they become weakly coupled, do not apply to a minority ion species.

It is also worth pointing out that, in systems with sufficiently hot electrons, the ion-ion collision time can be shorter than the electron-ion collision time, even for a fast ion. In such cases, after correlation heating supplies additional energy to the ions, the ion distribution re-Maxwellianises and this fills out the tail of the distribution on a faster timescale than the usual collisional timescale.

For all of these reasons, it may be that there is an advantage to heating the plasma even faster than the typical laser duration of $\sim\SI{20}{\pico\second}$ envisioned for fast ignition. This is within reach of modern ultrafast lasers, which can operate at the femtosecond level -- such intriguing possibilities merit further investigation.

\section{Conclusion}

In this paper, we used a novel expansion of the BBGKY hierarchy to rigorously derive the ion temperature increase that can be achieved using correlation heating in moderately coupled plasmas. The resulting formulas are compact and comprehensible, establishing a framework that makes it possible to quickly estimate how much potential energy could be extracted by suddenly stripping ions of their Debye clouds.

We began in \S\ref{sec:correlations} with a detailed review of the pair correlations in a two-temperature plasma, including a new, simple derivation that only uses elementary plasma physics concepts.

In a departure from conventional plasma kinetic theory, we focused on \textit{moderately coupled} plasmas by taking the mass ratio $m_e/m_i$ to be the smallest parameter in the problem, rather than $1/\Lambda$. Interchanging these two limits means the electron collision frequencies become faster than the ion plasma frequency, so the usual Bogoliubov timescale hierarchy is modified. This is crucial for understanding how the system relaxes on short timescales after rapid heating. We prove that the ions behave like a Yukawa one-component plasma, with electrons only entering via an effective screening length which can be calculated using conservation of energy.

The formalism is applied in \S\ref{sec:temps}--\ref{sec:distribution} to find the temperature changes that occur for both ions and electrons, the time-dependence of the ion kinetic energy, and the modification to the ion distribution function. The ion kinetic energy does not overshoot its final value, unlike in strongly coupled plasmas. Correlation heating is shown to be inefficient at heating fast ions in a moderately coupled plasma; however, we argue that this conclusion may not hold in strongly coupled systems, particularly if there if a light minority ion species present. This may be relevant for the fast-ignition approach to ICF, where a femtosecond laser pulse would likely be necessary to take advantage of correlation heating.

This work also uncovered potential advantages of integrating concepts from the fields of strongly coupled and ultracold plasmas into fusion research \cite[]{Bergeson2019}. We studied a process in which energy is stored in correlations and quickly converted into particle kinetic energy. An alternative possibility is to release the correlation energy by compressing the plasma \cite[]{Avinash2010,Avinash2014,Fetsch2023}. Similar mechanisms have been proposed for different types of correlation; for example, compressing a plasma with correlations arising from Langmuir waves can energise hot electrons \cite[]{Schmit2010,Schmit2013, Schmit2013b}, while compressing a plasma with correlations arising from turbulence can cause sudden dissipation of the turbulent energy as heat \cite[]{Davidovits2016,Davidovits2016sudden,Davidovits2017,Davidovits2019}. The heating discussed in this paper depends on intrinsic correlations that are present in any plasma in which electrons shield ions, and is therefore of universal interest. Other applications could be found as strongly coupled plasmas are further studied theoretically, numerically and in future experiments on ultracold plasmas.

\begin{acknowledgments}
The authors thank Matt Kunz for many helpful discussions. This work was supported by DOE Grant Nos. DE-SC0016072 and DE-AC02- 09CH11466, and DOE-NNSA Grant No. 83228–10966 [Prime No. DOE (NNSA) DE- NA0003764].

Competing interests: the authors declare none.
\end{acknowledgments}

\appendix

\section{Summary of Definitions and Notation}\label{app:defs}

This appendix explains the definitions and notation that we use throughout the paper relating to distribution functions and pair correlations.

\begin{enumerate}
	\item Let the position and velocity of particle $1$ be $\boldsymbol{r}_1$ and $\boldsymbol{v}_1$ respectively. We use the shorthand notation $(1)$ for the phase-space coordinates $(\boldsymbol{r}_1, \boldsymbol{v}_1)$ and we write $\int \rmd(1) = \int \rmd\boldsymbol{r}_1\rmd\boldsymbol{v}_1$ for integration over these coordinates. When a function $f$ depends on the phase-space coordinates of multiple particles, we use $f(12\ldots)$, and when integrating over the phase-space coordinates of multiple particles we write $\int \rmd(12\ldots)$. Also, we will represent integrals over the velocity of a particle, but not its position, by $\int \rmd[1] = \int \rmd\boldsymbol{v}_1$.
	
	\item The phase space distribution $\rho(12\ldots)=\rho(\boldsymbol{r}_1,\boldsymbol{v}_1,\boldsymbol{r}_2,\boldsymbol{v}_2,\ldots)$ is the probability density that each particle in the system has a specified position and velocity. It is normalised so that $\int \rmd(12\ldots)\rho(12\ldots) = 1$.  
	
	\item The one-particle distribution $f_\alpha(1)$, two-particle distribution $f_{\alpha\beta}(12)$, and higher-order reduced distribution functions are defined by
	\begin{align}
	f_\alpha(1) &= N_\alpha\int \rmd(23\ldots)\rho(12\ldots)\,,\\
	f_{\alpha\beta}(12) &= N_\alpha(N_\beta-\delta_{\alpha\beta})\int \rmd(34\ldots)\rho(12\ldots)\,,\\
	f_{\alpha\beta\gamma}(123) &= N_\alpha(N_\beta - \delta_{\alpha\beta})(N_\gamma - \delta_{\alpha\gamma} - \delta_{\beta\gamma}) \int \rmd(45\ldots)\rho(12\ldots)\,,\\
	&\nonumber\mathrm{etc.\,,}
	\end{align}
	where particle $1$ belongs to species $\alpha$, particle $2$ belongs to species $\beta$, particle $3$ belongs to species $\gamma$, and so on. Here, the Kronecker delta $\delta_{\alpha\beta}$ is zero unless the species indices $\alpha$ and $\beta$ are equal, and $N_\alpha$ is the number of particles of species $\alpha$. 
	
	\item If the particle positions and velocities are uncorrelated in a system in the thermodynamic limit, then $f_{\alpha\beta}(12) = f_\alpha(1) f_\beta(2)$. We therefore define pair correlations $g_{\alpha\beta}(12)$ by
	\begin{equation}
	f_{\alpha\beta}(12) = f_\alpha(1)f_\beta(2) + g_{\alpha\beta}(12)\,.
	\end{equation}
	Higher order correlations are defined using a Mayer cluster expansion scheme \cite[]{KrallTrivelpiece}; for example, the triple correlation $h_{\alpha\beta\gamma}(123)$ is given by
	\begin{align}
	f_{\alpha\beta\gamma}(123) = f_\alpha(1)f_\beta(2)f_\gamma(3) + f_\alpha(1)g_{\beta\gamma}(23) &+ f_\beta(2)g_{\gamma\alpha}(31)\\
	&\nonumber + f_\gamma(3)g_{\alpha\beta}(12) + h_{\alpha\beta\gamma}(123)\,.
	\end{align}
	
	\item In a homogeneous system, $f_\alpha(1) = f_\alpha(\boldsymbol{v}_1)$ must be independent of position $\boldsymbol{r}_1$. Also, $f_{\alpha\beta}(12)$ and $g_{\alpha\beta}(12)$ can only depend on the particle positions through the relative displacement $\boldsymbol{r}_1-\boldsymbol{r}_2$:
	\begin{equation}
	g_{\alpha\beta}(\boldsymbol{r}_1,\boldsymbol{v}_1, \boldsymbol{r}_2, \boldsymbol{v}_2) = g_{\alpha\beta}(\boldsymbol{r}_1 - \boldsymbol{r}_2,\boldsymbol{v}_1, \boldsymbol{v}_2)\,.
	\end{equation}
	Whenever a pair correlation is written using only three arguments, the assumption of homogeneity has been used and the first argument is the relative displacement.
	
	\item When we are interested in spatial correlations only,  we use the integrated pair correlation
	\begin{equation}
	G_{\alpha\beta}(\boldsymbol{r}) = \frac{1}{n_\alpha n_\beta}\int \rmd\boldsymbol{v}_1\rmd\boldsymbol{v}_2\; g_{\alpha\beta}(\boldsymbol{r}, \boldsymbol{v}_1, \boldsymbol{v}_2)\,.
	\end{equation}
	\item Fourier transforms are denoted using tildes. The Fourier transform of a function $F(\boldsymbol{r})$ is defined by
	\begin{equation}
	\widetilde{F}(\boldsymbol{k}) = \int \rmd\boldsymbol{r}\, F(\boldsymbol{r})\,\rme^{-\imag\boldsymbol{k\cdot r}}.
	\end{equation}
	The Fourier transform of the pair correlation with respect to the relative displacement $\boldsymbol{r}_1 - \boldsymbol{r}_2$ is denoted by the shorthand notation $\widetilde{g}_{\alpha\beta}(12) = \widetilde{g}_{\alpha\beta}(\boldsymbol{k},\boldsymbol{v}_1, \boldsymbol{v}_2)$.
\end{enumerate}

\section{Proof of Relations Between Pair Correlations and Fluctuations.}\label{app:fluc}

This appendix proves relations between pair correlations and Fourier-space density fluctuations. The charge density of species $\alpha$ at a given moment is
\begin{equation}
    \rho_\alpha(\boldsymbol{r}) = q_\alpha\sum_{i=1}^{N_\alpha} \delta(\boldsymbol{r}-\boldsymbol{r}_i)
\end{equation}
where the sum is over all particles, indexed by $i$, of species $\alpha$. The Fourier transform of this charge density is
\begin{equation}
    \widetilde{\rho}_\alpha(\boldsymbol{k}) = \sum_{i=1}^{N_\alpha} \rme^{-\imag\boldsymbol{k\cdot r}_i}\,.
\end{equation}
In a uniform, equilibrium plasma, the thermal average $\langle\widetilde{\rho}_\alpha(\boldsymbol{k})\rangle$ must vanish for all nonzero $\boldsymbol{k}$. However, $\widetilde{\rho}_\alpha(\boldsymbol{k})$ will have thermal fluctuations characterised by a nonzero $\langle|\widetilde{\rho}_\alpha(\boldsymbol{k})|^2\rangle$.

We now prove the following standard formulas, which hold for nonzero $\boldsymbol{k}$.
\begin{align}
\label{eq:struc}\langle|\widetilde{\rho}_\alpha(\boldsymbol{k})|^2\rangle &= Vn_\alpha q_\alpha^2 \left( 1 + n_\alpha \widetilde{G}_{\alpha\alpha}(\boldsymbol{k}) \right),\\
\label{eq:struc2}\langle\widetilde{\rho}_\alpha(\boldsymbol{k})\widetilde{\rho}_\beta^{\,*}(\boldsymbol{k})\rangle &= Vn_\alpha n_\beta q_\alpha q_\beta\widetilde{G}_{\alpha\beta}(\boldsymbol{k})\text{\hspace{10pt}when $\alpha\neq\beta$.}
\end{align}

The first identity \eqref{eq:struc} is the well-known relation between pair correlations and structure factors. It is derived as follows \cite[]{Hansen}:
\begin{align}
    \langle|\widetilde{\rho}_\alpha(\boldsymbol{k})|^2\rangle &= \left\langle q_\alpha^2\sum_{i,j=1}^{N_\alpha} \rme^{-\imag\boldsymbol{k\cdot}(\boldsymbol{r}_i-\boldsymbol{r}_j)} \right\rangle \\
    &\nonumber= \left\langle N_\alpha q_\alpha^2 + q_\alpha^2\sum_{\substack{i,j=1\\ j\neq i}}^{N_\alpha} \rme^{-\imag\boldsymbol{k\cdot}(\boldsymbol{r}_i-\boldsymbol{r}_j)} \right\rangle\\
    &\nonumber= N_\alpha q_\alpha^2 + N_\alpha(N_\alpha-1)q_\alpha^2\left\langle \rme^{-\imag\boldsymbol{k\cdot}(\boldsymbol{r}_i-\boldsymbol{r}_j)} \right\rangle\\
    &\nonumber= N_\alpha q_\alpha^2 + q_\alpha^2\int \rmd(12)\, f_{\alpha\alpha}(12)\, \rme^{-\imag\boldsymbol{k\cdot}(\boldsymbol{r}_i-\boldsymbol{r}_j)}\\
    &\nonumber= N_\alpha q_\alpha^2 + q_\alpha^2\int \rmd(12)\, g_{\alpha\alpha}(12)\, \rme^{-\imag\boldsymbol{k\cdot}(\boldsymbol{r}_i-\boldsymbol{r}_j)}\\
    &\nonumber= V n_\alpha q_\alpha^2 \left( 1 + n_\alpha \widetilde{G}_{\alpha\alpha}(\boldsymbol{k}) \right).
\end{align}
In going from the second to the third line we used the fact that all particles of species $\alpha$ are identical. In going from the third line to the fourth, we used $f_{\alpha\alpha}(12) = f_\alpha(1)f_\alpha(2) + g_{\alpha\alpha}(12)$ and the fact that $f_\alpha(1)f_\alpha(2)$ has no Fourier component at nonzero $\boldsymbol{k}$ in a homogeneous system.

The proof of the second identity \eqref{eq:struc2} is similar:
{\allowdisplaybreaks
\begin{align}
\langle\widetilde{\rho}_\alpha(\boldsymbol{k})\widetilde{\rho}_\beta^*(\boldsymbol{k})\rangle &= \left\langle q_\alpha q_\beta\sum_{i=1}^{N_\alpha}\sum_{j=1}^{N_\beta} \rme^{-\imag\boldsymbol{k\cdot}(\boldsymbol{r}_i-\boldsymbol{r}_j)} \right\rangle \\
    &\nonumber= N_\alpha N_\beta q_\alpha q_\beta\left\langle \rme^{-\imag\boldsymbol{k\cdot}(\boldsymbol{r}_i-\boldsymbol{r}_j)} \right\rangle\\
    &\nonumber=q_\alpha q_\beta\int \rmd(12)\, g_{\alpha\beta}(12)\, \rme^{-\imag\boldsymbol{k\cdot}(\boldsymbol{r}_i-\boldsymbol{r}_j)}\\
    &\nonumber= Vn_\alpha n_\beta q_\alpha q_\beta \widetilde{G}_{\alpha\beta}(\boldsymbol{k})\,.
\end{align}}

\section{Ion Energy Integral}\label{app:integral}

This appendix evaluates the integrals in \eqref{eq:K1} and \eqref{eq:K2}, repeated here, for the rate of change of ion kinetic energy density on the ion-plasma-frequency timescale:
\begin{align}\label{eq:i1}
\frac{\rmd K_i^{(1)}}{\rmd t} = \,&\frac{T_{i0}k_{i0}^4}{n_i^2}\int \frac{\rmd\boldsymbol{k}}{(2\upi)^3}\frac{1}{k^2+k_{D0}^2}\frac{1}{k^2+k_{e1}^2}\int \rmd\boldsymbol{v}_3\rmd\boldsymbol{v}_4 \,f_i(\boldsymbol{v}_3)f_i(\boldsymbol{v}_4)\\
\nonumber&\times\int \rmd\boldsymbol{v}_1\rmd\boldsymbol{v}_2\,(\imag\boldsymbol{k\cdot v}_1) \int_\mathcal{B}\frac{\rmd\omega}{2\upi}\int_{\mathcal{B}'}\frac{\rmd\omega'}{2\upi}\, \rme^{-\imag(\omega+\omega')t}\\
\nonumber&\times\left( \frac{\delta(\boldsymbol{v}_1-\boldsymbol{v}_3)}{\omega - \boldsymbol{k\cdot v}_1} + \frac{1}{n_i}\frac{k_{i0}^2}{k^2+k_{e1}^2}\frac{1}{\epsilon(\omega,\boldsymbol{k})}\frac{\boldsymbol{k\cdot v}_1}{\omega - \boldsymbol{k\cdot v}_1}\frac{1}{\omega - \boldsymbol{k\cdot v}_3} f_{i0}(\boldsymbol{v}_1)\right)\\
\nonumber&\times\left( \frac{\delta(\boldsymbol{v}_2-\boldsymbol{v}_4)}{\omega' + \boldsymbol{k\cdot v}_2} - \frac{1}{n_i}\frac{k_{i0}^2}{k^2+k_{e1}^2}\frac{1}{\epsilon(\omega',\boldsymbol{k})}\frac{\boldsymbol{k\cdot v}_2}{\omega' + \boldsymbol{k\cdot v}_2}\frac{1}{\omega' + \boldsymbol{k\cdot v}_4} f_{i0}(\boldsymbol{v}_2)\right),
\end{align}
\begin{align}\label{eq:i2}
\frac{\rmd K_i^{(2)}}{\rmd t} = \,&\frac{T_{i0}k_{i0}^4}{n_i^2}\int \frac{\rmd\boldsymbol{k}}{(2\upi)^3}\frac{1}{(k^2+k_{e1}^2)^2}\int \rmd\boldsymbol{v}_3\rmd\boldsymbol{v}_4 \,f_i(\boldsymbol{v}_3)f_i(\boldsymbol{v}_4)(i\!\boldsymbol{k\,\cdot}(\boldsymbol{v}_3-\boldsymbol{v}_4))\\
\nonumber&\times\int \rmd\boldsymbol{v}_1\rmd\boldsymbol{v}_2\,(\imag\boldsymbol{k\cdot v}_1) \int_0^t \rmd t'\int_\mathcal{B}\frac{\rmd\omega}{2\upi}\int_{\mathcal{B}'}\frac{\rmd\omega'}{2\upi}\, \rme^{-\imag(\omega+\omega')t'}\\
\nonumber&\times\left( \frac{\delta(\boldsymbol{v}_1-\boldsymbol{v}_3)}{\omega - \boldsymbol{k\cdot v}_1} + \frac{1}{n_i}\frac{k_{i0}^2}{k^2+k_{e1}^2}\frac{1}{\epsilon(\omega,\boldsymbol{k})}\frac{\boldsymbol{k\cdot v}_1}{\omega - \boldsymbol{k\cdot v}_1}\frac{1}{\omega - \boldsymbol{k\cdot v}_3} f_{i0}(\boldsymbol{v}_1)\right)\\
\nonumber&\times\left( \frac{\delta(\boldsymbol{v}_2-\boldsymbol{v}_4)}{\omega' + \boldsymbol{k\cdot v}_2} - \frac{1}{n_i}\frac{k_{i0}^2}{k^2+k_{e1}^2}\frac{1}{\epsilon(\omega',\boldsymbol{k})}\frac{\boldsymbol{k\cdot v}_2}{\omega' + \boldsymbol{k\cdot v}_2}\frac{1}{\omega' + \boldsymbol{k\cdot v}_4} f_{i0}(\boldsymbol{v}_2)\right).
\end{align}
First, some useful integrals involving the dielectric function are calculated, and some of their analytic properties are summarised; this will be important for solving various contour integrals later. Then, the multiple integrals in \eqref{eq:i1} and \eqref{eq:i2} are taken in order.

\subsection{Integrals Involving the Dielectric Function}

We begin by evaluating some integrals that will be used repeatedly in this appendix.

The plasma dielectric function was defined in \eqref{eq:dielectric} as
\begin{equation}
\epsilon(\omega ,\boldsymbol{k}) = 1 - \frac{1}{n_{\imag}}\frac{k_{i0}^2}{k^2+k_{e1}^2} \int_{\mathcal{L}} \mathrm{d}\boldsymbol{v} \, \frac{\boldsymbol{k}\cdot \boldsymbol{v}}{\omega -\boldsymbol{k}\cdot\boldsymbol{v}} \, f_{i0}(\boldsymbol{v})\,.\label{eq:dielectric2}
\end{equation}
Here, the subscript $\mathcal{L}$ means the velocity integrals should be taken using the usual Landau prescription.

We will need the integrals
\begin{align}
&\label{eq:J1}\int_{\mathcal{L}} \mathrm{d}\boldsymbol{v} \, \frac{\omega}{\omega -\boldsymbol{k}\cdot\boldsymbol{v}} \, f_{i0}(\boldsymbol{v}) = n_i\,\xi(\omega,\boldsymbol{k})\,, \\
&\label{eq:J2}\int_{\mathcal{L}} \mathrm{d}\boldsymbol{v} \, \frac{\omega}{\omega -\boldsymbol{k}\cdot\boldsymbol{v}}\left( \frac{\boldsymbol{k}\cdot \boldsymbol{v}}{\omega} \right) \, f_{i0}(\boldsymbol{v}) = n_i\,[\xi(\omega,\boldsymbol{k})-1]\,,\\
&\label{eq:J3}\int_{\mathcal{L}} \mathrm{d}\boldsymbol{v} \, \frac{\omega}{\omega -\boldsymbol{k}\cdot\boldsymbol{v}}\left( \frac{\boldsymbol{k}\cdot \boldsymbol{v}}{\omega} \right)^2 f_{i0}(\boldsymbol{v}) = n_i\,[\xi(\omega,\boldsymbol{k})-1]\,,
\end{align}
where
\begin{equation}
\xi(\omega,\boldsymbol{k}) =  1+\left(\frac{k^2 + k_{e1}^2}{k_{i0}^2}\right) (1 - \epsilon(\omega ,\boldsymbol{k}))\,.
\end{equation}
These formulas are simple to prove. Equation \eqref{eq:J2} is simply a rearrangement of the definition of $\epsilon(\omega,\boldsymbol{k})$ in \eqref{eq:dielectric2}. Equation \eqref{eq:J1} is obtained as follows:
\begin{align}
 \int_{\mathcal{L}} \mathrm{d}\boldsymbol{v} \, \frac{\omega}{\omega -\boldsymbol{k}\cdot\boldsymbol{v}} \, f_{i0}(\boldsymbol{v}) &= \int_{\mathcal{L}} \mathrm{d}\boldsymbol{v}\, \frac{\omega - \boldsymbol{k}\cdot \boldsymbol{v}}{\omega -\boldsymbol{k}\cdot\boldsymbol{v}} \, f_{i0}(\boldsymbol{v}) + \int_{\mathcal{L}} \mathrm{d}\boldsymbol{v}\, \frac{\boldsymbol{k}\cdot \boldsymbol{v}}{\omega -\boldsymbol{k}\cdot\boldsymbol{v}} \, f_{i0}(\boldsymbol{v}) \\
	\nonumber&= n_i + n_i\,[\xi(\omega,\boldsymbol{k})-1]\,.
\end{align}
Lastly, equation \eqref{eq:J3} comes from
\begin{align}
\int_{\mathcal{L}} \mathrm{d}\boldsymbol{v} \, \frac{\omega}{\omega -\boldsymbol{k}\cdot\boldsymbol{v}}\left( \frac{\boldsymbol{k}\cdot \boldsymbol{v}}{\omega} \right)^2 f_{i0}(\boldsymbol{v}) &= \int_{\mathcal{L}} \mathrm{d}\boldsymbol{v}\, \left(\frac{\boldsymbol{k}\cdot \boldsymbol{v} - \omega  + \omega }{\omega -\boldsymbol{k}\cdot\boldsymbol{v}}\right)\left(\frac{\boldsymbol{k}\cdot \boldsymbol{v}}{\omega}\right) f_{i0}(\boldsymbol{v})\\
	\nonumber&= \int_{\mathcal{L}} \mathrm{d}\boldsymbol{v}\, \frac{\boldsymbol{k}\cdot \boldsymbol{v}}{\omega -\boldsymbol{k}\cdot\boldsymbol{v}} \, f_{i0}(\boldsymbol{v})\\
	\nonumber&= n_i\, [\xi(\omega,\boldsymbol{k})-1]\,.
\end{align}

\subsection{Analytic Properties of $\epsilon(\omega,\boldsymbol{k})$ and $\xi(\omega,\boldsymbol{k})$}

We will need to evaluate many contour integrals in the complex $\omega$-plane involving the functions $\epsilon(\omega,\boldsymbol{k})$ and $\xi(\omega,\boldsymbol{k})$, so we briefly list some of their properties. These properties are easiest to prove using the closed-form expressions for these functions. By aligning the $z$-axis along $\boldsymbol{k}$, we have
\begin{align}
\xi(\omega,\boldsymbol{k}) &= \frac{1}{n_i}\int_{\mathcal{L}} \mathrm{d}\boldsymbol{v}\, \frac{\omega}{\omega -\boldsymbol{k}\cdot\boldsymbol{v}} \, f_{i0}(\boldsymbol{v})\\
    \nonumber&= -\frac{\omega}{k}\sqrt{\frac{m_i}{2\upi T_{i0}}}\int_{\mathcal{L}} \rmd v_z\, \frac{\rme^{-m_iv_z^2/2T_{i0}}}{v_z - \omega/k}\\
    \nonumber&= -\frac{\zeta}{\sqrt{\upi}}\int_{\mathcal{L}} \rmd u\, \frac{\rme^{-u^2}}{u - \zeta}\\
    \nonumber&= -\zeta Z(\zeta)\,,
\end{align}
where $\zeta = \omega/kv_{\mathrm{th}i}$, using the thermal velocity of the ions $v_{\mathrm{th}i} = (2T_{i0}/m_i)^{1/2}$, and $Z(\zeta)$ is the usual plasma dispersion function \cite[]{Faddeeva1954,Fried1961}:
\begin{equation}\label{eq:Z}
    Z(\zeta) = \frac{1}{\sqrt{\upi}}\int_{\mathcal{L}}\rmd u\,\frac{\rme^{-u^2}}{u-\zeta} = \imag\sqrt{\upi}\rme^{-\zeta^2}(1 + \imag\, \mathrm{erfi}(\zeta))\,.
\end{equation}
For $\epsilon(\omega,\boldsymbol{k})$ we then obtain
\begin{equation}\label{eq:epsilonclosedform}
    \epsilon(\omega,\boldsymbol{k}) = 1 + \frac{k_{i0}^2}{k^2+k_{e1}^2}(1 + \zeta Z(\zeta))\,.
\end{equation}
Using this expression for $\epsilon(\omega, \boldsymbol{k})$, we can now make the following observations.
\begin{enumerate}
\item Both  $\epsilon(\omega,\boldsymbol{k})$ and $\xi(\omega,\boldsymbol{k})$ are analytic functions of $\omega$.
\item The dielectric function $\epsilon(\omega, \boldsymbol{k}) $ has symmetries:
\begin{align}
\label{eq:epsomega}\epsilon(-\omega^*, \boldsymbol{k}) &= \epsilon(\omega, \boldsymbol{k})^*,\\
	\epsilon(\omega, \boldsymbol{k}) &= \epsilon(\omega, -\boldsymbol{k})\,.
\end{align}
Therefore, $\xi(\omega,\boldsymbol{k})$ has the same symmetries:
\begin{align}
\label{eq:xiomega}\xi(-\omega^*, \boldsymbol{k}) &= \xi(\omega, \boldsymbol{k})^*,\\
\xi(\omega, \boldsymbol{k}) &= \xi(\omega, -\boldsymbol{k})\,.
\end{align}
\item The behaviour of $\epsilon(\omega,\boldsymbol{k})$ and $\xi(\omega,\boldsymbol{k})$ as $\omega\to 0$ is:
\begin{align}
\label{eq:eps0}\epsilon(0,\boldsymbol{k}) &= 1 +\frac{k_{i0}^2}{k^2+k_{e1}^2}\,,\\
\xi(\omega,\boldsymbol{k}) &\sim -\imag\sqrt{\upi}\frac{\omega}{kv_{\mathrm{th}i}}\,,
\end{align}
using $Z(\zeta)\sim \imag\sqrt{\upi}$ as $\zeta\to 0$, from \eqref{eq:Z}. In particular, note that the singularity of $\xi(\omega,\boldsymbol{k})/\omega$ at $\omega=0$ is removable, so there is no pole of $\xi(\omega,\boldsymbol{k})/\omega$ at the origin.
\item The behaviour of $\epsilon(\omega,\boldsymbol{k})$ and $\xi(\omega,\boldsymbol{k})$ as $|\omega|\to\infty$ in the upper half-plane is
\begin{align}
\label{eq:elimit}\epsilon(\omega,\boldsymbol{k})&\sim 1 + O(1/\omega^2)\,,\\
\label{eq:xilimit} \xi(\omega,\boldsymbol{k})&\sim 1 + O(1/\omega^2)\,,
\end{align}
which is a consequence of the well-known asymptotic expansion \cite[]{Fried1961}
\begin{equation}\label{eq:Zasymptotics}
Z(\zeta)\sim \imag\sqrt{\upi}\sigma \rme^{-\zeta^2} -\frac{1}{\zeta}\left( 1 + \frac{1}{2\zeta^2} + \ldots \right)
\end{equation}
for $|\zeta|\gg 1$, where
\begin{equation}
\sigma = \begin{cases}0 \hspace{5mm}\text{ if $\operatorname{Im} \zeta > 0$}\\1 \hspace{5mm}\text{ if $\operatorname{Im} \zeta 
= 0 $}\\2 \hspace{5mm}\text{ if $\operatorname{Im} \zeta < 0. $}
\end{cases}
\end{equation}
\item The only zeros of $\epsilon(\omega,\boldsymbol{k})$ occur when $\omega$ lies in the lower half-plane. Therefore, the only poles of $1/\epsilon(\omega,\boldsymbol{k})$ are in the lower half-plane. This is necessary for the plasma to be stable and can be proved, for example, using Nyquist's method \cite[]{Nyquist1932,KrallTrivelpiece}.
\end{enumerate}

\subsubsection{$\boldsymbol{v}_2$ Integration}
In \eqref{eq:i1} and \eqref{eq:i2}, the $\boldsymbol{v}_1$ and $\boldsymbol{v}_2$ integrals can be evaluated immediately. Using \eqref{eq:J2} and the fact that $\xi(\omega,-\boldsymbol{k}) = \xi(\omega,\boldsymbol{k})$, see \eqref{eq:xiomega}, the $\boldsymbol{v}_2$ integral is
\begin{align}\label{eq:v2integral}
	\int \mathrm{d}\boldsymbol{v}_2\, \biggl( \frac{\delta(\boldsymbol{v}_2 - \boldsymbol{v}_4)}{\omega' + \boldsymbol{k}\cdot \boldsymbol{v}_4} &- \frac{1}{n_i} \frac{k_{i0}^2}{k^2 + k_{e1}^2}  \frac{1}{\epsilon(\omega', \boldsymbol{k})}\frac{\boldsymbol{k\cdot v}_2}{\omega + \boldsymbol{k\cdot v}_2}\frac{1}{\omega + \boldsymbol{k\cdot v}_4} f_{i0}(\boldsymbol{v}_2) \biggr)\\
	&\nonumber= \frac{1}{\omega' + \boldsymbol{k}\cdot \boldsymbol{v}_4} + \frac{1 - \epsilon(\omega', \boldsymbol{k})}{\epsilon(\omega', \boldsymbol{k})}\frac{1}{\omega' + \boldsymbol{k}\cdot \boldsymbol{v}_4}\\
	\nonumber&= \frac{1}{\epsilon(\omega', \boldsymbol{k})}\frac{1}{\omega' + \boldsymbol{k}\cdot \boldsymbol{v}_4} \,.
\end{align}

\subsubsection{$\boldsymbol{v}_1$ Integration}
Similarly, using \eqref{eq:J3}, the $\boldsymbol{v}_1$ integral is
\begin{align}
    \int \mathrm{d}\boldsymbol{v}_1\,(\boldsymbol{k\cdot v}_1) \biggl( \frac{\delta(\boldsymbol{v}_1 - \boldsymbol{v}_3)}{\omega - \boldsymbol{k}\cdot \boldsymbol{v}_3} &+ \frac{1}{n_i} \frac{k_{i0}^2}{k^2 + k_{e1}^2}  \frac{1}{\epsilon(\omega, \boldsymbol{k})}\frac{\boldsymbol{k\cdot v}_1}{\omega - \boldsymbol{k\cdot v}_1}\frac{1}{\omega - \boldsymbol{k\cdot v}_3} f_{i0}(\boldsymbol{v}_1) \biggr)\\
	&\nonumber= \frac{\boldsymbol{k\cdot v}_3}{\omega - \boldsymbol{k\cdot v}_3} + \frac{1 - \epsilon(\omega, \boldsymbol{k})}{\epsilon(\omega, \boldsymbol{k})}\frac{\omega}{\omega - \boldsymbol{k\cdot v}_3}\\
	&\nonumber= -1 + \frac{1}{\epsilon(\omega, \boldsymbol{k})}\frac{\omega}{\omega - \boldsymbol{k\cdot v}_3}\,.
\end{align}
Therefore, so far we have
\begin{align}
\frac{\rmd K_i^{(1)}}{\rmd t} = \,-\imag\,\frac{T_{i0}k_{i0}^4}{n_i^2}&\int \frac{\rmd\boldsymbol{k}}{(2\upi)^3}\frac{1}{k^2+k_{D0}^2}\frac{1}{k^2+k_{e1}^2}\int \rmd\boldsymbol{v}_3\rmd\boldsymbol{v}_4 \,f_i(\boldsymbol{v}_3)f_i(\boldsymbol{v}_4)\\
\nonumber\times&\int_\mathcal{B}\frac{\rmd\omega}{2\upi}\int_{\mathcal{B}'}\frac{\rmd\omega'}{2\upi}\, \rme^{-\imag(\omega+\omega')t}
\frac{1}{\epsilon(\omega', \boldsymbol{k})}\frac{1}{\omega' + \boldsymbol{k}\cdot \boldsymbol{v}_4}\\
\nonumber\times&\left( 1 - \frac{1}{\epsilon(\omega, \boldsymbol{k})}\frac{\omega}{\omega - \boldsymbol{k\cdot v}_3}\right),
\end{align}
\begin{align}
\frac{\rmd K_i^{(2)}}{\rmd t} = \,-\imag\frac{T_{i0}k_{i0}^4}{n_i^2}&\int \frac{\rmd\boldsymbol{k}}{(2\upi)^3}\frac{1}{(k^2+k_{e1}^2)^2}\int \rmd\boldsymbol{v}_3\rmd\boldsymbol{v}_4 \,f_i(\boldsymbol{v}_3)f_i(\boldsymbol{v}_4)(\imag\boldsymbol{k}\boldsymbol{\cdot}(\boldsymbol{v}_3-\boldsymbol{v}_4))\\
\nonumber\times &\int_0^t \rmd t'\int_\mathcal{B}\frac{\rmd\omega}{2\upi}\int_{\mathcal{B}'}\frac{\rmd\omega'}{2\upi}\, \rme^{-\imag(\omega+\omega')t'}\frac{1}{\epsilon(\omega', \boldsymbol{k})}\frac{1}{\omega' + \boldsymbol{k}\cdot \boldsymbol{v}_4}\\
\nonumber\times&\left( 1 - \frac{1}{\epsilon(\omega, \boldsymbol{k})}\frac{\omega}{\omega - \boldsymbol{k\cdot v}_3}\right).
\end{align}

\subsection{$\boldsymbol{v}_3$ and $\boldsymbol{v}_4$ Integration}

Next we can perform the $\boldsymbol{v}_3$ and $\boldsymbol{v}_4$ integrals, which are slightly different in each case. Starting with $\rmd K_i^{(1)}/\rmd t$, we use \eqref{eq:J1} to get
\begin{align}
    \int \rmd\boldsymbol{v}_3\rmd\boldsymbol{v}_4\,f_i(\boldsymbol{v}_3)f_i(\boldsymbol{v}_4)\,&\frac{1}{\epsilon(\omega', \boldsymbol{k})}\frac{1}{\omega' + \boldsymbol{k}\cdot \boldsymbol{v}_4}\left( 1 - \frac{1}{\epsilon(\omega, \boldsymbol{k})}\frac{\omega}{\omega - \boldsymbol{k\cdot v}_3} \right) \\
    \nonumber&= \int \rmd\boldsymbol{v}_3\,f_i(\boldsymbol{v}_3)\,\frac{n_i}{\omega'}\frac{\xi(\omega',\boldsymbol{k})}{\epsilon(\omega', \boldsymbol{k})}\left( 1 - \frac{1}{\epsilon(\omega, \boldsymbol{k})}\frac{\omega}{\omega - \boldsymbol{k\cdot v}_3} \right)\\
    \nonumber&= \frac{n_i^2}{\omega'}\frac{\xi(\omega',\boldsymbol{k})}{\epsilon(\omega', \boldsymbol{k})}\left( 1 - \frac{\xi(\omega,\boldsymbol{k})}{\epsilon(\omega, \boldsymbol{k})}\right).
\end{align}
Therefore,
\begin{align}\label{eq:IK1}
    \frac{\rmd K_i^{(1)}}{\rmd t} = -\imag\,T_{i0}k_{i0}^4\int \frac{\rmd\boldsymbol{k}}{(2\upi)^3} \frac{1}{k^2+k_{e1}^2}\frac{1}{k^2+k_{D0}^2} \int_\mathcal{B} \frac{\rmd\omega}{2\upi}\int_{\mathcal{B}'}\frac{\rmd\omega'}{2\upi}\, \rme^{-\imag(\omega+\omega')t}&\\
    \nonumber\times\frac{1}{\omega'}\frac{\xi(\omega',\boldsymbol{k})}{\epsilon(\omega', \boldsymbol{k})}\left( 1 - \frac{\xi(\omega,\boldsymbol{k})}{\epsilon(\omega, \boldsymbol{k})}\right)&.
\end{align}
In $\rmd K_i^{(2)}/\rmd t$, the $\boldsymbol{v}_4$ integral can be computed using \eqref{eq:J1} and \eqref{eq:J2} and the symmetry $\xi(\omega,-\boldsymbol{k}) = \xi(\omega,\boldsymbol{k})$ again:
\begin{equation}
    \int \rmd\boldsymbol{v}_4\, \frac{\boldsymbol{k\,\cdot\,}(\boldsymbol{v}_3-\boldsymbol{v}_4)}{\omega' + \boldsymbol{k\cdot v}_4}f_i(\boldsymbol{v}_4) = n_i \frac{\boldsymbol{k\cdot v}_3}{\omega'}\xi(\omega',\boldsymbol{k}) + n_i \,[\xi(\omega',\boldsymbol{k})-1]\,.
\end{equation}
Next we carry out the $\boldsymbol{v}_3$ integral. We need the following results:
\begin{equation}
    \int \rmd\boldsymbol{v}_3\, \frac{\boldsymbol{k\cdot v}_3}{\omega'}\biggl(  1 - \frac{1}{\epsilon(\omega, \boldsymbol{k})}\frac{\omega}{\omega - \boldsymbol{k\cdot v}_3} \biggr) f_i(\boldsymbol{v}_3) = -n_i \frac{\omega}{\omega'}\frac{\xi(\omega,\boldsymbol{k})-1}{\epsilon(\omega,\boldsymbol{k})}\,,
\end{equation}
which uses \eqref{eq:J2}, and
\begin{align}
   \nonumber \int \rmd\boldsymbol{v}_3\, \biggl(  1 - \frac{1}{\epsilon(\omega, \boldsymbol{k})}\frac{\omega}{\omega - \boldsymbol{k\cdot v}_3} \biggr) f_i(\boldsymbol{v}_3) = n_i - n_i \frac{\xi(\omega,\boldsymbol{k})}{\epsilon(\omega,\boldsymbol{k})}\,,
\end{align}
which uses \eqref{eq:J1}. Therefore,
\begin{align}
\frac{\rmd K_i^{(2)}}{\rmd t} = \,T_{i0}k_{i0}^4&\int \frac{\rmd\boldsymbol{k}}{(2\upi)^3}\frac{1}{(k^2+k_{e1}^2)^2}\int_0^t \rmd t'\int_\mathcal{B}\frac{\rmd\omega}{2\upi}\int_{\mathcal{B}'}\frac{\rmd\omega'}{2\upi}\, \rme^{-\imag(\omega+\omega')t'}\\
\nonumber&\times\left[-\frac{\omega}{\omega'}\frac{\xi(\omega',\boldsymbol{k})}{\epsilon(\omega', \boldsymbol{k})}\frac{\xi(\omega,\boldsymbol{k})-1}{\epsilon(\omega,\boldsymbol{k})} + \frac{\xi(\omega',\boldsymbol{k})-1}{\epsilon(\omega',\boldsymbol{k})}\left(1-\frac{\xi(\omega,\boldsymbol{k})}{\epsilon(\omega,\boldsymbol{k})} \right)\right].
\end{align}
If the contours are deformed so that $\mathcal{B}$ traces out the same curve in the $\omega$-plane as $\mathcal{B}'$ does in the $\omega'$-plane, then we can swap $\omega$ and $\omega'$ in the integrand. Using this symmetrisation on the second term in the square brackets leads to
\begin{align}
\frac{\rmd K_i^{(2)}}{\rmd t} = \,T_{i0}k_{i0}^4\int \frac{\rmd\boldsymbol{k}}{(2\upi)^3}\frac{1}{(k^2+k_{e1}^2)^2}\int_0^t \rmd t'\int_\mathcal{B}\frac{\rmd\omega}{2\upi}\int_{\mathcal{B}'}\frac{\rmd\omega'}{2\upi}\, \rme^{-\imag(\omega+\omega')t'}&\\
\nonumber\times\left( 1 - \frac{\omega + \omega'}{\omega'} \frac{\xi(\omega',\boldsymbol{k})}{\epsilon(\omega',\boldsymbol{k})} \right)\frac{\xi(\omega,\boldsymbol{k})-1}{\epsilon(\omega,\boldsymbol{k})}&\,.
\end{align}

\subsection{Time Integration}
\begin{figure}
\centering
\includegraphics{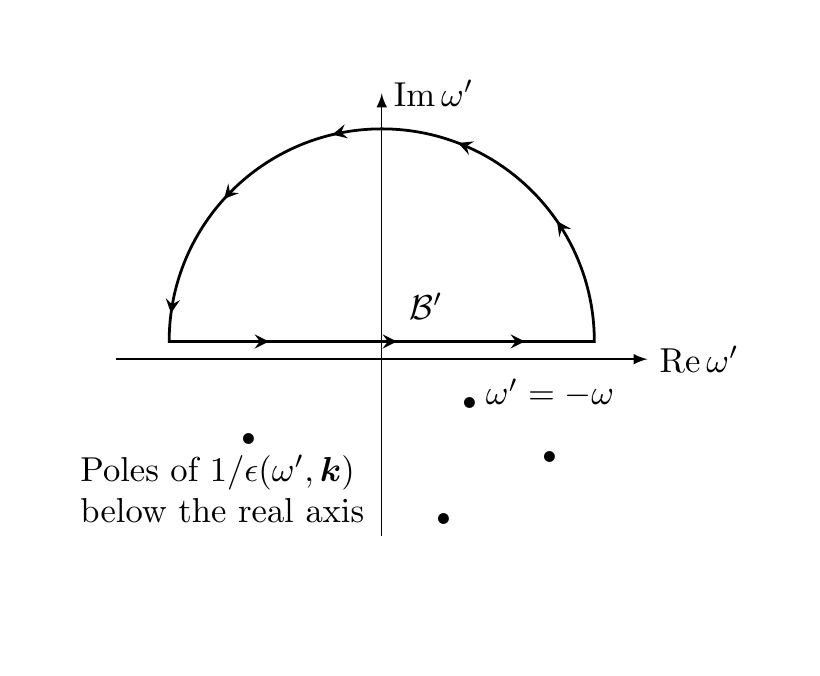}
\vspace*{-10mm}
\caption{Completing a contour in the upper half-plane.}
\label{contourA}
\end{figure}

We now take the time integral in $\rmd K_i^{(2)}/\rmd t$ to get
\begin{align}
\frac{\rmd K_i^{(2)}}{\rmd t} = \,T_{i0}k_{i0}^4\int \frac{\rmd\boldsymbol{k}}{(2\upi)^3}\frac{1}{(k^2+k_{e1}^2)^2}\int_\mathcal{B}\frac{\rmd\omega}{2\upi}\int_{\mathcal{B}'}\frac{\rmd\omega'}{2\upi} \frac{1 - \rme^{-\imag(\omega+\omega')t}}{\imag(\omega+\omega')}&\\
\nonumber\times\left( 1 - \frac{\omega + \omega'}{\omega'} \frac{\xi(\omega',\boldsymbol{k})}{\epsilon(\omega',\boldsymbol{k})} \right)\frac{\xi(\omega,\boldsymbol{k})-1}{\epsilon(\omega,\boldsymbol{k})}&\,.
\end{align}
This expression can be simplified. It is the sum of two terms,  $\rmd K_i^{(2a)}/\rmd t$ and  $\rmd K_i^{(2b)}/\rmd t$, defined by
\begin{align}
\frac{\rmd K_i^{(2a)}}{\rmd t} = \begin{aligned}[t]
{}T_{i0}k_{i0}^4\int \frac{\rmd\boldsymbol{k}}{(2\upi)^3}\frac{1}{(k^2+k_{e1}^2)^2}\int_\mathcal{B}\frac{\rmd\omega}{2\upi}\int_{\mathcal{B}'}\frac{\rmd\omega'}{2\upi} \frac{1}{\imag(\omega+\omega')}&\\
\times\left( 1 - \frac{\omega + \omega'}{\omega'} \frac{\xi(\omega',\boldsymbol{k})}{\epsilon(\omega',\boldsymbol{k})} \right)\frac{\xi(\omega,\boldsymbol{k})-1}{\epsilon(\omega,\boldsymbol{k})}&
\end{aligned}
\end{align}
and
\begin{align}
\label{eq:IK2}\frac{\rmd K_i^{(2b)}}{\rmd t} = \begin{aligned}[t]
{}-T_{i0}k_{i0}^4\int \frac{\rmd\boldsymbol{k}}{(2\upi)^3}\frac{1}{(k^2+k_{e1}^2)^2}\int_\mathcal{B}\frac{\rmd\omega}{2\upi}\int_{\mathcal{B}'}\frac{\rmd\omega'}{2\upi} \frac{\rme^{\imag(\omega+\omega')t}}{\imag(\omega+\omega')}\\
\times\left( 1 - \frac{\omega + \omega'}{\omega'} \frac{\xi(\omega',\boldsymbol{k})}{\epsilon(\omega',\boldsymbol{k})} \right)\frac{\xi(\omega,\boldsymbol{k})-1}{\epsilon(\omega,\boldsymbol{k})}\,.
\end{aligned}
\end{align}
We can show that $\rmd K_i^{(2a)}/\rmd t = 0$. The $\omega'$ integral can be completed using an arc in the upper half-plane as shown in figure \ref{contourA}. This arc does not contribute to the integral as its radius increases because $\epsilon(\omega',\boldsymbol{k})\sim 1$ and $\xi(\omega',\boldsymbol{k})\sim 1$ as $|\omega'|\to\infty$, according to \eqref{eq:elimit} and \eqref{eq:xilimit}, so
\begin{equation}
\frac{1}{\omega+\omega'}\left( 1 - \frac{\omega + \omega'}{\omega'} \frac{\xi(\omega',\boldsymbol{k})}{\epsilon(\omega',\boldsymbol{k})} \right) = O\left(\frac{1}{\omega'^2}\right).
\end{equation}
There is no pole enclosed, since $\omega' = -\omega$ never occurs when the contours $\mathcal{B}$ and $\mathcal{B}'$ both stay above the real axis. Therefore, by Cauchy's theorem, the integral for  $\rmd K_i^{(2a)}/\rmd t$ vanishes.
\begin{figure}
\centering
\includegraphics{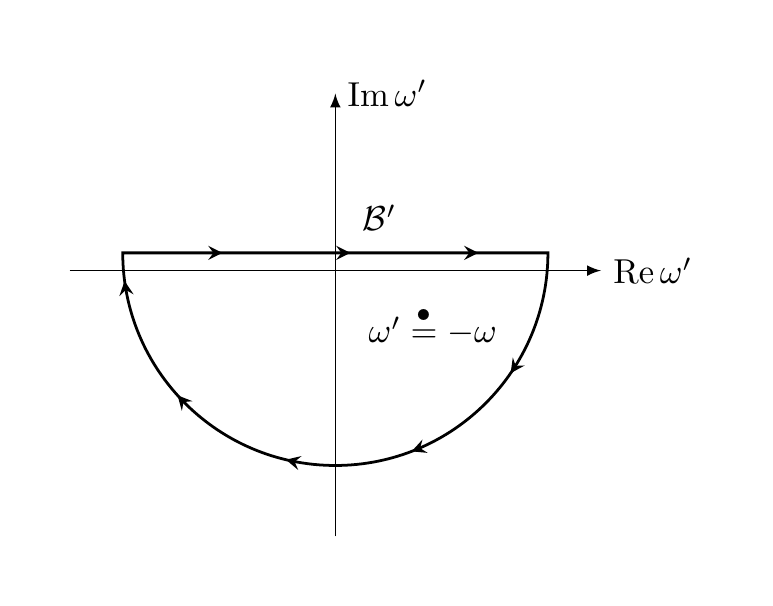}
\vspace*{-6mm}
\caption{Completing a contour in the lower half-plane.}
\label{contourB}
\end{figure}
Now we turn to  $\rmd K_i^{(2b)}/\rmd t$, which is written as a sum of two terms in \eqref{eq:IK2}. The first term is
\begin{align}\label{eq:Ki2zero}
-T_{i0}k_{i0}^4\int \frac{\rmd\boldsymbol{k}}{(2\upi)^3}\frac{1}{(k^2+k_{e1}^2)^2}\int_\mathcal{B}\frac{\rmd\omega}{2\upi}\int_{\mathcal{B}'}\frac{\rmd\omega'}{2\upi} \frac{\rme^{\imag(\omega+\omega')t}}{\imag(\omega+\omega')}\frac{\xi(\omega,\boldsymbol{k})-1}{\epsilon(\omega,\boldsymbol{k})}\,,
\end{align}
which is zero. To show this, we first complete the $\omega'$ contour in the lower half-plane, where the integrand is exponentially suppressed, so that a single pole is enclosed at $\omega' = -\omega$ (see figure \ref{contourB}). Then \eqref{eq:Ki2zero} becomes
\begin{align}
T_{i0}k_{i0}^4\int \frac{\rmd\boldsymbol{k}}{(2\upi)^3}\frac{1}{(k^2+k_{e1}^2)^2}\int_\mathcal{B}\frac{\rmd\omega}{2\upi}\frac{\xi(\omega,\boldsymbol{k})-1}{\epsilon(\omega,\boldsymbol{k})}
\end{align}
using the residue theorem. This equals zero because the 
 $\mathcal{B}$ contour can be completed using an arc in the upper half-plane, where there are no poles (similar to the contour in figure \ref{contourA}).
 
 Therefore, the nonzero part of $\rmd K_i^{(2)}/\rmd t$ is the remaining term in \eqref{eq:IK2},
\begin{align}\label{eq:IK22}
\frac{\rmd K_i^{(2)}}{\rmd t} = \,-\imag T_{i0}k_{i0}^4\int \frac{\rmd\boldsymbol{k}}{(2\upi)^3}\frac{1}{(k^2+k_{e1}^2)^2}\int_\mathcal{B}\frac{\rmd\omega}{2\upi}\int_{\mathcal{B}'}\frac{\rmd\omega'}{2\upi}\, \rme^{-\imag(\omega+\omega')t}&\\
\nonumber\times\frac{1}{\omega'}  \frac{\xi(\omega',\boldsymbol{k})}{\epsilon(\omega',\boldsymbol{k})}\frac{\xi(\omega,\boldsymbol{k})-1}{\epsilon(\omega,\boldsymbol{k})}&\,.
\end{align}

\subsection{Simplification}

\begin{figure}
\centering
\includegraphics{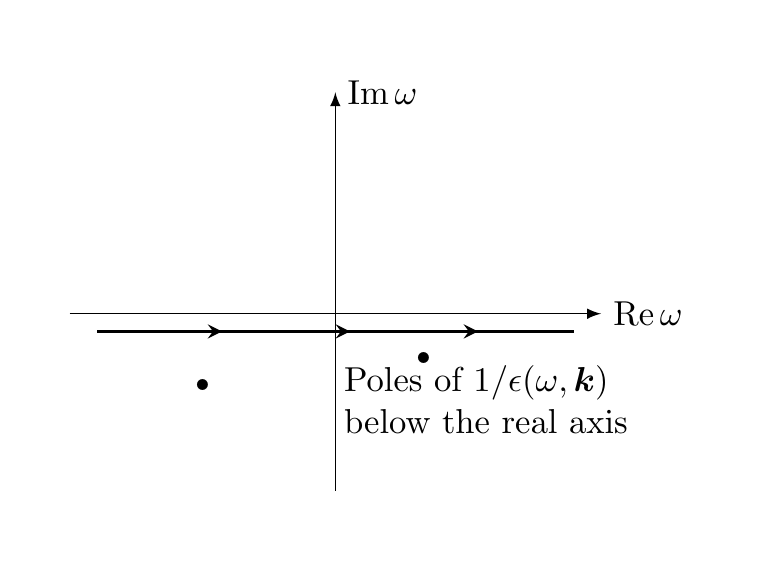}
\vspace*{-10mm}
\caption{Contours pushed below the real axis.}
\label{contourC}
\end{figure}

Using
\begin{equation}\label{eq:xiid}
1 - \frac{\xi(\omega,\boldsymbol{k})}{\epsilon(\omega,\boldsymbol{k})} = 1 - \frac{1}{\epsilon(\omega,\boldsymbol{k})} - \frac{k^2+k_{e1}^2}{k_{i0}^2}\left( \frac{1 - \epsilon(\omega,\boldsymbol{k})}{\epsilon(\omega,\boldsymbol{k})} \right) = - \frac{k^2+k_{D1}^2}{k_{i0}^2}\left( \frac{1 - \epsilon(\omega,\boldsymbol{k})}{\epsilon(\omega,\boldsymbol{k})} \right)
\end{equation}
in \eqref{eq:IK1}, we have
\begin{align}
    \frac{\rmd K_i^{(1)}}{\rmd t} = \imag T_{i0}k_{i0}^2\int \frac{\rmd\boldsymbol{k}}{(2\upi)^3} \frac{1}{k^2+k_{e1}^2}\frac{k^2+k_{D1}^2}{k^2+k_{D0}^2} \int_\mathcal{B} \frac{\rmd\omega}{2\upi}\int_{\mathcal{B}'}\frac{\rmd\omega'}{2\upi}\, \rme^{-\imag(\omega+\omega')t}&\\
    \nonumber \times\frac{1}{\omega'}\frac{\xi(\omega',\boldsymbol{k})}{\epsilon(\omega', \boldsymbol{k})} \frac{1 - \epsilon(\omega,\boldsymbol{k})}{\epsilon(\omega,\boldsymbol{k})}&\,.
\end{align}
Similarly, using
\begin{equation}
\xi(\omega,\boldsymbol{k})-1 = \left(\frac{k^2 + k_{e1}^2}{k_{i0}^2}\right) (1 - \epsilon(\omega ,\boldsymbol{k}))
\end{equation}
in \eqref{eq:IK22} leads to
 \begin{align}
\frac{\rmd K_i^{(2)}}{\rmd t} = \,\imag T_{i0}k_{i0}^2\int \frac{\rmd\boldsymbol{k}}{(2\upi)^3}\frac{1}{k^2+k_{e1}^2}\int_\mathcal{B}\frac{\rmd\omega}{2\upi}\int_{\mathcal{B}'}\frac{\rmd\omega'}{2\upi}\, \rme^{-\imag(\omega+\omega')t}&\\
\nonumber\times\frac{1}{\omega'}  \frac{\xi(\omega',\boldsymbol{k})}{\epsilon(\omega',\boldsymbol{k})}\frac{\epsilon(\omega,\boldsymbol{k})-1}{\epsilon(\omega,\boldsymbol{k})}&\,.
\end{align}
Now we have very similar expressions for $\rmd K_i^{(1)}/\rmd t$ and $\rmd K_i^{(2)}/\rmd t$. Combining them gives
\begin{align}
\nonumber\frac{\rmd K_i}{\rmd t} = \,\imag T _{i0}k_{i0}^2(k_{D0}^2-k_{D1}^2)\int \frac{\rmd\boldsymbol{k}}{(2\upi)^3}\frac{1}{k^2+k_{e1}^2}\frac{1}{k^2+k_{D0}^2}&\int_\mathcal{B}\frac{\rmd\omega}{2\upi}\int_{\mathcal{B}'}\frac{\rmd\omega'}{2\upi}\, \rme^{-\imag(\omega+\omega')t}\\
&\times\frac{1}{\omega'}  \frac{\xi(\omega',\boldsymbol{k})}{\epsilon(\omega',\boldsymbol{k})}\frac{\epsilon(\omega,\boldsymbol{k})-1}{\epsilon(\omega,\boldsymbol{k})}.
\end{align}
We can clearly see that, when there is no heating so that $k_{D0}=k_{D1}$, the two integrals cancel to give $\rmd K_i/\rmd t = 0$, as discussed in \S\ref{sec:timeevol}. Note also that the singularity at $\omega' = 0$ is removable, so we can push both contours $\mathcal{B}$ and $\mathcal{B}'$ below the real axis as shown in figure \ref{contourC}, provided they do not cross any of the poles of $1/\epsilon$.

\begin{figure}
\centering
\includegraphics{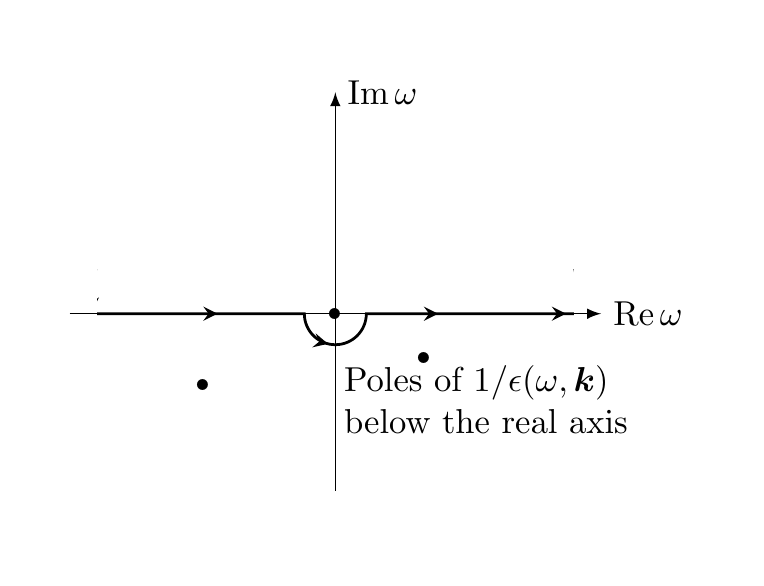}
\vspace*{-8mm}
\caption{Deformation of contour up to the real axis.}
\label{contourE}
\end{figure}

The next step is to use \eqref{eq:xiid} to eliminate $\xi$ and write the integral in terms of $\epsilon$ only, giving
\begin{align}
\frac{\rmd K_i}{\rmd t} = \,\imag T_{i0}k_{i0}^2(k_{D0}^2-k_{D1}^2)\int \frac{\rmd\boldsymbol{k}}{(2\upi)^3}\frac{1}{k^2+k_{e1}^2}\frac{1}{k^2+k_{D0}^2}\int_\mathcal{B}\frac{\rmd\omega}{2\upi}\int_{\mathcal{B}'}\frac{\rmd\omega'}{2\upi}\, \rme^{-\imag(\omega+\omega')t}&\\
\nonumber\times\frac{1}{\omega'}  \left( 1 + \frac{k^2 + k_{D1}^2}{k_{i0}^2} \frac{1-\epsilon(\omega',\boldsymbol{k})}{\epsilon(\omega',\boldsymbol{k})} \right)
\frac{\epsilon(\omega,\boldsymbol{k})-1}{\epsilon(\omega,\boldsymbol{k})}&\,.
\end{align}
We now use
\begin{equation}
\int_{\mathcal{B}'}\frac{\rmd\omega'}{2\upi}\, \rme^{-\imag(\omega+\omega')t}\frac{1}{\omega'} = 0\,,
\end{equation}
which is justified as follows. The contour can be closed in the lower half-plane, where the integrand is exponentially suppressed. It then encircles no poles, because we pushed $\mathcal{B}'$ below the real axis, so the integral vanishes. Therefore,
\begin{align}
\frac{\rmd K_i}{\rmd t} = \,-\imag T_{i0}(k_{D0}^2-k_{D1}^2)&\int \frac{\rmd\boldsymbol{k}}{(2\upi)^3}\frac{k^2 + k_{D1}^2}{(k^2+k_{e1}^2)(k^2+k_{D0}^2)}\\
\nonumber\times&\int_\mathcal{B}\frac{\rmd\omega}{2\upi}\int_{\mathcal{B}'}\frac{\rmd\omega'}{2\upi}\, \rme^{-\imag(\omega+\omega')t}
\frac{1}{\omega'}  \frac{\epsilon(\omega',\boldsymbol{k})-1}{\epsilon(\omega',\boldsymbol{k})} 
\frac{\epsilon(\omega,\boldsymbol{k})-1}{\epsilon(\omega,\boldsymbol{k})}\,.
\end{align}
This suggests we define a new function $\mathcal{F}(t,\boldsymbol{k})$ by
\begin{equation}
\mathcal{F}(t,\boldsymbol{k}) = \int_\mathcal{C}\frac{\rmd\omega}{2\upi}\, \rme^{-\imag\omega t}
\frac{1}{\omega}  \frac{\epsilon(\omega,\boldsymbol{k})-1}{\epsilon(\omega,\boldsymbol{k})}\,,
\end{equation}
where the contour $\mathcal{C}$ is below the real axis but passes above all the poles of $1/\epsilon$ as shown in figure \ref{contourD}. Then we have
\begin{align}
\label{eq:Fbump}\frac{\rmd K_i}{\rmd t} = \frac{\rmd}{\rmd t}\biggl[\frac{1}{2}&T_{i0}(k_{D0}^2-k_{D1}^2)\int \frac{\rmd\boldsymbol{k}}{(2\upi)^3}\frac{(k^2 + k_{D1}^2)\mathcal{F}(t,\boldsymbol{k})^2}{(k^2+k_{e1}^2)(k^2+k_{D0}^2)}\biggr].
\end{align}
Finally, we can prove that $\mathcal{F}(t,\boldsymbol{k})$ is purely imaginary. First, deform the contour $\mathcal{B}$ to the real axis as shown in figure \ref{contourE}, with a small bump around the pole at the origin. The small semicircular bump gives a contribution which is one half of the residue at $\omega=0$, while the straight portions of the contour give a principle value integral along the real line,
\begin{align}
\mathcal{F}(t,\boldsymbol{k}) =  \frac{\imag}{2}\left( \frac{k_{i0}^2}{k^2+k_{D1}^2} \right) + \dashint_{-\infty}^\infty \frac{\rmd\omega}{2\upi}\, \rme^{-\imag\omega t}
\frac{1}{\omega}  \frac{\epsilon(\omega,\boldsymbol{k})-1}{\epsilon(\omega,\boldsymbol{k})}\,,
\end{align}
where $\dashint_{-\infty}^\infty = \lim_{\epsilon\to 0}\int_{-\infty}^{-\epsilon} + \int_{\epsilon}^{\infty}$. The complex conjugate of this expression is
\begin{equation}
\mathcal{F}(t,\boldsymbol{k})^* = -\frac{\imag}{2}\left( \frac{k_{i0}^2}{k^2+k_{D1}^2} \right) + \dashint_{-\infty}^\infty \frac{\rmd\omega}{2\upi}\, \rme^{\imag\omega t}
\frac{1}{\omega}  \frac{\epsilon(\omega,\boldsymbol{k})^*-1}{\epsilon(\omega,\boldsymbol{k})^*}\,.
\end{equation}
Substituting $\omega\to-\omega$ and using $\epsilon(-\omega^*,\boldsymbol{k})^* = \epsilon(\omega,\boldsymbol{k})$, see \eqref{eq:epsomega}, we find
\begin{equation}
\mathcal{F}(t,\boldsymbol{k})^* = -\frac{\imag}{2}\left( \frac{k_{i0}^2}{k^2+k_{D1}^2} \right) - \dashint_{-\infty}^\infty \frac{\rmd\omega}{2\upi}\, \rme^{-\imag\omega t}
\frac{1}{\omega}  \frac{\epsilon(\omega,\boldsymbol{k})-1}{\epsilon(\omega,\boldsymbol{k})} = -\mathcal{F}(t,\boldsymbol{k})\,,
\end{equation}
so $\mathcal{F}(t,\boldsymbol{k})$ is imaginary as claimed.

The integral for the change in ion kinetic energy density may therefore be written compactly as
\begin{align}
\frac{\rmd K_i}{\rmd t} = -\frac{\rmd}{\rmd t}\biggl[\frac{1}{2}&T_{i0}(k_{D0}^2-k_{D1}^2)\int \frac{\rmd\boldsymbol{k}}{(2\upi)^3}\frac{(k^2 + k_{D1}^2)|\mathcal{F}(t,\boldsymbol{k})|^2}{(k^2+k_{e1}^2)(k^2+k_{D0}^2)}\biggr].
\end{align}

\section{Ion Distribution Integral}\label{app:integral2}

This appendix evaluates the integrals in \eqref{eq:delta1} and \eqref{eq:delta2}, repeated here, for the total change in the ion distribution function on the ion-plasma-frequency timescale:
\begin{align}
\delta f^{(1)}(\boldsymbol{v}_1) = \,&-\frac{T_{i0}k_{i0}^4}{n_i^2m_i}\int \frac{\rmd\boldsymbol{k}}{(2\upi)^3}\frac{1}{k^2+k_{D0}^2}\frac{1}{k^2+k_{e1}^2}\imag\boldsymbol{k}\boldsymbol{\cdot}\frac{\partial}{\partial \boldsymbol{v}_1}\int \rmd\boldsymbol{v}_3\rmd\boldsymbol{v}_4 \,f_i(\boldsymbol{v}_3)f_i(\boldsymbol{v}_4)\\
\nonumber&\times\int \rmd\boldsymbol{v}_2\, \int_0^\infty \rmd t\int_\mathcal{B}\frac{\rmd\omega}{2\upi}\int_{\mathcal{B}'}\frac{\rmd\omega'}{2\upi}\, \rme^{-\imag(\omega+\omega')t}\\
\nonumber&\times\left( \frac{\delta(\boldsymbol{v}_1-\boldsymbol{v}_3)}{\omega - \boldsymbol{k\cdot v}_1} + \frac{1}{n_i}\frac{k_{i0}^2}{k^2+k_{e1}^2}\frac{1}{\epsilon(\omega,\boldsymbol{k})}\frac{\boldsymbol{k\cdot v}_1}{\omega - \boldsymbol{k\cdot v}_1}\frac{1}{\omega - \boldsymbol{k\cdot v}_3} f_{i0}(\boldsymbol{v}_1)\right)\\
\nonumber&\times\left( \frac{\delta(\boldsymbol{v}_2-\boldsymbol{v}_4)}{\omega' + \boldsymbol{k\cdot v}_2} - \frac{1}{n_i}\frac{k_{i0}^2}{k^2+k_{e1}^2}\frac{1}{\epsilon(\omega',\boldsymbol{k})}\frac{\boldsymbol{k\cdot v}_2}{\omega' + \boldsymbol{k\cdot v}_2}\frac{1}{\omega' + \boldsymbol{k\cdot v}_4} f_{i0}(\boldsymbol{v}_2)\right),
\end{align}
\begin{align}
\delta f^{(2)}(\boldsymbol{v}_1) = \,&-\frac{T_{i0}k_{i0}^4}{n_i^2m_i}\int \frac{\rmd\boldsymbol{k}}{(2\upi)^3}\frac{1}{(k^2+k_{e1}^2)^2}\imag\boldsymbol{k}\boldsymbol{\cdot}\frac{\partial}{\partial \boldsymbol{v}_1}\int \rmd\boldsymbol{v}_3\rmd\boldsymbol{v}_4 \,f_i(\boldsymbol{v}_3)f_i(\boldsymbol{v}_4)\\
\nonumber&\times(\imag\boldsymbol{k}\boldsymbol{\cdot}(\boldsymbol{v}_3-\boldsymbol{v}_4))\int \rmd\boldsymbol{v}_2\,\int_0^\infty \rmd t \int_0^t \rmd t'\int_\mathcal{B}\frac{\rmd\omega}{2\upi}\int_{\mathcal{B}'}\frac{\rmd\omega'}{2\upi}\, \rme^{-\imag(\omega+\omega')t'}\\
\nonumber&\times\left( \frac{\delta(\boldsymbol{v}_1-\boldsymbol{v}_3)}{\omega - \boldsymbol{k\cdot v}_1} + \frac{1}{n_i}\frac{k_{i0}^2}{k^2+k_{e1}^2}\frac{1}{\epsilon(\omega,\boldsymbol{k})}\frac{\boldsymbol{k\cdot v}_1}{\omega - \boldsymbol{k\cdot v}_1}\frac{1}{\omega - \boldsymbol{k\cdot v}_3} f_{i0}(\boldsymbol{v}_1)\right)\\
\nonumber&\times\left( \frac{\delta(\boldsymbol{v}_2-\boldsymbol{v}_4)}{\omega' + \boldsymbol{k\cdot v}_2} - \frac{1}{n_i}\frac{k_{i0}^2}{k^2+k_{e1}^2}\frac{1}{\epsilon(\omega',\boldsymbol{k})}\frac{\boldsymbol{k\cdot v}_2}{\omega' + \boldsymbol{k\cdot v}_2}\frac{1}{\omega' + \boldsymbol{k\cdot v}_4} f_{i0}(\boldsymbol{v}_2)\right).
\end{align}
These integrals must cancel when there is no heating, so that $k_{D0} = k_{D1}$. However, $\delta f^{(2)}(\boldsymbol{v})$ does not depend on $k_{D0}$. Therefore, replacing $k_{D0}$ with $k_{D1}$ in $\delta f^{(1)}(\boldsymbol{v})$ gives an expression that must equal $-\delta f^{(2)}(\boldsymbol{v})$. Here, unlike in appendix \ref{app:integral}, we use this relationship to determine $\delta f^{(2)}(\boldsymbol{v})$ from $\delta f^{(1)}(\boldsymbol{v})$ instead of calculating both integrals explicitly.

As in appendix \ref{app:integral}, we evaluate the multiple integrals in order.

\subsection{$\boldsymbol{v}_2$ Integral}

We first perform the $\boldsymbol{v}_2$ integration using \eqref{eq:v2integral}, getting
\begin{align}
\delta f^{(1)}(\boldsymbol{v}_1) = \,&-\frac{T_{i0}k_{i0}^4}{n_i^2m_i}\int \frac{\rmd\boldsymbol{k}}{(2\upi)^3}\frac{1}{k^2+k_{D0}^2}\frac{1}{k^2+k_{e1}^2}\imag\boldsymbol{k}\boldsymbol{\cdot}\frac{\partial}{\partial \boldsymbol{v}_1}\int \rmd\boldsymbol{v}_3\rmd\boldsymbol{v}_4 \,f_i(\boldsymbol{v}_3)f_i(\boldsymbol{v}_4)\\
\nonumber&\times\int_0^\infty \rmd t\int_\mathcal{B}\frac{\rmd\omega}{2\upi}\int_{\mathcal{B}'}\frac{\rmd\omega'}{2\upi}\, \rme^{-\imag(\omega+\omega')t}\,\frac{1}{\epsilon(\omega', \boldsymbol{k})}\frac{1}{\omega' + \boldsymbol{k}\cdot \boldsymbol{v}_4}\\
\nonumber&\times\left( \frac{\delta(\boldsymbol{v}_1-\boldsymbol{v}_3)}{\omega - \boldsymbol{k\cdot v}_1} + \frac{1}{n_i}\frac{k_{i0}^2}{k^2+k_{e1}^2}\frac{1}{\epsilon(\omega,\boldsymbol{k})}\frac{\boldsymbol{k\cdot v}_1}{\omega - \boldsymbol{k\cdot v}_1}\frac{1}{\omega - \boldsymbol{k\cdot v}_3} f_{i0}(\boldsymbol{v}_1)\right).
\end{align}

\subsection{$\boldsymbol{v}_4$ Integral}

Next we evaluate the $\boldsymbol{v}_4$ integral using \eqref{eq:J1}:
\begin{align}
\delta f^{(1)}(\boldsymbol{v}_1) = \,&-\frac{T_{i0}k_{i0}^4}{n_im_i}\int \frac{\rmd\boldsymbol{k}}{(2\upi)^3}\frac{1}{k^2+k_{D0}^2}\frac{1}{k^2+k_{e1}^2}\imag\boldsymbol{k}\boldsymbol{\cdot}\frac{\partial}{\partial \boldsymbol{v}_1}\int \rmd\boldsymbol{v}_3 \,f_i(\boldsymbol{v}_3)\\
\nonumber&\times\int_0^\infty \rmd t\int_\mathcal{B}\frac{\rmd\omega}{2\upi}\int_{\mathcal{B}'}\frac{\rmd\omega'}{2\upi}\, \rme^{-\imag(\omega+\omega')t}\,\frac{1}{\omega'}\frac{\xi(\omega',\boldsymbol{k})}{\epsilon(\omega', \boldsymbol{k})}\\
\nonumber&\times\left( \frac{\delta(\boldsymbol{v}_1-\boldsymbol{v}_3)}{\omega - \boldsymbol{k\cdot v}_1} + \frac{1}{n_i}\frac{k_{i0}^2}{k^2+k_{e1}^2}\frac{1}{\epsilon(\omega,\boldsymbol{k})}\frac{\boldsymbol{k\cdot v}_1}{\omega - \boldsymbol{k\cdot v}_1}\frac{1}{\omega - \boldsymbol{k\cdot v}_3} f_{i0}(\boldsymbol{v}_1)\right).
\end{align}

\subsection{$\boldsymbol{v}_3$ Integral}

The $\boldsymbol{v}_3$ integral also requires \eqref{eq:J1}. We find
\begin{align}
\delta f^{(1)}(\boldsymbol{v}_1) = \,&-\frac{T_{i0}k_{i0}^4}{nim_i}\int \frac{\rmd\boldsymbol{k}}{(2\upi)^3}\frac{1}{k^2+k_{D0}^2}\frac{1}{k^2+k_{e1}^2}\imag\boldsymbol{k}\boldsymbol{\cdot}\frac{\partial}{\partial \boldsymbol{v}_1}\\
\nonumber&\times\int_0^\infty \rmd t\int_\mathcal{B}\frac{\rmd\omega}{2\upi}\int_{\mathcal{B}'}\frac{\rmd\omega'}{2\upi}\, \rme^{-\imag(\omega+\omega')t}\frac{1}{\omega'}\frac{\xi(\omega',\boldsymbol{k})}{\epsilon(\omega', \boldsymbol{k})}\\
\nonumber&\times\left( 1 + \frac{k_{i0}^2}{k^2+k_{e1}^2}\frac{\boldsymbol{k\cdot v}_1}{\omega}\frac{\xi(\omega,\boldsymbol{k})}{\epsilon(\omega,\boldsymbol{k})} \right)\frac{1}{\omega - \boldsymbol{k\cdot v}_1}f_{i0}(\boldsymbol{v}_1)&\,.
\end{align}

\subsection{Time Integral}

The Laplace-inversion contours $\mathcal{B}$ and $\mathcal{B}'$ have to pass above all poles of the integrand. Initially, there were ballistic poles on the real axis for $\omega$ and $\omega'$, so both contours had to remain at least partially above the real axis. Now, performing the velocity integrals has left an integrand that has only one ballistic pole at $\omega = \boldsymbol{k\cdot v}_1$, so the $\mathcal{B}'$ contour can be pushed below the real axis.

Suppose we deform the contours so that $\mathcal{B}$ is a horizontal line just above the real axis, while $\mathcal{B}'$ is a horizontal line that is above all the poles of $1/\epsilon$ but far enough below the real axis that $\omega + \omega'$ has a negative imaginary part. This allows us to take the time integral because the exponential now decays as $t\to \infty$. The result is
\begin{align}
\delta f^{(1)}(\boldsymbol{v}_1) = \,-\frac{T_{i0}k_{i0}^4}{n_im_i}\int \frac{\rmd\boldsymbol{k}}{(2\upi)^3}\frac{1}{k^2+k_{D0}^2}\frac{1}{k^2+k_{e1}^2}\imag\boldsymbol{k}\boldsymbol{\cdot}\frac{\partial}{\partial \boldsymbol{v}_1}\int_\mathcal{B}\frac{\rmd\omega}{2\upi}\int_{\mathcal{B}'}\frac{\rmd\omega'}{2\upi} \frac{1}{\imag(\omega + \omega')}&\\
\nonumber\times\frac{1}{\omega'}\frac{\xi(\omega',\boldsymbol{k})}{\epsilon(\omega', \boldsymbol{k})}\left( 1 + \frac{k_{i0}^2}{k^2+k_{e1}^2}\frac{\boldsymbol{k\cdot v}_1}{\omega}\frac{\xi(\omega,\boldsymbol{k})}{\epsilon(\omega,\boldsymbol{k})} \right)\frac{1}{\omega - \boldsymbol{k\cdot v}_1}f_{i0}(\boldsymbol{v}_1)&\,.
\end{align}

\subsection{$\omega$ Integral}

Now close the $\omega$ contour using a large arc in the upper half-plane, which does not contribute to the integral because the integrand decays like $1/\omega^2$ as $|\omega|\to\infty$. The only pole enclosed is at $\omega = -\omega'$, so the residue theorem gives
\begin{align}
\delta f^{(1)}(\boldsymbol{v}_1) = \,\frac{T_{i0}k_{i0}^4}{nim_i}\int \frac{\rmd\boldsymbol{k}}{(2\upi)^3}\frac{1}{k^2+k_{D0}^2}\frac{1}{k^2+k_{e1}^2}\imag\boldsymbol{k}\boldsymbol{\cdot}\frac{\partial}{\partial \boldsymbol{v}_1}\int_{\mathcal{B}'}\frac{\rmd\omega'}{2\upi}\frac{1}{\omega'}\frac{\xi(\omega',\boldsymbol{k})}{\epsilon(\omega', \boldsymbol{k})}&\\
\nonumber\times\left( 1 - \frac{k_{i0}^2}{k^2+k_{e1}^2}\frac{\boldsymbol{k\cdot v}_1}{\omega'}\frac{\xi(-\omega',\boldsymbol{k})}{\epsilon(-\omega',\boldsymbol{k})} \right)\frac{1}{\omega' + \boldsymbol{k\cdot v}_1}f_{i0}(\boldsymbol{v}_1)&\,.
\end{align}

\subsection{$\omega'$ Integral}

We now have a sum of two terms $\delta f^{(1a)}(\boldsymbol{v})$ and $\delta f^{(1b)}(\boldsymbol{v})$ defined by
\begin{align}
\delta f^{(1a)}(\boldsymbol{v}_1) &= \begin{aligned}[t]
{}\frac{T_{i0}k_{i0}^4}{n_im_i}&\int \frac{\rmd\boldsymbol{k}}{(2\upi)^3}\frac{1}{k^2+k_{D0}^2}\frac{1}{k^2+k_{e1}^2}\imag\boldsymbol{k}\boldsymbol{\cdot}\frac{\partial}{\partial \boldsymbol{v}_1}\\
\times&\int_{\mathcal{B}'}\frac{\rmd\omega'}{2\upi}\frac{1}{\omega'}\frac{\xi(\omega',\boldsymbol{k})}{\epsilon(\omega', \boldsymbol{k})}\frac{1}{\omega' + \boldsymbol{k\cdot v}_1}f_{i0}(\boldsymbol{v}_1)\,,
\end{aligned}\\
\delta f^{(1b)}(\boldsymbol{v}_1) &= \begin{aligned}[t]
{}-\frac{T_{i0}k_{i0}^6}{n_im_i}&\int \frac{\rmd\boldsymbol{k}}{(2\upi)^3}\frac{1}{k^2+k_{D0}^2}\frac{1}{(k^2+k_{e1}^2)^2}\,\imag\boldsymbol{k}\boldsymbol{\cdot}\frac{\partial}{\partial \boldsymbol{v}_1}\\
\times&\int_{\mathcal{B}'}\frac{\rmd\omega'}{2\upi}\frac{1}{\omega'^2}\frac{\xi(\omega',\boldsymbol{k})}{\epsilon(\omega', \boldsymbol{k})}\frac{\xi(-\omega',\boldsymbol{k})}{\epsilon(-\omega',\boldsymbol{k})} \frac{\boldsymbol{k\cdot v}_1}{\omega' + \boldsymbol{k\cdot v}_1}f_{i0}(\boldsymbol{v}_1)\,.
\end{aligned}
\end{align}
In $\delta f^{(1a)}(\boldsymbol{v})$ we close the $\omega'$ contour using a large arc in the upper half-plane as in figure \ref{contourA}, which does not contribute to the integral since the integrand decays like $1/\omega'^2$ as $|\omega'|\to\infty$. Once again, applying the residue theorem gives
\begin{align}
\delta f^{(1a)}(\boldsymbol{v}_1) = \frac{T_{i0}k_{i0}^4}{n_im_i}\int \frac{\rmd\boldsymbol{k}}{(2\upi)^3}\frac{1}{k^2+k_{D0}^2}\frac{1}{k^2+k_{e1}^2}\boldsymbol{k\,\cdot }\frac{\partial}{\partial \boldsymbol{v}_1}&\\
\times\left[\frac{\xi(-\boldsymbol{k\cdot v}_1,\boldsymbol{k})}{\epsilon(-\boldsymbol{k\cdot v}_1, \boldsymbol{k})}\frac{1}{\boldsymbol{k\cdot v}_1}f_{i0}(\boldsymbol{v}_1)\right]&,
\end{align}
since the only pole enclosed is at $\omega' = -\boldsymbol{k}\cdot\boldsymbol{v}_1$. 

In $\delta f^{(1b)}(\boldsymbol{v})$, we deform the contour $\mathcal{B}'$ up to the real axis, with a small bump below the pole at $\omega' = -\boldsymbol{k}\cdot\boldsymbol{v}_1$, similar to the contour in figure \ref{contourE}. The straight portion of the contour gives a principal value integral along the real axis, as in \eqref{eq:Fbump}, while the small semicircular bump gives a contribution that is one half of the residue at $\omega' = -\boldsymbol{k}\cdot\boldsymbol{v}_1$. So, we have
\begin{align}
\delta f^{(1b)}(\boldsymbol{v}_1) = \,&-\frac{T_{i0}k_{i0}^6}{n_im_i}\int \frac{\rmd\boldsymbol{k}}{(2\upi)^3}\frac{1}{k^2+k_{D0}^2}\frac{1}{(k^2+k_{e1}^2)^2}\boldsymbol{k\,\cdot }\frac{\partial}{\partial \boldsymbol{v}_1}\\
&\nonumber\phantom{\;\frac{T_{i0}k_{i0}^6}{n_im_i}}\times\dashint_{-\infty}^\infty\,\frac{\rmd\omega'}{2\upi}\frac{1}{\omega'^2}\frac{\xi(\omega',\boldsymbol{k})}{\epsilon(\omega', \boldsymbol{k})}\frac{\xi(-\omega',\boldsymbol{k})}{\epsilon(-\omega',\boldsymbol{k})} \frac{\boldsymbol{k\cdot v}_1}{\omega' + \boldsymbol{k\cdot v}_1}f_{i0}(\boldsymbol{v}_1)\\
&\nonumber+ \frac{T_{i0}k_{i0}^6}{2m_i}\int \frac{\rmd\boldsymbol{k}}{(2\upi)^3}\frac{1}{k^2+k_{D0}^2}\frac{1}{(k^2+k_{e1}^2)^2}\,\boldsymbol{k\,\cdot }\frac{\partial}{\partial \boldsymbol{v}_1}\\
\nonumber&\phantom{l\frac{T_{i0}k_{i0}^6}{n_im_i}}\times\left[\frac{\xi(-\boldsymbol{k}\cdot\boldsymbol{v}_1,\boldsymbol{k})}{\epsilon(-\boldsymbol{k}\cdot\boldsymbol{v}_1, \boldsymbol{k})}\frac{\xi(\boldsymbol{k}\cdot\boldsymbol{v}_1,\boldsymbol{k})}{\epsilon(\boldsymbol{k}\cdot\boldsymbol{v}_1,\boldsymbol{k})}\frac{1}{\boldsymbol{k}\cdot\boldsymbol{v}_1}f_{i0}(\boldsymbol{v}_1)\right].
\end{align}
The first term vanishes because the integral changes sign under $\boldsymbol{k}\to -\boldsymbol{k}$ and $\omega' \to -\omega'$. Therefore, 
\begin{align}
\delta f^{(1b)}(\boldsymbol{v}_1) =  \frac{T_{i0}k_{i0}^6}{2n_im_i}\int \frac{\rmd\boldsymbol{k}}{(2\upi)^3}\frac{1}{k^2+k_{D0}^2}\frac{1}{(k^2+k_{e1}^2)^2}\boldsymbol{k\,\cdot }\frac{\partial}{\partial \boldsymbol{v}_1}&\\
\nonumber\times\left[\frac{\xi(-\boldsymbol{k}\cdot\boldsymbol{v}_1,\boldsymbol{k})}{\epsilon(-\boldsymbol{k}\cdot\boldsymbol{v}_1,\boldsymbol{k})}\frac{\xi(\boldsymbol{k}\cdot\boldsymbol{v}_1,\boldsymbol{k})}{\epsilon(\boldsymbol{k}\cdot\boldsymbol{v}_1,\boldsymbol{k})}\frac{1}{\boldsymbol{k}\cdot\boldsymbol{v}_1}f_{i0}(\boldsymbol{v}_1)\right]&.
\end{align}
Now we combine $\delta f^{(1a)}(\boldsymbol{v})$ and $\delta f^{(1b)}(\boldsymbol{v})$ again. 
\begin{align}
\delta f^{(1)}(\boldsymbol{v}_1) = \frac{T_{i0}k_{i0}^6}{2n_im_i}&\int \frac{\rmd\boldsymbol{k}}{(2\upi)^3}\frac{1}{k^2+k_{D0}^2}\frac{1}{(k^2+k_{e1}^2)^2}\boldsymbol{k\,\cdot }\frac{\partial}{\partial \boldsymbol{v}_1}\\
\nonumber&\times \biggl[\biggl(\frac{\xi(-\boldsymbol{k}\cdot\boldsymbol{v}_1,\boldsymbol{k})}{\epsilon(-\boldsymbol{k}\cdot\boldsymbol{v}_1, \boldsymbol{k})}\frac{\xi(\boldsymbol{k}\cdot\boldsymbol{v}_1,\boldsymbol{k})}{\epsilon(\boldsymbol{k}\cdot\boldsymbol{v}_1,\boldsymbol{k})} + \frac{k^2+k_{e1}^2}{k_{i0}^2}\frac{\xi(\boldsymbol{k}\cdot\boldsymbol{v}_1,\boldsymbol{k})}{\epsilon(\boldsymbol{k}\cdot\boldsymbol{v}_1,\boldsymbol{k})}\\
\nonumber &\phantom{blahblahblah}+ \frac{k^2+k_{e1}^2}{k_{i0}^2}\frac{\xi(-\boldsymbol{k}\cdot\boldsymbol{v}_1,\boldsymbol{k})}{\epsilon(-\boldsymbol{k}\cdot\boldsymbol{v}_1,\boldsymbol{k})}\biggr)\frac{1}{\boldsymbol{k}\cdot\boldsymbol{v}_1}f_{i0}(\boldsymbol{v}_1)\biggr].
\end{align}
Completing the square using ${\epsilon(-\boldsymbol{k\cdot v}, \boldsymbol{k}) = \epsilon(\boldsymbol{k\cdot v}, \boldsymbol{k})^*}$ and ${\xi(-\boldsymbol{k\cdot v}, \boldsymbol{k}) = \xi(\boldsymbol{k\cdot v}, \boldsymbol{k})^*}$, see \eqref{eq:epsomega} and \eqref{eq:xiomega}, gives
\begin{align}
\delta f^{(1)}(\boldsymbol{v}_1) = \frac{T_{i0}k_{i0}^6}{2n_im_i}&\int \frac{\rmd\boldsymbol{k}}{(2\upi)^3}\frac{1}{k^2+k_{D0}^2}\frac{1}{(k^2+k_{e1}^2)^2}\boldsymbol{k\,\cdot }\frac{\partial}{\partial \boldsymbol{v}_1}\\
\nonumber\times&\left[\left(\biggl|\frac{\xi(\boldsymbol{k}\cdot\boldsymbol{v}_1,\boldsymbol{k})}{\epsilon(\boldsymbol{k}\cdot\boldsymbol{v}_1, \boldsymbol{k})} + \frac{k^2+k_{e1}^2}{k_{i0}^2}\biggr|^2 - \left(\frac{k^2+k_{e1}^2}{k_{i0}^2}\right)^2\right)\frac{1}{\boldsymbol{k}\cdot\boldsymbol{v}_1}f_{i0}(\boldsymbol{v}_1)\right].
\end{align}
Then, using \eqref{eq:xiid} to eliminate the function $\xi$ and express the integral using $\epsilon$ only, we find
\begin{align}
\delta f^{(1)}(\boldsymbol{v}_1) = \frac{T_{i0}k_{i0}^2}{2n_im_i}&\int \frac{\rmd\boldsymbol{k}}{(2\upi)^3}\frac{1}{k^2+k_{D0}^2}\frac{1}{(k^2+k_{e1}^2)^2}\boldsymbol{k\,\cdot }\frac{\partial}{\partial \boldsymbol{v}_1}\\
\nonumber\times&\left[\left(\biggl|\frac{k^2+k_{D1}^2}{\epsilon(\boldsymbol{k}\cdot\boldsymbol{v}_1, \boldsymbol{k})}\biggr|^2 - \left(k^2+k_{e1}^2\right)^2\right)\frac{1}{\boldsymbol{k}\cdot\boldsymbol{v}_1}f_{i0}(\boldsymbol{v}_1)\right].
\end{align}
Taking the velocity derivative outside the integral, and using \eqref{eq:eps0}, we can write this as:
\begin{align}
\delta f^{(1)}(\boldsymbol{v}_1) = \frac{T_{i0}k_{i0}^2}{2n_im_i}\frac{\partial}{\partial \boldsymbol{v}_1}\boldsymbol{\cdot}\left[\int \frac{\rmd\boldsymbol{k}}{(2\upi)^3}\frac{|\epsilon(0,\boldsymbol{k})/\epsilon(\boldsymbol{k}\cdot\boldsymbol{v}, \boldsymbol{k})|^2-1}{k^2+k_{D0}^2}\frac{\boldsymbol{k}}{\boldsymbol{k}\cdot\boldsymbol{v}_1}f_{i0}(\boldsymbol{v}_1)\right].
\end{align}
The integral is now a vector which must point along $\boldsymbol{v}_1$, since there is no preferred direction in velocity space. Therefore, it must unchanged if it is projected along $\boldsymbol{v}_1$:
\begin{align}
\delta f^{(1)}(\boldsymbol{v}_1) = \frac{T_{i0}k_{i0}^2}{2n_im_i}\frac{\partial}{\partial \boldsymbol{v}_1}\boldsymbol{\cdot}\left[\frac{\boldsymbol{v}_1f_{i0}(\boldsymbol{v}_1)}{\boldsymbol{v}_1^2}\int \frac{\rmd\boldsymbol{k}}{(2\upi)^3}\frac{|\epsilon(0,\boldsymbol{k})/\epsilon(\boldsymbol{k}\cdot\boldsymbol{v}, \boldsymbol{k})|^2-1}{k^2+k_{D0}^2}\right].
\end{align}
Finally, by replacing $k_{D0}$ with $k_{D1}$ and multiplying by $-1$ we obtain $\delta f^{(2)}(\boldsymbol{v})$ as discussed previously. Therefore, using $\delta f(\boldsymbol{v}) = \delta f^{(1)}(\boldsymbol{v}) + \delta f^{(2)}(\boldsymbol{v})$, we find
\begin{align}
\delta f(\boldsymbol{v}) = \frac{T_{i0}k_{i0}^2}{2n_im_i}(k_{D0}^2-k_{D1}^2)\frac{\partial}{\partial \boldsymbol{v}}\boldsymbol{\cdot}\left[\frac{\boldsymbol{v}f_{i0}(\boldsymbol{v})}{\boldsymbol{v}^2}\int \frac{\rmd\boldsymbol{k}}{(2\upi)^3}\frac{1-|\epsilon(0,\boldsymbol{k})/\epsilon(\boldsymbol{k}\cdot\boldsymbol{v}, \boldsymbol{k})|^2}{(k^2+k_{D0}^2)(k^2+k_{D1}^2)}\right].
\end{align}

\bibliographystyle{jpp}
\bibliography{correlationheating}
\end{document}